\documentclass[12pt,a4paper,final]{iopart}

\usepackage{iopams}  
\usepackage{graphicx}
\usepackage[breaklinks=true,colorlinks=true,linkcolor=blue,urlcolor=blue,citecolor=blue]{hyperref}

\usepackage{braket}
\expandafter\let\csname equation*\endcsname\relax
\expandafter\let\csname endequation*\endcsname\relax
\usepackage{amsmath,amsfonts,amssymb,mathrsfs }
\usepackage[toc,page]{appendix}
\usepackage{varwidth}
\allowdisplaybreaks
\usepackage[usenames]{color}
\usepackage{multirow}
\usepackage{array}
\usepackage{physics}
\usepackage[utf8]{inputenc}
\usepackage{float,diagbox}
\usepackage{varwidth}
\usepackage{subcaption}
\usepackage{makecell}

\begin{document}

\newcommand{\RW}[1]{{\color{green}[RW: #1]}}
\def\tA{\tilde{A}}
\def\bG{\mathbb{G}}
\newcommand{\hA}{\mathcal{H}_A}
\newcommand{\hB}{\mathcal{H}_B}
\newcommand{\hlA}{\mathcal{H}_{\tilde A}}
\newcommand{\tlA}{\tilde{A}}
\newcommand{\tlB}{\tilde{B}}
\newcommand{\tlC}{\tilde{C}}
\newcommand{\tla}{\tilde{a}}
\newcommand{\bIB}{\mathbb{I}_B}
\newcommand{\bIA}{\mathbb{I}_A}
\newcommand{\negaAB}{negativity susceptibility }
\newcommand{\negaABshort}{S}
\newcommand{\ot}{\otimes}
\newcommand*{\dt}[1]{
  \accentset{\mbox{\large\bfseries .}}{#1}}
\newcommand*{\ddt}[1]{
  \accentset{\mbox{\large\bfseries .\hspace{-0.25ex}.}}{#1}}

\def\pra{Phys. Rev. A}
\def\prl{Phys. Rev. Lett.}

\title[The Transfer of Entanglement Negativity at the Onset of  Interactions]{The Transfer of Entanglement Negativity at the Onset of Interactions}

\author{Robin Yunfei Wen$^{1,2}$ and Achim Kempf$^{1,3,4}$} 

\address{$^1$Department of Applied Mathematics, University of Waterloo, Waterloo, Ontario, N2L 3G1, Canada}

\address{$^2$California Institute of Technology, 1200 E. California Boulevard, Pasadena, CA 91125, USA}

\address{$^3$Department of Physics, University of Waterloo, Waterloo, Ontario, N2L 3G1, Canada}

\address{$^4$Institute for Quantum Computing, University of Waterloo, Waterloo, Ontario, N2L 3G1, Canada}

\ead{y52wen@uwaterloo.ca}

\begin{abstract}
Quantum information, in the form of entanglement with an ancilla, can be transmitted  to a third system through interaction. Here, we investigate this process of entanglement transmission perturbatively in time. Using the entanglement monotone negativity, we determine how the proclivity of an interaction to either generate, transfer or lose entanglement depends on the choice of Hamiltonians and initial states. These three proclivities are captured by Hamiltonian- and state-dependent quantities that we call negativity susceptibility, negativity transmissibility and negativity vulnerability respectively. These notions could serve, for example, as cost functions in quantum technologies such as machine-learned quantum error correction. 
\end{abstract}

\tableofcontents
\markboth{The Transfer of Entanglement Negativity at the Onset of  Interactions}{}

\section{Introduction}
\subsection{Motivation}
A key consequence of the existence of entanglement is that quantum information can be stored not only locally within individual quantum systems but also non-locally among quantum systems. This phenomenon is ubiquitous since, for large composite quantum systems, most of the Hilbert space consists of almost maximally entangled states \cite{tqs}, i.e., states whose information content is encoded in this sense nonlocally.

In the present paper, we investigate the transfer of such nonlocal quantum information, i.e., of entanglement, at the onset of interactions. 
To this end, we consider a system $A$ that is entangled with and purified by an ancilla  system $\tilde{A}$. We then let $A$ interact with a system $B$. Throughout this interaction, $\tilde{A}$ always maintains its entanglement with the total system $AB$, but $\tilde{A}$ may lose entanglement with $A$ while gaining entanglement with $B$.  
We ask how this entanglement transfer depends on the states and Hamiltonians involved. 

We are, therefore, considering the interaction between two systems, $A$ and $B$, as creating a quantum channel. The channel established by the interaction has a quantum channel capacity, a notion that has no classical analog \cite{97LloydQCC,11Wilde,18GyongyosiQCCReview,17CuevasEDQCC}. Interactions do of course also possess classical channel capacity \cite{1948ShannonCommunication,06CoverThomasInformationTheory}, but we here exclusively focus on the ability of interactions to transfer entanglement. 

The study of the transfer of quantum information in interactions is of both fundamental and practical importance. 
On the fundamental level, each interaction Hamiltonian in the standard model can be viewed as providing a means for the transfer of entanglement and, therefore, for the transfer of delocalized quantum information. The study of how the various fundamental interactions transfer entanglement can therefore provide insights into how not only classical but also quantum information flows in nature. Work in this direction may even yield a completely information-theoretic description of the fundamental interactions. 

The transfer of quantum information in interactions is also of practical importance for quantum technologies, such
as quantum computation \cite{10NielsenChuangText,2010LaddQCReview,17BoyerDQC1,18KlcoDigitizedSFQC}, quantum communication \cite{10YuanPhotonQCCommR,2021ChenQComm,18StephanieQI} and quantum sensing \cite{17DegenQS,17RosskpfQS,16HuntemannIonClock}, where fined-tuned control over the transfer of entanglement in interactions is crucial \cite{10DongQControll,17Gonzalez-Henao}. For example, it is likely to be significantly more feasible to build small rather than large error-corrected quantum processors. In this case, it would be desirable to be able to link a large number of small quantum processors. 
These modules could then perform as one large processor - if entanglement can be transferred and thereby spread among the modules in a controlled way. The capability of controlled entanglement transfer would be necessary in order to be able to access all of the high-dimensional Hilbert space obtained by tensoring the Hilbert spaces of the modules.  

In prior work,  \cite{20EmilyFirstPaper,21EmilySecondPaper}, we investigated the transfer of entanglement, i.e., the dynamics of the quantum channel capacity, by studying the dynamics of coherent information. Use of the notion of coherent information is advantageous because coherent information is the basis for the conventional definition of quantum channel capacity \cite{11Wilde,18GyongyosiQCCReview}. However, the notion of coherent information  has the drawback that an individual coherent information value is hard to interpret since it is not an entanglement monotone. Also, the quantum channel capacity is obtained by an elaborate optimization over coherent information and this optimization tends to be infeasible most circumstances \cite{11Wilde,18GyongyosiQCCReview}. 

In the present work, we therefore  quantify the transfer of entanglement at the onset of interactions by using the notion of entanglement negativity \cite{02EMNegativity} rather than the notion of coherent information. The negativity has the advantage of being calculable in practice and it possesses a direct interpretation as an entanglement monotone. 
While our studies here are mostly perturbative, we also address some non-perturbative phenomena. In particular, given the existence of the phenomenon of sudden death of entanglement \cite{09YuSuddenDeathEntanglement}, we will here address the related question in which circumstances at the onset of an interaction there can be a finite wait time for entanglement transmission.

\subsection{Organization of this Paper}
We introduce our system setup in Sec.~\ref{sec:system setup} and summarize the key features for the entanglement dynamics of our system in Table~\ref{tab:dynamics-entanglement-summary} and \ref{tab:tlA-B-summary} of Sec.~\ref{sec:summary}. In Sec.~\ref{sec:N and P}, we review the negativity as an entanglement measure and explain the framework for our perturbative calculations. In Sec.~\ref{sec:mixed, finite time} and \ref{sec:pure, perturbative}, we analyze the perturbative dynamics of negativity for the bipartite system between $A$ and $B$ and the bipartite system between $\tilde{A}$ and $B$ at the onset of interaction under different initial conditions (${\rm det}(\rho_B)\neq 0$ versus ${\rm det}(\rho_B)=0$). We also introduce the notions of \negaAB and negativity transmissibility in Sec.~\ref{sec:pure, perturbative}. We focus on the the perturbative dynamics of negativity for the bipartite system between $\tlA$ and $A$ and introduce the notion of negativity vulnerability in Sec.~\ref{sec:negativity-perturbed-tlA-A}, and we study the behavior of entanglement delocalization in the rest of Sec.~\ref{sec:delocalization}. We next go beyond perturbative calculations and study whether $\tlA$ and $B$ can become entangled throughout the course of time evolution with different forms of total Hamiltonians between $A$ and $B$ in Sec.~\ref{sec:tlA-B-separable}. We conclude in Sec.~\ref{sec:conclusion} and discuss future work to apply and extend our results in the context of quantum technologies and fundamental physics.

\subsection{System Setup}\label{sec:system setup}

\begin{figure}[htbp!]
\centerline{\includegraphics[width=0.4\hsize]{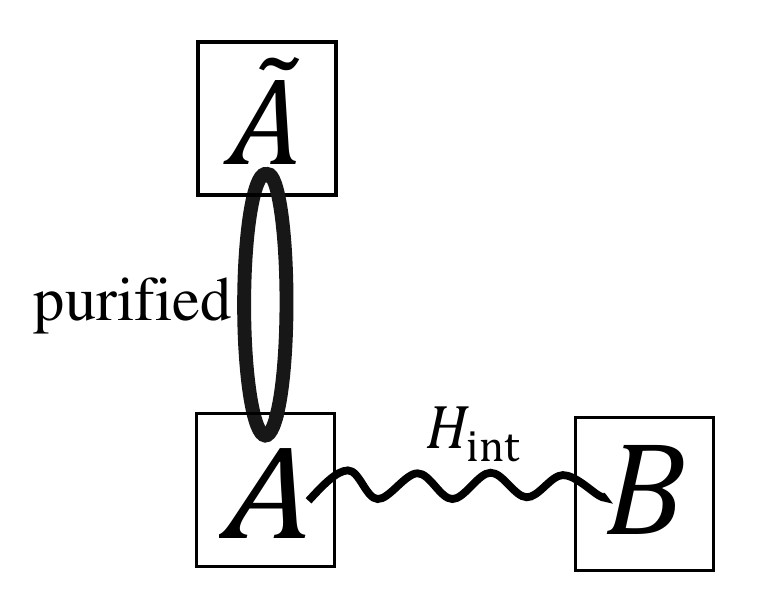}}
\caption{The tripartite system studied in this work. Boxes represent systems, and we adopt the convention that systems whose boxes are linked by a solid ellipse are entangled with another. Systems whose boxes are connected by a wiggly line are interacting. Here, the diagram shows a setup in which a system $A$ is initially entangled with (and in this case purified by) a system $\tlA$, while $B$ is unentangled with both $A$ and $\tlA$. $A$ and $B$ are starting to interact.}
\label{fig:system-setup}
\end{figure}

We consider a simple arrangement of three finite-dimensional systems, $A$, $\tlA$ and $B$, wherein $A$ and $\tlA$ are initially entangled such that $\tlA$ purifies $A$. We denote the dimensions of $\hA$ and $\hB$ by $d_A$ and $d_B$ respectively. We assume that $\hA$ and $\mathcal{H}_{\tlA}$ possess the same finite dimension $d_A$. Both $A$ and $\tlA$ will individually be mixed due to their mutual purification. $B$ is initially unentangled with both $A$ and $\tlA$. Also, $B$ can be initially either mixed or pure. We assume that only the two systems $A$ and $B$ interact. Our aim is to study the dynamics of bipartite entanglement between the different subsystems within the setup. The setup is shown in Fig.~\ref{fig:system-setup}. It is analogous to the setup used in \cite{21EmilySecondPaper} which studied the dynamics of coherent information at the onset of interactions. 

With $\tlA A$ initially pure, we have $\rho_{\tlA A}=\rho_{\tlA A}(0)=\ket*{\omega}\bra*{\omega}$, where $\ket*{\omega}\in\hlA\ot\hA$. By the Schmidt decomposition, we know $\ket*{\omega}=\sum_i \alpha_i \ket*{\tla_i}\ot\ket*{a_i}$, where $\{\alpha_i\}$ are non-negative. $\{\ket*{\tla_i}\}$ and $\{\ket*{a_i}\}$ are bases for $\hlA$ and $\hA$ respectively. Then 
\begin{align}
    \rho_{\tlA A}=\sum_{i,j}\alpha_i\alpha_{j}\ket*{\tla_i}\bra*{\tla_j}\ot\ket*{a_i}\bra*{a_j}.
    \label{eq:AtlA-rho0}
\end{align}
The partial trace gives us the initial density matrices of the subsystems $A$ and $\tlA$:
\begin{align}
    \rho_A&=\Tr_{\tlA}[\rho_{\tlA A}]=\sum_{i}\alpha_i^2\ket*{a_i}\bra*{a_i}=\sum_{i}\lambda_A^{i}\ket*{a_i}\bra*{a_i} \label{eq:rho-A},\\
    \rho_{\tlA}&=\Tr_{A}[\rho_{\tlA A}]=\sum_{i}\alpha_i^2\ket*{\tla_i}\bra*{\tla_i}=\sum_{i}\lambda_A^{i}\ket*{\tla_i}\bra*{\tla_i}\label{eq:rho-tlA},
\end{align}
where $\lambda_A^{i}=\alpha_i^2$. Eqs,~\ref{eq:rho-A} and \ref{eq:rho-tlA} are  the eigen-decomposition of the initial density matrices $\rho_A$ and $\rho_{\tlA}$, which share the same eigenvalues $\{\lambda_A^i\}$. In our setup, we further assume ${\rm det}(\rho_A)\neq 0$, that is all eigenvalues of $\rho_A$ and $\rho_{\tlA}$ are non-vanishing with $\lambda_A^{i}>0$ for all $i$. This assumption greatly simplifies our perturbative calculations in Sec.~\ref{sec:pure, perturbative} and \ref{sec:negativity-perturbed-tlA-A}. The assumption of ${\rm det}(\rho_A)\neq 0$ is also reasonable since $A$ and $\tlA$ purifies each other. Since $B$ is initially unentangled with $\tlA$ or $A$, the total tripartite system has the initial state:
\begin{align}
    \rho_{\rm tri}&=\rho_{\tlA A}\ot\rho_B=\sum_{i,j,k}\alpha_i\alpha_{j}\lambda_B^k\ket*{\tla_i}\bra*{\tla_j}\ot\ket*{a_i}\bra*{a_j}\ot\ket*{b_k}\bra*{b_k},
    \label{eq:tripartite-rho0}
\end{align}
where $\rho_B=\sum_{k}\lambda_B^k\ket*{b_k}\bra*{b_k}$. We thereby also know the initial density matrices for $AB$ and $\tlA B$:
\begin{align}
    \rho_{AB}&=\rho_{A}\ot\rho_{B}=\sum_{i,k}\lambda_A^i\lambda_B^k\ket*{a_i}\bra*{a_i}\ot\ket*{b_k}\bra*{b_k}\label{eq:AB-rho0},\\
    \rho_{\tlA B}&=\rho_{\tlA}\ot\rho_{B}=\sum_{i,k}\lambda_A^i\lambda_B^k\ket*{\tla_i}\bra*{\tla_i}\ot\ket*{b_k}\bra*{b_k}\label{eq:tlAB-rho0}.
\end{align}
The exact time evolution of the total tripartite system $\tlA AB$ is given by 
\begin{align}
    \rho_{\rm tri}(t)=e^{-it\hat{H}_{{\rm tri}}}\rho_{\rm tri}e^{it\hat{H}_{{\rm tri}}}
    \label{eq:time-evolution-exact}
\end{align}
where $\hat{H}_{{\rm tri}}$ is the Hamiltonian of our tripartite system. Throughout our work, we assume $\hat{H}_{{\rm tri}}$ is time-independent with the following form:
\begin{align}
\hat{H}_{{\rm tri}}&=\hat{E}\ot\mathbb{I}_A\ot\mathbb{I}_B+\mathbb{I}_{\tlA}\ot \hat{H}_{{\rm tot}}\label{eq:tri-Hamiltonian}\\
\hat{H}_{{\rm tot}}&=\hat{C}\ot\mathbb{I}_B+\hat{H}_{{\rm int}}+\mathbb{I}_A\ot\hat{D}=\sum_{q}\hat{A}^q\ot\hat{B}^q
\label{eq:total-Hamiltonian}
\end{align}
Here, $\hat{E}$ is the free Hamiltonian for $\tlA$ and $\hat{H}_{{\rm tot}}$ is the total Hamiltonian between $A$ and $B$, which includes both the interaction Hamiltonian $\hat{H}_{{\rm int}}$ and the free Hamiltonians $\hat{C}$ for the subsystem $A$ and $\hat{D}$ for the subsystem $B$. The term $\hat{A}^q\ot\hat{B}^q$ can absorb the respective free Hamiltonians for $A$ and $B$. In parts of Sec.~\ref{sec:pure, perturbative}, \ref{sec:negativity-perturbed-tlA-A}, and \ref{sec:tlA-B-separable}, we will discuss the special case of a product-form interaction Hamiltonian $\hat{H}_{{\rm int}}=\hat{A}\ot \hat{B}$. Such product Hamiltonian are also the focus of \cite{21EmilySecondPaper} for studying the dynamics of coherent information. 

Perturbatively upto second-order, we can expand $\rho(t)$ in Eq.~\ref{eq:time-evolution-exact} with respect to $t=0$: 
\begin{align}
&\rho_{\rm tri}(t)=\rho_{\rm tri}^{(0)}+t\rho_{\rm tri}^{(1)}+t^2\rho_{\rm tri}^{(2)}+O(t^3)\\
    &=\rho_{\rm tri}+i(\rho_{\rm tri}\hat{H}_{{\rm tri}}-\hat{H}_{{\rm tri}}\rho_{\rm tri})t+(\hat{H}_{{\rm tri}}\rho_{\rm tri}\hat{H}_{{\rm tri}}-\frac{1}{2}\hat{H}_{{\rm tri}}^2\rho_{\rm tri}-\frac{1}{2}\rho_{\rm tri}\hat{H}_{{\rm tri}}^2)t^2+O(t^3).
    \label{eq:time-evolution-perturbed}
\end{align}
We will use the exact formula in Eq.~\ref{eq:time-evolution-exact} to study the conditions for generating entanglement between $\tlA$ and $B$ in Sec.~\ref{sec:tlA-B-separable} and to perform numerical calculations for particular systems. We use the perturbative expansion in Eq.~\ref{eq:time-evolution-perturbed} to study the entanglement dynamics within our setup at the onset of interaction in Sec.~\ref{sec:mixed, finite time}-\ref{sec:delocalization}. When $\hat{H}$ is finite dimensional, we can in principle use Eq.~\ref{eq:time-evolution-exact} to perform analytical calculation for $\rho_{\rm tri}(t)$, but finding the analytical formulas for the eigenvalues of the partial transpose of $\rho_{\rm tri}(t)$ is highly non-trivial for a system under a generic total Hamiltonian between $A$ and $B$, which justifies the pertubative analysis. 

\subsection{Results Summary}\label{sec:summary}

\begin{table}[htbp!]
\resizebox{\columnwidth}{!}{
\begin{tabular}{|c|l@{}|} \hline
Bipartite Systems & \begin{tabular}{c}
Dynamics of Entanglement \end{tabular} \\ 
\hline
\begin{tabular}{c}
    $A; B$ \\ ${\rm det}(\rho_B)\neq 0$
\end{tabular} & 
\begin{tabular}{l}
Negativity initially vanishes for a finite amount of time;\\
Guaranteed no entanglement  for a finite amount of time if \\ 
\qquad $d_Ad_B\leq 6$ or $\Tr[\rho_A^2]\Tr[\rho_B^2]< 1/(d_Ad_B-1)$ (Sec.~\ref{sec:mixed, finite time}). \end{tabular}\\  \hline 
\begin{tabular}{c}
    $A; B$ \\ ${\rm det}(\rho_B)=0$
\end{tabular} & 
\begin{tabular}{l}
Negativity susceptibility is given in Eq.~\ref{eq:negativity-A-B-2-summary}-\ref{eq:FAB-B-matrix-summary};\\
They become entangled at the onset for generic $\hat{H}_{\rm tot}$ (Sec.~\ref{sec:negativity-perturbed-A-B}).
\end{tabular}\\  \hline 
\begin{tabular}{c}
    $\tilde{A};A$
\end{tabular} & 
\begin{tabular}{l}
Negativity vulnerability is given in Eq.~\ref{eq:negativity-tlA-A-2-summary};\\ 
The amplitude-based variance $G_A$ is introduced in Eq.~\ref{eq:GA};\\
Entanglement diminishes at the onset for generic $\hat{H}_{\rm tot}$ (Sec.~\ref{sec:negativity-perturbed-tlA-A}).
\end{tabular}\\
\hline
\begin{tabular}{c}
$\tlA;AB$
\end{tabular}&
\begin{tabular}{l}
The degree of entanglement remains constant;\\ 
Delocalization occurs when $\tlA;B$ unentangled (Sec.~\ref{sec:tlA-AB-delocalization}).
\end{tabular}\\
\hline
\begin{tabular}{c}
    $B;\tlA A$ 
\end{tabular} &
\begin{tabular}{l}
Become entangled at the onset under a generic $\hat{H}_{\rm tot}$;\\ 
Delocalization occurs for an initial period when ${\rm det}(\rho_B)\neq 0$ (Sec.~\ref{sec:B-tlAA-delocalization}).
\end{tabular}\\
\hline
$\tlA; B$ &
\begin{tabular}{l}
Results are summarized in Table.~\ref{tab:tlA-B-summary}.
\end{tabular}
\\ \hline
\begin{tabular}{c}
    All \\ Bipartite Systems
\end{tabular} &
\begin{tabular}{l}
$1^{st}$-order perturbation of negativity initially vanishes  (Eq.~\ref{eq:negativity-A-B-1-summary}, \ref{eq:negativity-tlA-B-1-summary}, \ref{eq:negativity-tlA-A-1-summary}).\\ 
Free Hamiltonians do not change negativity up to second order\\ \qquad (Sec.~\ref{sec:free-negativity-A-B}, \ref{sec:free-negativity-tlA-B}, and \ref{sec:free-negativity-tlA-A}).
\end{tabular}\\  \hline 

\end{tabular}
}
\caption{The summary for the key features of the dynamics of entanglement for all bipartite subsystems (excluding $A;\tlA B$) under our tripartite system setup described in Sec.~\ref{sec:system setup}. We used the notation $\tlA;A$ (or, e.g., $\tlA;AB)$ to indicate that we are treating $\tlA$ and $A$ (or , e.g., $\tlA$ and $AB$) as the two subsystems between which the entanglement will be considered. We have assumed ${\rm det}(\rho_A)\neq 0$ all our results except ones presented in Sec.~\ref{sec:tlA-B-separable}. We have defined $d_A:={\rm dim}(\hA)$ and $d_B:={\rm dim}(\hB)$. 
}\label{tab:dynamics-entanglement-summary}
\end{table}

In this subsection, we summarize our findings for the entanglement dynamics of our system setup. We use the notation $\tlA;A$ (or, e.g., $\tlA;AB)$ to indicate that we are treating $\tlA$ and $A$ (or, e.g., $\tlA$ and $AB$) as the two subsystems between which the entanglement will be considered. We list the key features, matched with their corresponding sections and equations in the text, for the entanglement dynamics of all bipartite systems (excluding $A;\tlA B)$ in Table~\ref{tab:dynamics-entanglement-summary}.

\begin{table}[htbp!]
\centering
\resizebox{\columnwidth}{!}{
\begin{tabular}{|l|@{}l@{}|} \hline
\begin{tabular}{c}
    $\hat{H}_{{\rm tot}}=$\\
    $\hat{A}\ot\hat{B}+$\\
    $\bIA\ot\hat{D}$
\end{tabular} & 
\begin{tabular}{l}
$\tlA$ and $B$ remain unentangled throughout the time evolution (Sec.~\ref{sec:pf-nf-tlA-B-separable}, \ref{sec:pf-fb-tla-b-separable}).\\
Quantum discord can be immediately generated for qubits with $\rho_B$ pure (Sec.~\ref{sec:quantum-discord}).
\end{tabular}
\tabularnewline 
\Xhline{3.5\arrayrulewidth}
\begin{tabular}{c}
$\hat{H}_{{\rm tot}}\neq$\\
$\hat{A}\ot\hat{B}+$\\
$\bIA\ot\hat{D}$
\end{tabular}
&
\begin{tabular}{l|@{}l@{}}
\begin{tabular}{c}
${\rm det}(\rho_B)$\\
$\neq 0$ 
\end{tabular} &
\begin{tabular}{l}
Negativity initially vanishes for a finite amount of time (Sec.~\ref{sec:mixed, finite time}).\\ 
$\tlA$ and $B$ are initially unentangled for a finite amount of time when \\ \qquad $d_Ad_B\leq 6$ or $\Tr[\rho_{\tlA}^2]\Tr[\rho_B^2]< 1/(d_Ad_B-1)$ (Sec.~\ref{sec:mixed, finite time}).\\
$\tlA$ and $B$ can get entangled after a finite amount of time (Sec.~\ref{sec:m-tla-b-entangled}).
\end{tabular}
\\
\Xhline{3.5\arrayrulewidth}
\begin{tabular}{c}
${\rm det}(\rho_B)$\\
$=0$
\end{tabular} &
\begin{tabular}{l|l}
\begin{tabular}{c}
$\hat{H}_{{\rm int}}=$\\
$\hat{A}\ot\hat{B}$
\end{tabular} &
\begin{tabular}{l}
Negativity vanishes at the $2^{nd}$-order at the onset (Sec.~\ref{sec:pure-pf-tlAB-unentangled-perturbed}). \\ 
$\tlA$ and $B$ can get entangled when $[\hat{C},\hat{A}]\neq 0$ (Sec.~\ref{sec:pf-fa-tla-b-entangled}). \end{tabular}
\\
\Xhline{3.5\arrayrulewidth}
\begin{tabular}{c}
$\hat{H}_{{\rm int}}\neq$\\
$\hat{A}\ot\hat{B}$
\end{tabular}
 &
\begin{tabular}{l}
Negativity transmissibility is given in Eq.~\ref{eq:negativity-tlA-B-2-summary}-\ref{eq:FtlAB-tlA-matrix-summary};\\ $\tlA$ and $B$ can get entangled at the onset (Sec.~\ref{sec:negativity-perturbed-tlA-B}).
\end{tabular} 
\end{tabular}
\end{tabular}
\\ \hline
\end{tabular}
}
\caption{The summary for the features of the entanglement dynamics between $\tlA$ and $B$ depending on the initial density matrix $\rho_B$ for the subsystem $B$ and the forms of the interaction Hamiltonian $\hat{H}_{{\rm int}}$ and the total Hamiltonian $\hat{H}_{{\rm tot}}$ (including $\hat{H}_{{\rm int}}$ and the free Hamiltonians $\hat{C}$ for the subsystem A and $\hat{D}$ for the subsystem $B$) within $AB$. We have assumed ${\rm det}(\rho_A)\neq 0$ all our results except ones presented in Sec.~\ref{sec:tlA-B-separable}, and we defined $d_A:={\rm dim}(\hA)$ and $d_B:={\rm dim}(\hB)$.}\label{tab:tlA-B-summary}
\end{table}

One of the key goals of this work is to study the quantum channel capacity, that is we want to know under what conditions we can transfer the initial entanglement within $\tlA;A$ to $\tlA;B$ as well as the efficiency of such transfer, so the entanglement between $\tlA$ and $B$ is of particular importance. We summarize the features of the entanglement dynamics for $\tlA;B$ under different initial conditions in Table~\ref{tab:tlA-B-summary}. We give a more direct description for the four cases of the entanglement dynamics listed in Table~\ref{tab:tlA-B-summary}. The first case represents a product-form interaction Hamiltonian between $A$ and $B$ with or without the free Hamiltonian on subsystem $B$ (either $\hat{D}\neq0$ or $\hat{D}=0)$. The second case represents a total Hamiltonian in a form different from $\hat{A}\ot\hat{B}+\bIA\ot\hat{D}$ under the condition ${\rm det}(\rho_B)\neq 0$. The third case represents a product-form interaction Hamiltonian $\hat{A}\ot\hat{B}$ in addition to a non-trivial free Hamiltonian $\hat{C}$ on subsystem $A$ under ${\rm det}(\rho_B)=0$. The fourth case represents a generic interaction Hamiltonian of multiple terms (not in the product form) between $A$ and $B$ under the condition ${\rm det}(\rho_B)=0$.

Regardless of the initial state of $B$, we see that a product-form total Hamiltonian between $A$ and $B$ can never transfer any $\tlA;A$ entanglement to $\tlA;B$, and the inclusion of a free Hamiltonian $\hat{D}$ in subsystem $B$ will not help with the entanglement transfer. The inclusion of free Hamiltonian $\hat{C}$ on subsystem $A$ (in addition to the product-form interaction Hamiltonian $\hat{A}\ot\hat{B}$) can entangle, albeit weakly at the onset, $\tlA$ and $B$. When ${\rm det}(\rho_B)\neq 0$, negativity for $\tlA;B$ will vanish for a finite amount of time at the onset. Only when ${\rm det}(\rho_B)=0$ and the interaction Hamiltonian $\hat{H}_{{\rm tot}}$ contains multiple non-trivial terms can the entanglement be efficiently transferred from $\tlA;A$ to $\tlA;B$ at the onset of an interaction. In such a case, we can measure the speed of the entanglement transfer at the onset with the negativity transmissibility introduced in Eq.~\ref{eq:negativity-tlA-B-2-summary}.

\section{Negativity and Perturbation}\label{sec:N and P}
In this section, we first review the basic results on entanglement and negativity, the entanglement measure we will use for our study. We then setup the framework for our perturbative calculation of the dynamics of negativity.

\subsection{The Entanglement Measure  Negativity}\label{sec:entanglement}

A system at a state $\rho$ on $\mathcal{H}_{1}\otimes \mathcal{H}_{2}$ is said to be separable over $\mathcal{H}_1$ and $\mathcal{H}_2$ if and only if there exist $\{\rho_1^{k}\}$ and $\{\rho_{2}^{k}\}$ which are the respective density matrices on $\mathcal{H}_{1}$ and $\mathcal{H}_{2}$ such that
\begin{equation}
    \rho=\sum_{k}{p_k}\rho_1^{k}\ot\rho_2^k, \text{ where }\sum_{k}{p_k}=1 \text{ and } p_k\geq 0 \text{ for all } k. 
    \label{eq:separable}
\end{equation}
The above definition includes the case of the bipartite product state $\rho=\rho_1\ot\rho_2$. If $\rho$ is not separable as described in Eq.~\ref{eq:separable}, then it is defined to be entangled. 

For a pure bipartite state $\rho=\ket{\omega}\bra{\omega}$, we can use the n-R\'enyi entropy\footnote{when $n\to 1$, R\'enyi entropy becomes the von-Neumann entropy} to quantify the amount of entanglement \cite{1961renyientropy}. \cite{20EmilyFirstPaper} has studied the dynamics of n-R\'enyi entropy between $A$ and $B$ at the onset of interaction under our system setup. When the bipartite state $\rho$ is mixed, n-R\'enyi entropy is no longer a proper measure for entanglement. We consider a function a proper measurement of entanglement when the function satisfies the definitions of entanglement monotone \cite{00EMproposal}, a non-negative function whose values do not increase under local operations and classical communication (LOCC). Unlike many proposed bipartite entanglement monotones which are computationally intractable \cite{05EMReview}, negativity has a relatively simple expression which allows both analytical and numerical calculation, so we use negativity for our work.

Negativity is defined as the absolute sum of all negative eigenvalues for the partial transpose of the bipartite density matrix \cite{02EMNegativity}. For an arbitrary bipartite density matrix $\displaystyle \rho = \sum_{ijkl} \rho^{ik}_{jl} |i\rangle \langle k | \otimes |j\rangle \langle l|$ over $\mathcal{H}_1\ot\mathcal{H}_2$, the partial transpose with respect to the first system is defined as 
\begin{align}
    \rho^{T_1} &:= (T \otimes I) (\rho) = \sum_{ijkl} \rho^{ik} _{jl} (|i\rangle \langle k |)^T \otimes |j\rangle \langle l|
\\&= \sum_{ijkl} \rho^{ik} _{jl} |k^*\rangle \langle i^* | \otimes |j\rangle \langle l| = \sum_{ijkl} \rho^{ki} _{jl} |i^*\rangle \langle k^*| \otimes |j\rangle \langle l|,
\label{eq:partial-transpose}
\end{align}
where $T$ is the transpose operator \cite{96PeresSeparability}. The partial transpose with respect to the first system flips the order of indices associated to the first system. Partial transpose preserves the trace of the density matrix where $\Tr\left[\rho^{T_1}\right]=\Tr\left[\rho\right]=1$. For the definition of negativity, taking partial transpose with respect to the first or second system does not matter, so we will always transpose the first system in our work for consistency.

Let $\lambda_i^{T_1}$ be the eigenvalues of $\rho^{T_1}$. Then the negativity for the bipartite state $\rho$ can be written as \cite{02EMNegativity}: 
\begin{align}
   \mathcal{N}(\rho)&:=\sum_{\lambda_i^{T_1}<0}-\lambda_i^{T_1}=\frac{\sum_i |\lambda_i^{T_1}|-1}{2}\label{eq:negativity-eigen}\\
   &=\frac{||\rho^{T_1}||_1-1}{2}=\frac{\Tr[\sqrt{\rho^{T_1 \dag}\rho^{T_1}}]-1}{2}.
   \label{eq:negativity-norm}
\end{align} 
The two expressions in Eq.~\ref{eq:negativity-eigen} are equivalent since $\Tr\left[\rho^{T_1}\right]=1$. In Eq.~\ref{eq:negativity-norm}, both $||\rho^{T_1}||_1$ and $\Tr[\sqrt{\rho^{T_1 \dag}\rho^{T_1}}]$ represent the sum of singular values of $\rho^{T_1}$. Since $\rho^{T_1}$ is Hermitian, singular values of $\rho^{T_1}$ are simply the absolute values of eigenvalues of $\rho^{T_1}$, which explains the equivalence between Eq.~\ref{eq:negativity-eigen} and \ref{eq:negativity-norm}.

The positivity of $\rho^{T_1}$, equivalent to $\mathcal{N}(\rho)=0$, implies bipartite separability \cite{96PeresSeparability}, which is known as the Peres-Horodecki (PPT) criterion. However, $\mathcal{N}(\rho)=0$ is generally not a sufficient condition for separability \cite{96HorodeckiSeparability}. In the special case where the bipartite system $\mathcal{H}_1\ot\mathcal{H}_2$ has the dimension $2\times 2$ or $2\times 3$, the PPT criterion is both necessary and sufficient for separability \cite{96HorodeckiSeparability}, which will have implications in Sec.~\ref{sec:mixed, finite time}. 

Negativity can be easily calculated when the bipartite state $\rho$ is pure. Under our system setup, the initial density matrix for $\tlA A$ is indeed pure according to Eq.~\ref{eq:AtlA-rho0}. The partial transposition of $\rho_{\tlA A}$ gives us \begin{equation}
\rho_{\tlA;A}^{T_1}=\sum_{i,j}\alpha_i\alpha_{j}\ket*{\tla^*_j}\bra*{\tla^*_i}\ot\ket*{a_i}\bra*{a_j}
\label{eq:pt-tlA-A-density-0}.
\end{equation}
One can verify that $\frac{\sqrt{2}}{2}\ket*{\tla^*_u}\ot\ket{a_v}\pm\frac{\sqrt{2}}{2}\ket*{\tla^*_v}\ot\ket{a_u}$ are eigenvectors of $\rho_{\tlA;A}^{T_1}$ with corresponding eigenvalues $\pm\alpha_u\alpha_v$ for $u<v$. $\ket*{\tla_u}\ot\ket{a_u}$ are also eigenvectors with corresponding eigenvalues $\alpha_u^2$ for all $u$. When $\alpha_u\neq 0$ for all $u$ (we have assumed ${\rm det}(\rho_A)\neq 0$), $\{\alpha_u^2,\pm\alpha_u\alpha_v\}$ are the spectra of $\rho_{\tlA;A}^{T_1}$. We introduce the following notations for all of the eigenvalues and eigenvectors of $\rho_{\tlA;A}^{T_1}$, which will be used for the calculations in Appendix~\ref{sec:negativity-tlA-A-calculation}:
\begin{align}
\lambda_{uv\pm,\tlA; A}^{T_1}&=\pm\alpha_u\alpha_v,\qquad u<v\label{eq:pt-tlA-A-eigenvalue-0-uv}\\
\lambda_{u,\tlA; A}^{T_1}&=\alpha_u^2=\lambda_A^u, \label{eq:pt-tlA-A-eigenvalue-0-u}\\
\ket*{\tla^*_{u}\alpha_{v},\pm}&=\frac{\sqrt{2}}{2}\ket*{\tla^*_u}\ot\ket{a_v}\pm\frac{\sqrt{2}}{2}\ket*{\tla^*_v}\ot\ket{a_u},\qquad u<v\label{eq:pt-tlA-A-eigenvector-0-u-v}\\
\ket*{\tla_{u}^*\alpha_{u}}&=\ket*{\tla^*_u}\ot\ket{a_u}.\label{eq:pt-tlA-A-eigenvector-0-u}
\end{align}
According to the definition of negativity in Eq.~\ref{eq:negativity-eigen}, only eigenvectors $\ket*{\tla^*_{u}\alpha_{v},-}$ with eigenvalues $-\alpha_u\alpha_v$ contribute to negativity, so
\begin{align}
 \mathcal{N}(\rho_{\tlA; A})=\sum_{u<v} \alpha_u\alpha_v,
 \label{eq:tlA-A-negativity-0}
\end{align}
which applies to any pure bipartite state. 

When $\rho$ is mixed, we can easily perform numerical calculations for eigenvalues of $\rho^{T_1}$ and obtain the dynamics of negativity for specific examples. However, it is difficult to find the general expressions for the eigenvalues of $\rho^{T_1}$. Under our system setup, the three bipartite density matrices $\rho_{AB}(t)$, $\rho_{\tlA A}(t)$, and $\rho_{\tlA B}(t)$ are all expected to become mixed under the interaction Hamiltonian in Eq.~\ref{eq:tri-Hamiltonian}. We will therefore resort to perturbation theory to find the general analytical expressions for the dynamics of negativity. 

With the Hamiltonian of our tripartite system taking the form $\hat{H}_{{\rm tri}}=\hat{E}\ot\mathbb{I}_{AB}+\mathbb{I}_{\tlA}\ot \hat{H}_{{\rm tot}}$ according to Eq.~\ref{eq:tri-Hamiltonian} and \ref{eq:total-Hamiltonian}, we know that the free Hamiltonian $\hat{E}$ on the system $\tlA$ will not impact the separability or the negativity of any bipartite systems in our setup. Taking the bipartite system $\tlA;B$ as an example,
\begin{align}
&\rho_{\tlA B}^{\rm with\; \hat{E}}(t)=\Tr_A[U(t)\rho_{\tlA A}\ot\rho_B U^{\dag}(t)]\\
&=\Tr_A[\exp\{-it(\hat{E}\ot\mathbb{I}_{AB}+\mathbb{I}_{\tlA}\ot \hat{H}_{{\rm tot}})\}\rho_{\tlA A}\ot\rho_B\exp\{it(\hat{E}\ot\mathbb{I}_{AB}+\mathbb{I}_{\tlA}\ot \hat{H}_{{\rm tot}})\}]\\
&=\exp\{-it\hat{E}\}\Tr_A[\exp\{-it\hat{H}_{{\rm tot}}\}\rho_{\tlA A}\ot\rho_B\exp\{it \hat{H}_{{\rm tot}}\}]\exp\{it\hat{E}\}\\
&=\exp\{-it\hat{E}\}\rho_{\tlA B}^{\rm without\; \hat{E}}(t)\exp\{it\hat{E}\},
\label{eq:ftlA-no-impact-on-tlAB}
\end{align}
where we can decompose $\exp\{it(\hat{E}\ot\mathbb{I}_{AB}+\mathbb{I}_{\tlA}\ot \hat{H}_{{\rm tot}})\}$ into $\exp\{it\mathbb{I}_{\tlA}\ot \hat{H}_{{\rm tot}}\}\exp\{it\hat{E}\ot\mathbb{I}_{AB}$ since $\hat{E}\ot\mathbb{I}_{AB}$ and $\mathbb{I}_{\tlA}\ot \hat{H}_{{\rm tot}}$ commute. Eq.~\ref{eq:ftlA-no-impact-on-tlAB} shows that the separability of $\rho_{\tlA B}(t)$ remains the same with or without the free Hamiltonian $\hat{E}$ for the system $\tlA$. The negativity $\mathcal{N}_{\tlA B}(t)$ is unaffected since the local unitary operation $\exp\{-it\hat{E}\}$ can be considered a rotation of basis on $\mathcal{H}_{\tlA}$, thereby leaving the $\rho^{ki}_{jl}$ components in Eq.~\ref{eq:partial-transpose} unchanged. We therefore ignore the free Hamiltonian $\hat{E}$ on the system $\tlA$ and only consider the Hamiltonian $\hat{H}_{{\rm tot}}$ on $AB$ in Eq.~\ref{eq:total-Hamiltonian} for the rest of the work.

\subsection{Perturbation of Negativity: A General Framework}\label{sec:PoN}

Consider a general bipartite state with time evolution $\rho(t)$. Let the initial density matrix be $\rho(t_0)=\rho_0$, and the eigenvalues and eigenvectors of $\rho_0^{T_1}$ be $\lambda_n^{T_1}$ and $\ket*{v_{n}^{T_1}}$ respectively. To find the perturbation expressions of $\mathcal{N}(\rho(t))$ with respect to $t=t_0$, we can first perturbatively expand each eigenvalue $\lambda_n^{T_1}(t)$ of $\rho^{T_1}(t)$ with respect to $t_0$. We therefore need to know the eigenvalue perturbations of $\rho^{T_1}(t)$. The general eigenvalue perturbation problem for a Hermitian operator upto second-order is reviewed in Appendix~\ref{sec:EP}, which will be extensively used in Sec.~\ref{sec:pure, perturbative} and \ref{sec:negativity-perturbed-tlA-A}. \cite{18CresswellNegativityExpansion} provides an alternative formalism for the perturbative expansion of negativity using patterned matrix calculus, which is less intuitive for calculating the negativity dynamics at the onset of our system compared to the eigenvalue perturbation approach.  

Using notations in Appendix~\ref{sec:EP}, we write the eigenvalue perturbation problem of $\rho^{T_1}(t)$ upto second-order in the following forms:
\begin{align}
    \rho^{T_1}(t)&=\rho^{T_1(0)}+t\rho^{T_1(1)}+t^2\rho^{T_1(2)}+O(t^3),\label{eq:perturb-density-pt}\\
    \lambda_{n}^{T_1}(t)&=\lambda_{n}^{T_1(0)}+\lambda_{n}^{T_1(1)}t+\lambda_{n}^{T_1(2)}t^2+O(t^3),\label{eq:perturb-eigen-pt}\\
    \ket*{v_{n}^{T_1}(t)}&=\ket*{v_{n}^{T_1(0)}}+t\ket*{v_{n}^{T_1(1)}}+t^2\ket*{v_{n}^{T_1(2)}}+O(t^3)\label{eq:perturb-vector-pt},
\end{align}
where $\lambda_n^{T_1(i)}$ and $\ket*{v_{n}^{T_1(i)}}$ are the $i^{\rm th}$-order correction for the $n^{\rm th}$ eigenvalue and eigenvector of $\rho^{T_1}(t_0)$. In Eq.~\ref{eq:perturb-eigen-pt}, the zeroth order term is simply the partial transpose of the initial density matrix $\rho^{T_1(0)}=\rho_0^{T_1}$, so $\lambda_{n}^{T_1(0)}=\lambda_{n}^{T_1}$. The perturbative expansions in Eq.~\ref{eq:perturb-density-pt}-\ref{eq:perturb-vector-pt} are valid, since $\rho(t)$ follows the general form of Eq.~\ref{eq:time-evolution-exact}, which is analytic with respect to $t$. Therefore, $\rho^{T_1}(t)$ and $\lambda_{n}^{T_1}(t)$ are also analytic. Results in Appendix~\ref{sec:EP} will allow us to find analytical expressions for $\lambda_{n}^{T_1(1)}$ and $\lambda_{n}^{T_1(2)}$. Using Eq.~\ref{eq:negativity-eigen}, negativity can be perturbatively expanded to second-order in two equivalent ways:
\begin{align}
\mathcal{N}(\rho(t))&=\mathcal{N}(\rho_0)+\mathcal{N}^{(1)}t+\mathcal{N}^{(2)}t^2\label{eq:negativity-perturb}\\
&=\sum_{\lambda_n^{T_1}(t)<0}-\lambda_n^{T_1}(t)=\sum_{\lambda_{n}^{T_1}(t)<0}-(\lambda_{n}^{T_1}+\lambda_{n}^{T_1(1)}t+\lambda_{n}^{T_1(2)}t^2)+O(t^3)\label{eq:negativity-perturb-negative}\\
&=\frac{\sum_n|\lambda_n^{T_1}(t)|-1}{2}=\frac{\sum_n|\lambda_{n}^{T_1}+\lambda_{n}^{T_1(1)}t+\lambda_{n}^{T_1(2)}t^2+O(t^3)|-1}{2}\label{eq:negativity-perturb-absolute}
\end{align}
From Eq.~\ref{eq:negativity-perturb-absolute}, we see that the perturbative expansion in Eq.~\ref{eq:negativity-perturb} will only be valid for a finite amount of time at the onset when no $\lambda_{n}^{T_1}(t)$ changes sign from their original values $\lambda_{n}^{T_1}$. The absolute value function is not analytic across the origin, so $\lambda_{n}^{T_1}(t)$ can not cross zero for Eq.~\ref{eq:negativity-perturb} to be valid. The initial eigenvalues $\lambda_{n}^{T_1}(t_0)=\lambda_{n}^{T_1}$ can be zero, since the absolute value functions are analytic at the origin from one side.

We use Eq.~\ref{eq:negativity-perturb-negative} to analyze how each $\lambda_{n}^{T_1}(t)$ contributes to the negativity at the onset. It is clear that the summation condition $\lambda_{n}^{T_1}(t)<0$ depends on both the initial eigenvalue $\lambda_{n}^{T_1}$ and whether $\lambda_{n}^{T_1}(t)$ increases or decreases at the leading order at the onset of interaction. The evolution of $\lambda_{n}^{T_1}(t)$ is continuous due to analyticity, so there can be no sudden jump or fall of eigenvalues of $\rho^{T_1}(t)$. When initially $\lambda_{n}^{T_1}>0$, it will take a finite amount of time for $\lambda_{n}^{T_1}(t)$ to decrease below 0 if it indeed decreases. Therefore, $\lambda_{n}^{T_1}(t)$ can not contribute to negativity for a finite amount of time when $\lambda_{n}^{T_1}>0$, and the perturbation terms $\lambda_{n}^{T_1(1)}$ and $\lambda_{n}^{T_1(2)}$ can be ignored. When $\lambda_{n}^{T_1}(t)$ eventually decrease below 0, the perturbation expansion in Eq.~\ref{eq:negativity-perturb} breaks down. 

When initially $\lambda_{n}^{T_1}<0$, it will similarly take a finite amount of time for $\lambda_{n}^{T_1}(t)$ to increase above 0 if it does increase. In this case, $\lambda_{n}^{T_1(1)}$ and $\lambda_{n}^{T_1(2)}$ will directly increase or decrease the negativity perturbation terms $\mathcal{N}^{(1)}$ and $\mathcal{N}^{(2)}$ until the negativity expansion in Eq.~\ref{eq:negativity-perturb} breaks down when one of the eigenvalues of $\rho^{T_1}(t)$ eventually changes sign. When initially $\lambda_{n}^{T_1}=0$, then the leading non-zero perturbation term will decide whether $\lambda_{n}^{T_1}(t)$ contributes to negativity at the onset. Suppose $\lambda_{n}^{T_1(u)}$ is the leading non-zero term in Eq.~\ref{eq:perturb-eigen-pt} at the $u^{th}$ order. When $\lambda_{n}^{T_1(u)}>0$, $\lambda_{n}^{T_1}(t)$ will increase and remain greater than zero for a finite time until the subsequent perturbation terms start to dominate the behavior of $\lambda_{n}^{T_1}(t)$, so $\lambda_{n}^{T_1}(t)$ will not affect $\mathcal{N}(\rho(t))$ for a finite amount of time. On the contrary, when the leading term $\lambda_{n}^{T_1(u)}<0$, $\lambda_{n}^{T_1}(t)$ will immediately drop below 0 and directly increase the negativity. 

In summary, positive $\lambda_{n}^{T_1}$ of $\rho_0^{T_1}$ is irrelevant for negativity perturbation at $t_0$. Vanishing $\lambda_{n}^{T_1}$ can only increase negativity if the leading-order perturbation $\lambda_{n}^{T_1(u)}<0$, while negative $\lambda_{n}^{T_1}$ can either increase or decrease negativity. This dependence of the negativity dynamics on the initial sign of $\lambda_{n}^{T_1}$ is important for the bipartite systems $A;B$ and $\tlA;B$ under our system setup. In particular, the negativity dynamics at the onset will have different features depending on whether ${\rm det}(\rho_B)$ vanishes or not, which will be the focus of following Sec.~\ref{sec:mixed, finite time} and Sec.~\ref{sec:pure, perturbative}. In comparison, the dynamics of negativity for the bipartite system $\tlA; A$ do not exhibit such different characteristics under whether the determinant of $\rho_B$ vanishes or not.

\section{The Case \texorpdfstring{\boldmath{${\rm det}(\rho_B)\neq0$}}{detnot0}: Negativity of \texorpdfstring{\boldmath{$A;B$}}{A;B} and \texorpdfstring{\boldmath{$\tlA;B$}}{A;B} Vanishing for Finite Time}\label{sec:mixed, finite time}

We now study the dynamics of negativity for the bipartite systems $A;B$ and $\tlA;B$ at the onset of interaction when the initial state $\rho_B$ has non-vanishing determinant, that is all eigenvalues of $\rho_{B}$ are non-zero. We observe that the negativity for $A;B$ and $\tlA;B$ must vanish for a finite amount of time when ${\rm det}(\rho_B)\neq 0$. When the dimensions of the bipartite Hilbert space $\mathcal{H}_{AB}$ ($\mathcal{H}_{\tlA B}$ has the same dimension due to purification) is smaller than or equal to 6, we can conclude that $A$ and $B$ (also $\tlA$ and $B$) will remain unentangled for a finite amount of time by the PPT criterion \cite{96HorodeckiSeparability} assuming the non-vanishing determinant of $\rho_B$.

From our system setup in Eq.~\ref{eq:AB-rho0} and \ref{eq:tlAB-rho0}, we know $A;B$ and $\tlA;B$ are initially in the product state, so $\mathcal{N}(\rho_{A;B})=\mathcal{N}(\rho_{\tlA;B})=0$. Using notations introduced in Sec.~\ref{sec:PoN}, the zeroth order of the partial transpose density matrices and their respective eigenvalues are given below in Eq.~\ref{eq:pt-A-B-density-0}-\ref{eq:pt-tlA-B-eigen-0}. Due to the eigenvalues' dependence on both Hilbert spaces $\hA$ and $\hB$ (with the dimensions $d_A$ and $d_B$ respectively), we use double indices $ik$ (where $1\leq i\leq d_A$ and $1\leq k \leq d_B$) instead of the single index $n$ (where $1\leq n\leq d_Ad_B$) to track the eigenvalues of $\rho_{A;B}^{T_1(0)}$ and $\rho_{A;B}^{T_1(0)}$. We define the following operations to relate the double and single index notations: $i:=\lfloor n/d_B \rfloor+1$ and $k:=(n\; {\rm mod }
\;d_B)+1$.
\begin{align}
\rho_{A;B}^{T_1(0)}&=(\rho_A)^{T}\ot\rho_B=\left(\sum_{i}\lambda_A^i\ket*{a^{*}_i}\bra*{a^{*}_i}\right)\ot\left(\sum_{k}\lambda_B^k\ket*{b_k}\bra*{b_k}\right)
\label{eq:pt-A-B-density-0}\\
\rho_{\tlA;B}^{T_1(0)}&=(\rho_{\tlA})^{T}\ot\rho_B=\left(\sum_{i}\lambda_A^i\ket*{\tla^{*}_i}\bra*{\tla^{*}_i}\right)\ot\left(\sum_{k}\lambda_B^k\ket*{b_k}\bra*{b_k}\right)
\label{eq:pt-tlA-B-density-0}\\
\lambda_{ik,A;B}^{T_1(0)}&=\lambda_{ik,\tlA;B}^{T_1(0)}=\lambda_A^{i}\lambda_B^{k}
\label{eq:pt-tlA-B-eigen-0}
\end{align}

Under our system setup, we have assumed the determinants of $\rho_A$ and $\rho_{\tlA}$ are non-vanishing, so $\lambda_A^{i}>0$ for all $i$. If we also assume the non-vanishing determinant condition for $\rho_B$, then $\lambda_B^{k}>0$ for all $k$, so $\lambda_{ik,\tlA;B}^{T_1}=\lambda_{ik,A;B}^{T_1}=\lambda_A^{i}\lambda_B^{k}>0$ for all $i,k$, and all eigenvalues of $\rho_{A;B}^{T_1}$ and $\rho_{\tlA;B}^{T_1}$ are greater than 0. According to Sec.~\ref{sec:PoN}, we know that any initially positive eigenvalues $\lambda_n^{T_1}$ of $\rho^{T_1}$ can not contribute to the negativity for a finite amount of time until $\lambda_n^{T_1}(t)$ decreases below $0$. It follows that the negativity for both $A;B$ and $\tlA;B$ systems will remain $0$ for a finite amount of time after the start of interaction between $A$ and $B$.

When the bipartite systems $A;B$ and $\tlA;B$ have the dimensions $2\times2$, $2\times3$, or $3\times2$, the vanishing negativity implies separability according to the PPT criterion. As a result, both $A;B$ and $\tlA;B$ bipartite systems will remain unentangled for a finite amount of time. When $d_Ad_B\leq 6$, we can conclude that no interaction Hamiltonians between $A$ and $B$ can entangle $A$ and $B$ or $\tlA$ and $B$ for a finite amount of time until one of the eigenvalues of the partial transpose density matrices decreases below 0. In the perspective of entanglement transmission, no entanglement can be immediately transferred from $\tlA;A$ to $\tlA;B$. 

When $d_Ad_B>6$, we can still strengthen the result of vanishing negativity and conclude that the entanglement will kick in  after a finite amount of time in the situation where the purity of the initial density matrix $\rho_A\ot\rho_B$ satisfies $\Tr[\rho_A^2]\Tr[\rho_B^2]< 1/(d_Ad_B-1)$. Intuitively, this mixedness condition implies that $\rho_A\ot\rho_B$ is sufficiently close to the maximally mixed state $\frac{1}{d_A}\bIA\ot\frac{1}{d_B}\bIB$. According to Corollary 3  of \cite{02GurvitsLSBaMMBS}, there exists a largest separable ball around the maximally mixed bipartite state $\frac{1}{d_1}\mathbb{I}_1\ot\frac{1}{d_2}\mathbb{I}_2=\frac{1}{d}\mathbb{I}$ such that if $||\rho-\frac{1}{d}\mathbb{I}||_2\leq 1/\sqrt{d(d-1)}=:r$ under the 2-norm, then $\rho$ is separable, and $r$ is the largest such constant. This condition on radius in 2-norm is equivalent to the condition on purity $\Tr[\rho^2]\leq 1/(d-1)$. Therefore, when $\Tr[\rho_A^2]\Tr[\rho_B^2]< 1/(d_Ad_B-1)$ for the $A;B$ system in our setup, let $a:=||\rho_A\ot\rho_B-\frac{1}{d_Ad_B}\mathbb{I}_A\ot\bIB||_2<r=1/\sqrt{d_Ad_B(d_Ad_B-1)}$, then there exists a separable closed ball $\overline{d}_{\rho_A\ot\rho_B}^2(r-a)$ with radius $r-a$ under 2-norm centered around the initial state $\rho_A\ot\rho_B$. It will take a finite amount of time for $\rho_{AB}(t)$ to evolve beyond the boundary of the separable ball $\overline{d}_{\rho_A\ot\rho_B}^2(r-a)$, so it will also take a finite amount of time for the subsystems $A$ and $B$ to become entangled. The bipartite system $\tlA;B$ follows the same argument since $\rho_A$ and $\rho_{\tlA}$ are initially the same upto some unitary transformations due to the purification between $\tlA$ and $A$ under our system setup. 

However, not all initial states $\rho_A\ot\rho_B$ (or $\rho_{\tlA}\ot\rho_{B}$) satisfy the above mixedness condition for separability or equivalently reside in the largest separable ball around the maximally mixed state. Take the two qutrits described in Eq.~\ref{eq:A-qutrit} and \ref{eq:qutrit-B-mixed} in Appendix.~\ref{sec:qutrit} as an example, clearly $\Tr[\rho_A^2]\Tr[\rho_B^2]=0.1587>1/8$. In such situations where $d_Ad_B>6$ and $\Tr[\rho_A^2]\Tr[\rho_B^2]>1/(d_Ad_B-1)$, we will need to resort to negativity, and we can still conclude the entanglement generation within $\tlA;B$ or $A;B$ will be slow initially due to the vanishing negativity. In general, non-zero determinant for $\rho_B$ does not allow fast entanglement generation within $A;B$ or fast entanglement transmission from $\tlA;A$ to $\tlA;B$. We will explore the case where $\rho_B$ has zero eigenvalues in the next section, and we will see that $A;B$ and $\tlA;B$ can become immediately entangled with increasing negativity at the onset under certain forms of $\hat{H}_{{\rm tot}}$. 
\begin{figure}[htbp!]
\centerline{\includegraphics[width=0.6\hsize]{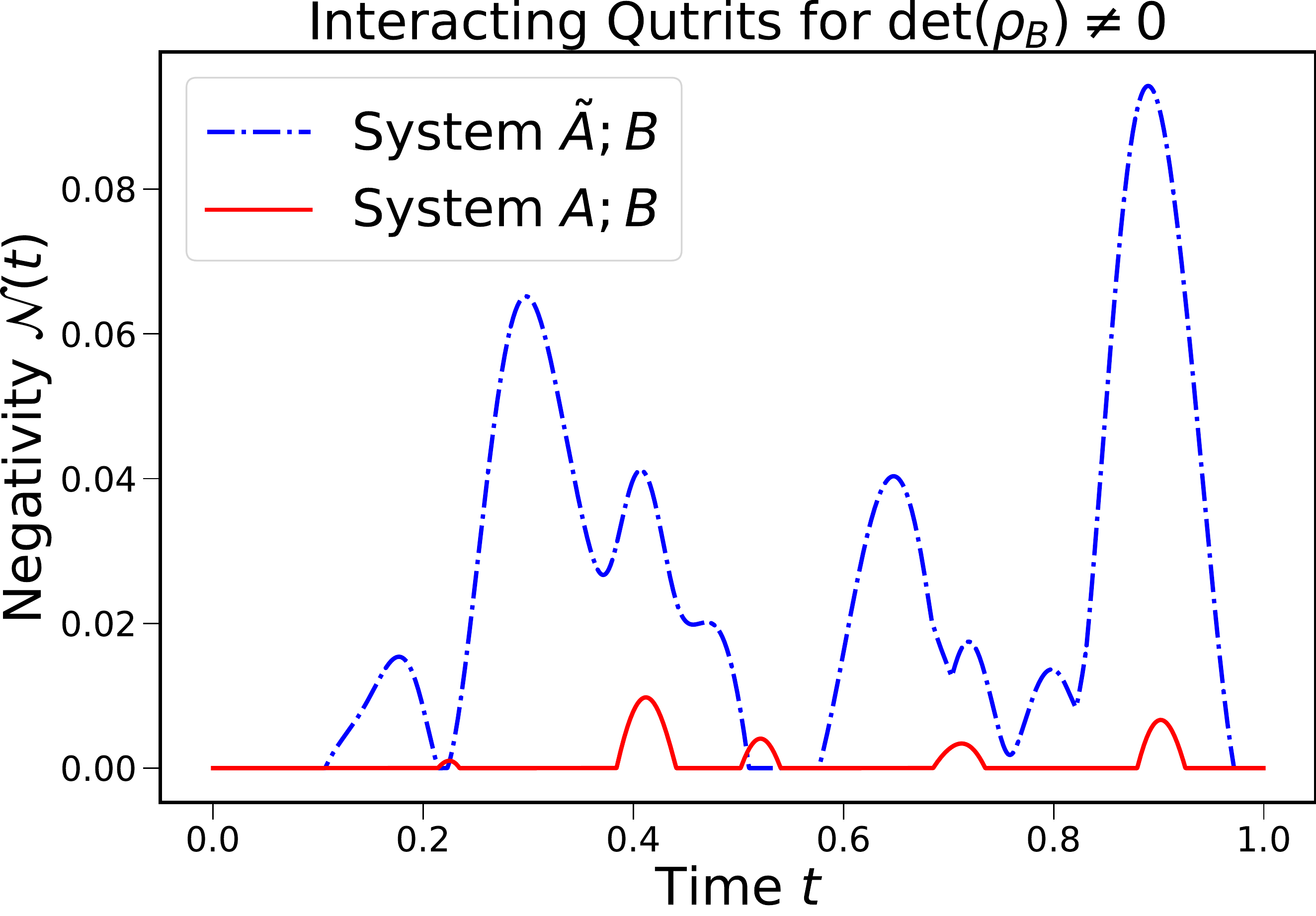}}
\caption{The dynamics of negativity for the $A;B$ and $\tlA;B$ systems assuming $\det(\rho_B)\neq 0$ and $\det(\rho_A)\neq 0$. We see the negativity for both $\tlA;B$ and $A;B$ initially vanishes for a finite amount of time. All subsystems $\tlA, A$, and $B$ are qutrits, and we choose a generic total Hamiltonian of the form $\hat{H}_{{\rm tot}}=\hat{A}_1\ot \hat{B}_1+\hat{A}_2\ot \hat{B}_2$. The Hamiltonians and initial density matrices are given in Eq.~\ref{eq:A-qutrit}-\ref{eq:B_2-qutrit} in Appendix~\ref{sec:qutrit}.}
\label{fig:mixed_qutrit_AB_tlAB_initial}
\end{figure}

Because negativity vanishes for a finite amount of time, we are not able to use perturbation theory to probe into the dynamics of entanglement after one of the eigenvalues of the partial transpose density matrix $\lambda_i^{T_1}(t)$ decreases below $0$. We therefore use a numerical example of interacting qutrits described through Eq.~\ref{eq:A-qutrit}-\ref{eq:B_2-qutrit} in Appendix~\ref{sec:qutrit} to explore the full dynamics. When $\rho_A$ and $\rho_B$ have non-vanishing determinants, the phenomenon of vanishing negativity for a finite amount of time can be clearly observed at the initial time for the systems $A;B$ and $\tlA ;B$ in Fig.~\ref{fig:mixed_qutrit_AB_tlAB_initial}. We see that negativity can suddenly increase above zero or decrease to zero with non-continuous derivatives, demonstrating that the evolution of negativity (entanglement measures in general) can be non-analytic. As seen in both $\tlA;B$ and $A;B$ systems in Fig.~\ref{fig:mixed_qutrit_AB_tlAB_initial}, negativity can be suddenly generated or destructed in a non-analytical way multiple times throughout the time evolution, which is similar to the entanglement sudden death (ESD) or anti-ESD phenomenon discussed in the literature \cite{09YuSuddenDeathEntanglement}. The sudden birth or death of the negativity can be explained by the non-analytic transition of the signs of $\lambda_n^{T_1}(t)$. Any change on $\lambda^{T_1}_{n}(t)$ can only impact negativity when $\lambda_n^{T_1}(t)\leq 0$, causing the non-analytic evolution of the negativity.

The phenomenon of vanishing negativity for a finite amount of time can occur in a more general setting, not limited by our system setup. As long as the initial partial transpose density matrix $\rho^{T_1}$ for the bipartite system has all strictly positive eigenvalues, the negativity for the bipartite system will vanish for a finite amount of time regardless of the Hamiltonians acting on the system. One special feature about our system setup is that both $\tlA;B$ and $A;B$ are initially in product states, which might have important implications for the separability of the bipartite systems. We have so far been able to conclude that $A;B$ or $\tlA;B$ remains separable for a finite amount of time only when $d_Ad_B\leq 6$ based on the PPT criterion \cite{96HorodeckiSeparability} or $\Tr[\rho_A^2]\Tr[\rho_B^2]<1/(d_Ad_B-1)$ based on the largest separable ball around the maximally mixed bipartite state \cite{02GurvitsLSBaMMBS}. Based on the vanishing negativity phenomenon we observed in this section, we make the following conjecture that there generally exists a separable ball around any mixed bipartite product state $\rho_1\ot\rho_2$ when ${\rm det}(\rho_1){\rm det}(\rho_2)\neq 0$. When ${\rm det}(\rho_1)=0$ or ${\rm det}(\rho_2)=0$, we will show in the next section that the bipartite system can become immediately entangled, so no separable ball around the state $\rho_1\ot\rho_2$ can exist. We therefore further conjecture that the largest radius of the separable ball around any bipartite product state $\rho_1\ot\rho_2$ will be related to ${\rm det}(\rho_1)$ and ${\rm det}(\rho_2)$. If our conjecture is proven to be true, we can straightforwardly conclude it will take a finite amount of time for $A$ and $B$ or $\tlA$ and $B$ to become entangled when ${\rm det}(\rho_A){\rm det}(\rho_B)\neq 0$.

\section{The Case \texorpdfstring{\boldmath{${\rm det}(\rho_B)=0$}}{det=0}: Perturbative Results for Negativity of \texorpdfstring{\boldmath{$A;B$}}{A;B} and \texorpdfstring{\boldmath{$\tlA;B$}}{A;B}}\label{sec:pure, perturbative}

In the previous section, we have observed that negativity of $A;B$ or $\tlA;B$ remains vanishing for a finite amount of time when $\rho_B$ has initially non-zero determinant. In this section, we explore whether it is possible to have immediate, fast entanglement generation within $A;B$ or $\tlA;B$ at the onset of interaction when $\rho_B$ has vanishing determinant, that is $\rho_B$ possessing some zero eigenvalues. Using the perturbation formalism introduced in Sec.~\ref{sec:PoN}, we will calculate the first and second-order derivative of negativity for the bipartite systems $A;B$ and $\tlA;B$ at the onset of interaction assuming ${\rm det}(\rho_B)=0$. We find that the first-order derivative of the negativity for both systems vanishes under any generic total Hamiltonian between $A$ and $B$. The free Hamiltonians on $A$ or $B$ will not contribute to the second-order derivative of negativity for both systems. For a product interaction Hamiltonian with $\hat{H}_{{\rm tot}}=\hat{A}\ot\hat{B}$, $A$ and $B$ can become immediately entangled at the onset, while the second-order derivative of the negativity for $\tlA;B$ still vanishes. It takes a non-product interaction Hamiltonian consisting of multiple terms to contribute to the second-order derivative of the negativity for $\tlA; B$ at the onset of interaction. 

To calculate negativity perturbatively at the onset, we first need to find $\rho_{\rm tri}^{(0)}$, $\rho_{\rm tri}^{(1)}$, and $\rho_{\rm tri}^{(2)}$, the perturbative terms for the total density matrix of the tripartite system defined in Eq.~\ref{eq:time-evolution-perturbed} under the total Hamiltonian given in Eq.~\ref{eq:total-Hamiltonian}. We next take the appropriate partial trace to find pertubative expressions for the density matrix of each of the bipartite systems $A;B$, $\tlA;B$, and $\tlA;A$, which will then be used to calculate negativity perturbation. The calculation results for these perturbative density matrices are shown in Appendix~\ref{sec:perturbation calculation density matrix}. 

\subsection{Negativity of System \texorpdfstring{$A;B$}{A;B}}
\label{sec:negativity-perturbed-A-B}

With the formalism in Sec.~\ref{sec:PoN}, we need to find the first three perturbation terms of the partial transpose density matrix $\rho^{T_1}_{A;B}(t)$. The zeroth order term $\rho_{A;B}^{T_1(0)}$ and its associated eigenvalues $\lambda_{ik,A;B}^{T_1(0)}$ were given in Eq.~\ref{eq:pt-A-B-density-0} and \ref{eq:pt-tlA-B-eigen-0}. The first and second-order terms $\rho_{AB}^{T_1(1)}$ and $\rho_{AB}^{T_1(2)}$ can be obtained by taking the partial transpose of Eq.~\ref{eq:A-B-density-1} and \ref{eq:A-B-density-2} derived in Appendix~\ref{sec:perturbation calculation density matrix}. We summarize the results below after changing some indices:
\begin{align}
&\lambda_{ik,A;B}^{T_1(0)}=\lambda_{ik,A;B}^{T_1}=\lambda_A^{i}\lambda_B^{k}
\label{eq:pt-A-B-eigen-0(second appearance)}\\
&\rho_{A;B}^{T_1(0)}=\left(\sum_{i}\lambda_A^i\ket*{a^{*}_i}\bra*{a^{*}_i}\right)\ot\left(\sum_{k}\lambda_B^k\ket*{b_k}\bra*{b_k}\right)\\
&\rho_{A;B}^{T_1(1)}=i\sum_{p,i,j,k,l}A^p_{ki}B^p_{jl}\left(\lambda_A^k\lambda_B^j-\lambda_A^i\lambda_B^l\right)\ket*{a^*_i}\bra*{a^*_k}\ot\ket*{b_j}\bra*{b_l}
\label{eq:pt-A-B-density-1}\\
&\rho_{A;B}^{T_1(2)}=\sum_{p,q,i,j,k,l,m,u} A^p_{km}A^q_{mi}B^p_{ju}B^q_{ul}\left( \lambda_A^{m}\lambda_B^{u}-\frac{1}{2}\lambda_A^{k}\lambda_B^{j}-\frac{1}{2}\lambda_A^{i}\lambda_B^{l}\right)\ket{a^*_i}\bra{a^*_k}\otimes\ket{b_j}\bra{b_l}
\label{eq:pt-A-B-density-2}
\end{align}

Under the assumption of ${\rm det}(\rho_{B})=0$, we can separate the spectrum of $\rho_B$ into two parts: $\lambda_B^{k}\neq 0$ for $1\leq k\leq N$ and $\lambda_B^{k}=0$ for $k\geq N+1$. We define two projection operators $\hat{P}_B^{N}$ and $\hat{P}_B^{D}$ that decompose $\hB$ into two subspaces $\mathcal{N}_B$ and $\mathcal{D}_B$ respectively:
\begin{center}
\begin{minipage}[b]{.4\textwidth}
\vspace{-0.2cm}
\begin{equation}
    \hat{P}_{B}^N:=\sum_{k=1}^{N}\ket*{b_k}\bra*{b_k}\label{eq:non-degenerate-projection}
\end{equation}
\end{minipage}
\hfill
\begin{minipage}[b]{.4\textwidth}
\vspace{-\baselineskip}
\begin{equation}
 \mathcal{N}_{B}:=\hat{P}_B^N \hB\label{eq:non-degenerate-B}
\end{equation}
\end{minipage}
\hspace*{1cm}
\end{center}

\begin{center}
\begin{minipage}[b]{.4\textwidth}
\vspace{-0.2cm}
\begin{equation}
    \hat{P}_{B}^D:=\sum_{k>N}\ket*{b_k}\bra*{b_k}\label{eq:degenerate-projection}
\end{equation}
\end{minipage}
\hfill
\begin{minipage}[b]{.4\textwidth}
\vspace{-\baselineskip}
\begin{equation}
\mathcal{D}_{B}:=\hat{P}_B^D\hB\label{eq:degenerate-B}
\end{equation}
\end{minipage}
\hspace*{1cm}
\end{center}

Therefore, $\rho_B$ has non-zero eigenvalues in $\mathcal{N}_B$ and zero eigenvalues in $\mathcal{D}_{B}$. With the assumption of ${\rm det}(\rho_A)=0$ under our setup, we know $\lambda_{ik;A;B}^{T_1(0)}> 0$ for any $i$ and $1\leq k\leq N$, which will not contribute to the negativity perturbation according to Sec.~\ref{sec:PoN}. On the contrary, $\lambda_{ik,A;B}^{T_1(0)}=0$ for any $i$ and $k\geq N+1$, which can potentially contribute to negativity if the leading order change of $\lambda_{ik,A;B}^{T_1}(t)$ is negative. However, $\lambda_{ik,A;B}^{T_1(0)}=0$ is a degenerate eigenvalue: eigenvalues of $\rho^{T_1(0)}_{A;B}$ vanish in the entire subspace $\hA\ot \mathcal{D}_B$. We therefore need to use the degenerate eigenvalue perturbation theory, of which the results are reviewed in Appendix~\ref{sec:DSOMP}. We apply the formulas of Eq.~\ref{eq:1-reduction-matrix} and \ref{eq:2-reduction-matrix-3} to find $\lambda_{ik,A;B}^{T_1(1)}$ and $\lambda_{ik,A;B}^{T_1(2)}$ for $k>N$, thereby obtaining expressions for $\mathcal{N}^{(1)}_{A;B}$ and $\mathcal{N}^{(2)}_{A;B}$. The detailed steps are shown in Appendix~\ref{sec:negativity-A-B-calculation}. We summarize the results below:
\begin{align}
    \mathcal{N}_{A;B}^{(1)}&=0,
    \label{eq:negativity-A-B-1-summary}\\
   \negaABshort(\rho_A,\rho_B,\hat{H}_{{\rm tot}}):=\mathcal{N}_{A;B}^{(2)}&=\frac{1}{2}\left(\Tr\left[\sqrt{\hat{F}_{A;B}^{\dagger} \hat{F}_{A;B}}\right]-\Tr\left[ \hat{F}_{A;B}\right]\right)\label{eq:negativity-A-B-2-summary},
\end{align}
where we have made the following definitions:
\begin{align}
\hat{F}_{A;B}:&=\sum_{p,q}\hat{F}_A^{p,q}\ot\hat{F}_B^{p,q} \label{eq:F-A-B-summary}\\
\hat{F}_A^{p,q}:&=\sum_{i,k,m}\left(\lambda_A^{m}A^p_{km}A^q_{mi}-\frac{\lambda_A^{i}\lambda_A^{k}}{\lambda_A^m}A^q_{km}A^p_{mi}\right)\ket{a^*_i}\bra{a^*_k}\label{eq:FAB-A-dirac-summary}\\
    &=(\hat{A}^{q}\rho_A\hat{A}^{p}-\rho_A \hat{A}^{p}\rho_A^{-1}\hat{A}^{q}\rho_A)^*
    \label{eq:FAB-A-matrix-summary}\\
\hat{F}_{B}^{p,q}:&=\sum_{j>N,l>N,t}B^p_{jt}\lambda_ B^tB^q_{tl}\ket{b_{j}}\bra{b_{l}}\label{eq:FAB-B-dirac-summary}\\
    &=\hat{P}_B^D\hat{B}^p\rho_B\hat{B}^q\hat{P}_B^D\label{eq:FAB-B-matrix-summary}
\end{align}

According to Eq.~\ref{eq:negativity-A-B-1-summary}, the first-order derivative for the negativity of $A;B$ vanishes at the onset. A similar result of vanishing first-order derivative at the onset is observed for $n$-R\'enyi entropy of the $A;B$ system in \cite{20EmilyFirstPaper} under the same system setup. As seen in the numerical examples of Fig.~\ref{fig:pure_qutrit_AB_perturb_initial} of Appendix~\ref{sec:verification}, our perturbative expressions from Eq.~\ref{eq:negativity-A-B-1-summary}-\ref{eq:FAB-B-matrix-summary} agree with the numerical calculations of negativity at the onset, and the second-order perturbation $\mathcal{N}_{A;B}^{(2)}$ in Eq.~\ref{eq:negativity-A-B-2-summary} indicates the initial speed of entanglement generation between $A$ and $B$. We can in general use the expression of $\mathcal{N}_{A;B}^{(2)}$ in Eq.~\ref{eq:negativity-A-B-2-summary}, the leading-order negativity perturbation, to measure how fast $A$ and $B$ are becoming entangled at the onset under a generic total Hamiltonian. We define $\mathcal{N}_{A;B}^{(2)}$ in Eq.~\ref{eq:negativity-A-B-2-summary} as the \negaAB ($S$), which is a function of the initial states $\rho_A$, $\rho_B$ and the interaction Hamiltonian $\hat{H}_{\rm int}$ between the two subsystems $A$ and $B$. $S(\rho_A,\rho_B,\hat{H}_{{\rm tot}})$ quantifies how the initial states and the interaction Hamiltonian are susceptible to the generation of negativity between $A$ and $B$. Compared to the definition of negativity in Eq.~\ref{eq:negativity-norm}, \negaAB is also dependent on the matrix 1-norm and has essentially the same structure as the negativity with $\rho^{T_1}$ replaced by a more complex operator $\hat{F
}_{A;B}$ to include both the states and 
Hamiltonians. In principle, we can use this new notion of \negaAB for better control of interaction between two quantum systems at the initial stage. With a known Hamiltonian $\hat{H}_{int}$, we can either maximize or minimize $\negaABshort(\rho_A,\rho_B)$ in Eq.~\ref{eq:negativity-A-B-2-summary} by adjusting the initial states of $A$ and $B$ through numerical or analytical procedures to maximize or minimize the entanglement generation between $A$ and $B$ at the onset. We can also obtain the optimal $\hat{H}_{{\rm tot}}$ by maximizing or minimizing $ \negaABshort(\hat{H}_{{\rm tot}})$, depending on whether it is intended to promote entanglement transmission, e.g., to increase a quantum channel capacity or whether it is intended to prevent entanglement transmission, e.g., to suppress decoherence \cite{Shor1995,Monz2009}. We next study the properties of the \negaAB under some special circumstances.

\subsubsection{Free Hamiltonians do not contribute to the \negaAB \texorpdfstring{$\mathcal{N}_{A;B}^{(2)}$}{negaAB}.}
\label{sec:free-negativity-A-B}

For results in Eq.~\ref{eq:negativity-A-B-2-summary}-\ref{eq:FAB-B-matrix-summary}, we have assumed a generic interaction Hamiltonian between $A$ and $B$ with $\hat{H}_{{\rm tot}}=\sum_{p}\hat{A}^p\ot\hat{B}^p$, which incorporates both $\hat{C}\ot\mathbb{I}_B$ and $\mathbb{I}_A\ot\hat{D}$, the free Hamiltonians of $A$ and $B$. We now examine the effects of free Hamiltonians separately from the true interaction terms, and we will show that free Hamiltonians can not contribute to $\mathcal{N}_{A;B}^{(2)}$.

We first consider the free Hamiltonian for $A$. We can reorganize $\hat{H}_{{\rm tot}}$ as:
\begin{align}
    \hat{H}_{{\rm tot}}=\hat{C}\ot\bIB+\sum_{p\geq 2}\hat{A}^p\ot\hat{B}^p
    \label{eq:H-int-A-free}
\end{align}
Then using Eq.~\ref{eq:F-A-B-summary}, \ref{eq:FAB-A-matrix-summary} and \ref{eq:FAB-B-matrix-summary}, we can express $\hat{F}_{A;B}$ as:
\begin{align}
\hat{F}_{A;B}&=\sum_{p\geq 2}(\hat{C}\rho_A\hat{A}^{p}-\rho_A \hat{A}^{p}\rho_A^{-1}\hat{C}\rho_A)^*\ot\hat{P}_B^D\hat{B}^p\rho_B\bIB\hat{P}_B^D\nonumber\\  
&\qquad+\sum_{q\geq 2}(\hat{A}^{q}\rho_A\hat{C}-\rho_A \hat{C}\rho_A^{-1}\hat{A}^{q}\rho_A)^*\ot\hat{P}_B^D\bIB\rho_B\hat{B}^q\hat{P}_B^D\nonumber\\
&\qquad+(\hat{C}\rho_A\hat{C}-\rho_A \hat{C}\rho_A^{-1}\hat{C}\rho_A)^*\ot\hat{P}_B^D\bIB\rho_B\bIB\hat{P}_B^D\nonumber\\
&\qquad+\sum_{p\geq 2,q\geq 2}(\hat{A}^{q}\rho_A\hat{A}^{p}-\rho_A \hat{A}^{p}\rho_A^{-1}\hat{A}^{q}\rho_A)^*\ot\hat{P}_B^D\hat{B}^p\rho_B\hat{B}^q\hat{P}_B^D
\label{eq:FAB-C}
\end{align}
With $\rho_B$ has zero eigenvalues in the subspace $\mathcal{D}_B$, then
\begin{equation}
    \rho_B\bIB\hat{P}_B^D=\sum_{k=1}^N\lambda_B^k\ket*{b_k}\bra*{b_k}\sum_{u>N}\ket*{b_u}\bra*{b_u}=0.
    \label{eq:FAB-C-B}
\end{equation}
Similarly, $\hat{P}_B^D\bIB\rho_B=0$, so the first three terms of Eq.~\ref{eq:FAB-C} vanish, and $\hat{F}_{A;B}$ do not contain the free Hamiltonian $\hat{C}$. Since $\mathcal{N}_{A;B}^{(2)}$ only depends on $\hat{F}_{A;B}$ according to Eq.~\ref{eq:negativity-A-B-2-summary}, the free Hamiltonian term $\hat{C}\ot\bIB$ do not contribute to $\mathcal{N}_{A;B}^{(2)}$. 

Let us now consider the term $\bIA\ot\hat{D}$. We similarly reorganize the interaction Hamiltonian:
\begin{align}
    \hat{H}_{{\rm tot}}=\bIA\ot\hat{D}+\sum_{p\geq 2}\hat{A}^p\ot\hat{B}^p
    \label{eq:H-int-B-free}
\end{align}
Using Eq.~\ref{eq:F-A-B-summary}, \ref{eq:FAB-A-matrix-summary} and \ref{eq:FAB-B-matrix-summary}, we can express $\hat{F}_{A;B}$ as:
\begin{align}
\hat{F}_{A;B}&=\sum_{p\geq 2}(\bIA\rho_A\hat{A}^{p}-\rho_A \hat{A}^{p}\rho_A^{-1}\bIA\rho_A)^*\ot\hat{P}_B^D\hat{B}^p\rho_B\hat{D}\hat{P}_B^D\nonumber\\  
&\qquad+\sum_{q\geq 2}(\hat{A}^{q}\rho_A\bIA-\rho_A \bIA\rho_A^{-1}\hat{A}^{q}\rho_A)^*\ot\hat{P}_B^D\hat{D}\rho_B\hat{B}^q\hat{P}_B^D\nonumber\\
&\qquad+(\bIA\rho_A\bIA-\rho_A \bIA\rho_A^{-1}\bIA\rho_A)^*\ot\hat{P}_B^D\hat{D}\rho_B\hat{D}\hat{P}_B^D\nonumber\\
&\qquad+\sum_{p\geq 2,q\geq 2}(\hat{A}^{q}\rho_A\hat{A}^{p}-\rho_A \hat{A}^{p}\rho_A^{-1}\hat{A}^{q}\rho_A)^*\ot\hat{P}_B^D\hat{B}^p\rho_B\hat{B}^q\hat{P}_B^D
\label{eq:FAB-D}
\end{align}
We see that the first term in Eq.~\ref{eq:FAB-D} vanishes, since $\bIA\rho_A\hat{A}^{p}-\rho_A \hat{A}^{p}\rho_A^{-1}\bIA\rho_A=\rho_A\hat{A}^{p}-\rho_A\hat{A}^{p}=0$. Similarly, the second and third term of Eq.~\ref{eq:FAB-D} also vanish, so $\hat{F}_{A;B}$ do not contain the free Hamiltonian $\hat{D}$, which implies the free Hamiltonian on $B$ can not contribute to $\mathcal{N}_{A;B}^{(2)}$. 

Hence, we have shown that terms of the form $\hat{C}\ot\bIB$ and $\bIA\ot\hat{D}$ (i.e. free Hamiltonians of $A$ and $B$) do not contribute to the first and second-order time derivatives of the negativity for $A;B$, which is confirmed  by our numerical example in Fig.~\ref{fig:pure_qutrit_AB_perturb_initial}. We can therefore ignore the free Hamiltonians of $A$ and $B$ when we study the dynamics of entanglement between $A$ and $B$ at the onset of interaction to second-order. A similar phenomenon is observed for the first and second-order time-derivatives of the n-purity for $A;B$ in \cite{21EmilySecondPaper} under the same setup. Both negativity and $n$-purity provide evidence that free Hamiltonians do not immediately impact the quantum correlation between $A$ and $B$ at the onset. This, curiously, also means that resonance phenomena, which require free Hamiltonians, do not occur upto the second-order, a phenomenon that will be interesting to investigate further. 

\subsubsection{\texorpdfstring{$A$}{A} and \texorpdfstring{$B$}{B} can become immediately entangled under \texorpdfstring{$\hat{H}_{{\rm tot}}=\hat{A}\ot\hat{B}$}{product Hamiltonian}.} \label{sec:pure-pf-AB-entangled}

We here show the \negaAB $\mathcal{N}_{A;B}^{(2)}$ is positive under a generic product Hamiltonian $\hat{H}_{{\rm tot}}=\hat{A}\ot\hat{B}$. The positive negativity perturbation at the second-order suggests $A$ and $B$ become immediately entangled at the onset of interaction when ${\rm det}(\rho_B)=0$. This is in stark contrast to the case with ${\rm det}(\rho_B)\neq0$ studied in Sec.~\ref{sec:mixed, finite time}, where the negativity for $A;B$ will remain zero for a finite amount of time regardless of the interaction Hamiltonian, which prohibits fast entanglement generation within $A;B$. 

When the total Hamiltonian takes the product form $\hat{H}_{{\rm tot}}=\hat{A}\ot\hat{B}$, Eq.~\ref{eq:F-A-B-summary}-\ref{eq:FAB-B-matrix-summary} simplifies to:
\begin{align}
\hat{F}_{A;B}:&=\hat{F}_A\ot\hat{F}_B \label{eq:F-A-B-product}\\
\hat{F}_A:&=\sum_{i,k,m}\left(\lambda_A^{m}A_{km}A_{mi}-\frac{\lambda_A^{i}\lambda_A^{k}}{\lambda_A^m}A_{km}A_{mi}\right)\ket{a^*_i}\bra{a^*_k}\label{eq:FAB-A-dirac-product}\\
    &=(\hat{A}\rho_A\hat{A}-\rho_A \hat{A}\rho_A^{-1}\hat{A}\rho_A)^*
    \label{eq:FAB-A-matrix-product}\\
\hat{F}_{B}:&=\sum_{j>N,l>N,t}B_{jt}\lambda_ B^tB_{tl}\ket{b_{j}}\bra{b_{l}}\label{eq:FAB-B-dirac-product}\\
    &=\hat{P}_B^D\hat{B}\rho_B\hat{B}\hat{P}_B^D=\sum_{i}\lambda_B^i\left(\hat{P}_B^D\hat{B}\ket{b_i}\right)\left(\bra{b_i}\hat{B}\hat{P}_B^{D}\right)\label{eq:FAB-B-matrix-product}
\end{align}

We see that $\hat{F}_{A;B}$ is now in the product form according to Eq.~\ref{eq:F-A-B-product}, which suggests the eigenvectors of $\hat{F}_{A;B}$ will be in the product state, and we can obtain eigenvalues of $\hat{F}_{A;B}$ by diagonalizing $\hat{F}_A$ and $\hat{F}_B$ separately. According to Eq.~\ref{eq:negativity-A-B-2-summary}, $\mathcal{N}_{A;B}^{(2)}$ is the absolute value of the sum of all negative eigenvalues of $\hat{F}_{A;B}$. As seen in Eq.~\ref{eq:FAB-B-matrix-product}, $\hat{F}_B$ is positive (semi-)definite as the sum of positive (semi-)definite matrices, so the signs of eigenvalues of $\hat{F}_{A;B}$ only depends on 
$\hat{F}_A$, and the \negaAB in Eq.~\ref{eq:negativity-A-B-2-summary} can be simplified to:
\begin{align}
   \negaABshort(\rho_A,\rho_B,\hat{A}\ot\hat{B})&:=\mathcal{N}_{A;B}^{(2)}=\frac{\Tr[\hat{F}_B]}{2}\left(\Tr\left[\sqrt{\hat{F}_{A}^{\dagger} \hat{F}_{A}}\right]-\Tr\left[ \hat{F}_{A}\right]\right),
   \label{eq:S-AB-product}
\end{align}
where 
\begin{align}
\Tr \left[\hat{F}_{B}\right]=\Tr [\hat{P}_B^D\hat{B}\rho_B\hat{B}\hat{P}_B^D]=\sum_{y>N,t}|B_{yt}|^2\lambda_B^t=\sum_{t\leq N,y>N}|B_{yt}|^2\lambda_B^t\geq 0.
\label{eq:trace-FB-product}
\end{align}
We can see that $\Tr_{}[\hat{F}_{B}]$ is in fact the second moment of $\hat{P}_B^D\hat{B}$ under $\rho_B$, and $\Tr_{}[\hat{F}_{B}]>0$ as long as there exists some $t\leq N, y\geq N$ such that $\bra*{b_y}\hat{B}\ket*{b_t}\neq 0$, which is equivalent to $\hat{P}_B^{D}B\hat{P}_B^{N}\neq 0$. Through demonstrating $\Tr_{}[\hat{F}_{A}]<0$, we can show some eigenvalues of $\hat{F}_{AB}$ will guaranteed to be negative, which implies $\mathcal{N}_{A;B}^{(2)}>0$. Using Eq.~\ref{eq:FAB-A-dirac-product}, we have: 
\begin{align}
\Tr\left[\hat{F}_{A}\right]&=\sum_{u,m}\left\{\lambda_A^{m}-\frac{(\lambda_A^{u})^2}{\lambda_A^m}\right\}|A_{um}|^2=\sum_{u,m\neq u}\left\{\lambda_A^{m}-\frac{(\lambda_A^{u})^2}{\lambda_A^m}\right\}|A_{um}|^2\\
&=\sum_{u}\sum_{m\geq u+1}\left\{\left(\lambda_A^{m}-\frac{(\lambda_A^{u})^2}{\lambda_A^m}\right)|A_{um}|^2+\left(\lambda_A^{u}-\frac{(\lambda_A^{m})^2}{\lambda_A^u}\right)|A_{mu}|^2\right\}\\
&=-\sum_{u}\sum_{m\geq u+1}|A_{mu}|^2\frac{
\lambda_A^{m}+\lambda_A^{u}}{\lambda_A^m\lambda_A^u}
\left(\lambda_A^m-\lambda_A^u\right)^2\leq 0
\label{eq:trace-FA-product}
\end{align}

In Eq.~\ref{eq:trace-FA-product}, $\Tr_{}[\hat{F}_{A}]<0$ as long as there exists some $m,u$ such that $\bra*{a_u}\hat{A}\ket*{a_m}\neq 0$ and $\lambda_A^m\neq \lambda_A^u$, which is equivalent to $[\hat{A},\rho_A]\neq 0$. When $\hat{A}$ and $\rho_A$ commute, we have $\Tr_{}[\hat{F}_{A}]=0$, and we do not know whether $\hat{F}_A$ has negative eigenvalues or not, so we can not determine whether $\mathcal{N}_{A;B}^{(2)}$ vanishes. We can conclude that when $\hat{P}_B^{D}B\hat{P}_B^{N}\neq 0$ and $[\hat{A},\rho_A]\neq 0$, $\mathcal{N}_{A;B}^{(2)}>0$, which implies that $A$ and $B$ will become immediately entangled at the onset. Most generic choices of $\hat{A}$ and $\hat{B}$ will satisfy the above requirements and ensure $\Tr_{}[\hat{F}_{AB}]<0$, which implies $\mathcal{N}_{AB}^{(2)}>0$. In fact, when $\Tr_{}[\hat{F}_{AB}]=\Tr_{}[\hat{F}_{A}]\Tr_{}[\hat{F}_{B}]<0$, the absolute value of this trace provides a lower bound for the negativity susceptibility $\mathcal{N}_{A;B}^{(2)}$.

\subsubsection*{}
\begin{figure}[htbp]
\centerline{\includegraphics[width=0.63\hsize]{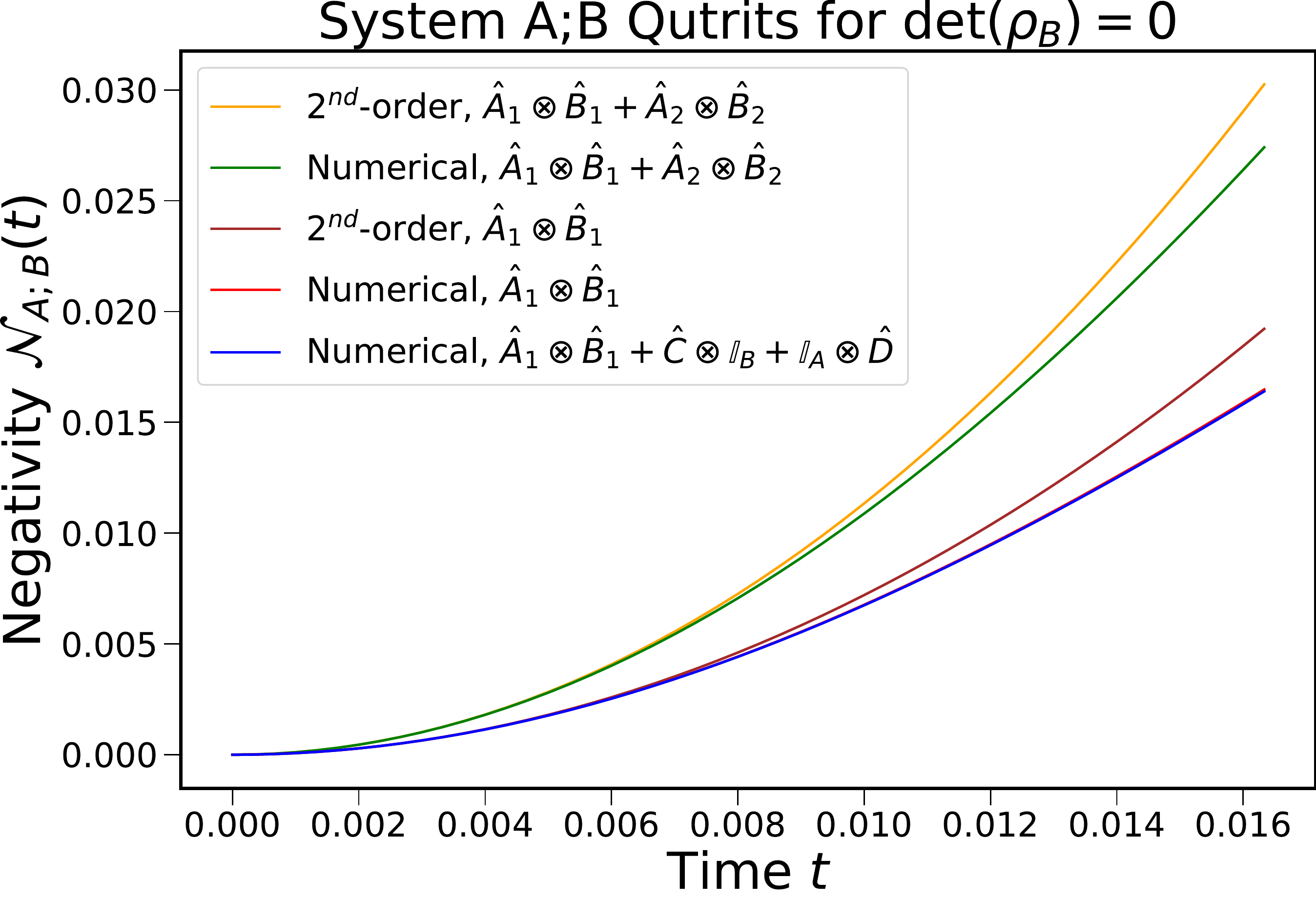}}
\caption{The dynamics of negativity for $A;B$ assuming $\det(\rho_B)=0$ under our setup. We plot both the numerical calculations and the perturbative results to the $2^{nd}$-order using Eq.~\ref{eq:negativity-A-B-1-summary}-\ref{eq:FAB-B-matrix-summary}. The curves for numerical results under $H_{\rm int}=\hat{A}_1\ot\hat{B}_1$ and $H_{\rm int}=\hat{A}_1\ot\hat{B}_1+\hat{C}\ot\bIB+\bIA\ot\hat{D}$ overlap, which verifies the result in Sec.~\ref{sec:free-negativity-A-B} that free Hamiltonians $\hat{C}$ and $\hat{D}$ do no contribute to the $2^{nd}$-order negativity perturbation at the onset.}
\label{fig:pure_qutrit_AB_perturb_initial}
\end{figure}

We conclude that product Hamiltonian $\hat{H}_{\rm tot}=\hat{A}\ot\hat{B}$ (such that $\hat{P}_B^{D}\hat{B}\hat{P}_B^{N}\neq 0$ and $[\hat{A},\rho_A]\neq 0$) produces non-vanishing second-order derivative for the negativity of system $A;B$, and we expect a generic total Hamiltonian $\hat{H}_{\rm tot}$ with multiple terms to have non-vanishing $\mathcal{N}_{A;B}^{(2)}$ as well according to Eq.~\ref{eq:negativity-A-B-2-summary}. $A$ and $B$ can indeed become immediately entangled at the start of the interaction when ${\rm det}(\rho_{B})=0$, which is verified by the numerical examples in Fig.~\ref{fig:pure_qutrit_AB_perturb_initial}. In contrast, Sec.~\ref{sec:mixed, finite time} shows that the negativity for $A;B$ will remain $0$ for a finite amount of time when ${\rm det}(\rho_{B})\neq0$. We see that zero eigenvalues in $\rho_{B}$ causes substantial difference in the dynamic of $\mathcal{N}_{A;B}(t)$ and allows fast entanglement generation within $A;B$. In the case that both $A$ and $B$ are qubits ($\rho_A$ is assumed to be mixed), a mixed $\rho_B$ prevents entanglement generation with $A;B$ for a finite amount of time, while $A$ and $B$ can become immediately entangled when $\rho_B$ is pure. Therefore, we need to choose $\rho_B$ pure when we want to generate entanglement between $A$ and $B$ as fast as possible, while we can prevent $A$ and $B$ from becoming entangled for a finite amount of time when $\rho_B$ is mixed regardless of the interaction Hamiltonian between the two subsystems. 

A similar phenomenon is observed for the von-Neumann entropy of $A;B$ in \cite{21EmilySecondPaper} under the same setup: zero eigenvalues in $\rho_B$ cause divergence in the second-order derivative of the von-Neumann entropy at the onset of interaction. A divergent second-order derivative suggests the fastest increase of von-Neumann entropy, which only occurs when $\rho_B$ has zero eigenvalues or ${\rm det}(\rho_B)=0$. Unlike negativity, von-Neumann entropy is not a proper measure for bipartite entanglement, but it does provide further evidence for the importance of zero eigenvalues of $\rho_B$ in the dynamics of quantum correlation between $A$ and $B$. $\rho_B$ with zero eigenvalues is more prone to the generation of quantum correlation between $A$ and $B$.

Our perturbative calculation of $\mathcal{N}_{A;B}^{(2)}$ in this section is more general than our system setup and is independent of the subsystem $\tlA$. Eq.~\ref{eq:negativity-A-B-2-summary} applies to any bipartite systems with a generic interaction Hamiltonian and an initial state $\rho_A\ot\rho_B$ where ${\rm det}(\rho_A)\neq 0$ and  ${\rm det}(\rho_B)=0$ or ${\rm det}(\rho_A)=0$ and  ${\rm det}(\rho_B)\neq 0$. Our different assumptions on ${\rm det}(\rho_A)$ and ${\rm det}(\rho_B)$ lead to the asymmetry between $A$ and $B$ in $\hat{F}_{A;B}$ in Eq.~\ref{eq:F-A-B-summary}-\ref{eq:FAB-B-matrix-summary}. The case where both ${\rm det}(\rho_A)$ and ${\rm det}(\rho_B)$ are non-zero is covered in Sec.~\ref{sec:mixed, finite time}, where we have found that negativity will vanish for a finite amount of time. The case where both ${\rm det}(\rho_A)$ and ${\rm det}(\rho_B)$ vanish is not covered by this work, but in the special case that both $\rho_1$ and $\rho_2$ are pure, we can simply use the second-order derivative of the n-R\'enyi entropy $\ddot{H}_n|_{t=t_0}$ to indicate the change of entanglement at the onset, since n-R\'enyi entropy is a proper entanglement measure when the bipartite system is pure. From \cite{21EmilySecondPaper}, we have
\begin{align}
\ddot{H}_n|_{t=t_0}=\frac{4(\Delta A)^2(\Delta B)^2}{n-1},
\label{eq:renyi-derivative}
\end{align}
when both $A$ and $B$ are initially pure. The speed of entanglement generation within $A;B$ is simply proportional to the variance terms of both subsystems.

\subsection{Negativity of System \texorpdfstring{$\tilde{A};B$}{A;B}}
\label{sec:negativity-perturbed-tlA-B}

We now focus on the bipartite system $\tlA;B$ under our setup described in Sec.~\ref{sec:system setup}. In Sec.~\ref{sec:mixed, finite time}, we have shown that the negativity between $\tlA$ and $B$ remains zero for a finite amount of time if ${\rm det}(\rho_B)\neq 0$ with the assumption of ${\rm det}(\rho_A)\neq 0$ under our system setup. We here study how fast a generic interaction Hamiltonian between $A$ and $B$ can transfer entanglement from $\tlA;A$ to $\tlA;B$ at the onset assuming ${\rm det}(\rho_B)=0$. The calculation of the first and second order perturbations of the negativity of $\tlA;B$ is similar to the case for the $A;B$ system in the previous section. We leave the detailed steps to Appendix~\ref{sec:negativity-tlA-B-calculation} and summarize  $\mathcal{N}^{(1)}_{\tlA;B}$ and $\mathcal{N}^{(2)}_{\tlA;B}$ below:
\begin{align}
    \mathcal{N}_{\tlA;B}^{(1)}&=0
    \label{eq:negativity-tlA-B-1-summary}\\
    T(\rho_{\tlA A},\rho_B,\hat{H}_{{\rm tot}}):=\mathcal{N}_{\tlA;B}^{(2)}&=\frac{1}{2}\left(\Tr\left[\sqrt{\hat{F}_{\tlA ;B}^{\dagger} \hat{F}_{\tlA ;B}}\right]-Tr\left[ \hat{F}_{\tlA ;B}\right]\right)\label{eq:negativity-tlA-B-2-summary},
\end{align}
where we make the following new definitions:
\begin{align}
\hat{F}_{\tlA; B}:&=\sum_{p,q}\hat{F}_{\tlA}^{p,q}\ot\hat{F}_B^{p,q} \label{eq:F-tlA-B-summary}\\
\hat{F}_{\tlA}^{p,q}:&=\sum_{i,j,m} \alpha_i\alpha_{j}\left(A^q_{im}A^p_{mj}-A^p_{im}A^q_{mj}\right)\ket*{\tla^*_i}\bra*{\tla^*_j}
\label{eq:FtlAB-tlA-dirac-summary}\\
    &=\left\{\sum_{i,j}\left(\bra*{a_j}\hat{A}^{p}\hat{A}^{q}-\hat{A}^{q}\hat{A}^{p}\ket*{a_i}\right)\alpha_i\alpha_j\ket*{\tla_i}\bra*{\tla_j}\right\}^*,
    \label{eq:FtlAB-tlA-matrix-summary}
\end{align}
and $\hat{F}_{B}^{p,q}$ has already been given in Eq.~\ref{eq:FAB-B-dirac-summary} and \ref{eq:FAB-B-matrix-summary}. If we express $\{\ket{a_i}\}$ as the coordinate basis in $\hA$, that is $\ket{a_i}\equiv\vec{e}_i$, and $\{\ket{
\tla_i}\}$ as the coordinate basis in $\hlA$, then $\hat{F}^{p,q}_{\tlA}$ simplifies to
\begin{align}
    \hat{F}^{p,q}_{\tlA}\equiv (\hat{A}^q\hat{A}^p-\hat{A}^p\hat{A}^q)\circ R_{A},
     \label{eq:FtlAB-tlA-matrix-coordinate-basis}
\end{align}
where $\circ$ represents the element-wise product, and we define the $R_{A}$ matrix:
\begin{align}
 R_{A}:=\sum_{i,j}\alpha_i\alpha_j\ket{\alpha_i}\bra{\alpha_j}\equiv\sum_{i,j}\alpha_i\alpha_j\vec{e}_i\vec{e}_j^T,
\end{align}
which is completely specified by any given initial state $\rho_A$ for the subsystem A. Eq.~\ref{eq:FtlAB-tlA-matrix-coordinate-basis} is a coordinate-dependent expression. According to Eq.~\ref{eq:negativity-tlA-B-1-summary}, the first-order derivative of negativity for $\tlA; B$ vanishes at the onset of interaction, similar to the negativity for $A;B$. The second-order derivative $\mathcal{N}_{\tlA;B}^{(2)}$ in Eq.~\ref{eq:negativity-tlA-B-2-summary} shares a similar structure to $\mathcal{N}_{A;B}^{(2)}$ in Eq.~\ref{eq:negativity-A-B-2-summary}. We define $\mathcal{N}_{\tlA;B}^{(2)}$ in Eq.~\ref{eq:negativity-tlA-B-2-summary} as the negativity transmissibility  $T(\rho_{\tlA A},\rho_B,\hat{H}_{{\rm tot}})$, which can be used to measure how fast the entanglement from $\tlA; A$ can be transferred to $\tlA;B$ through an interaction Hamiltonian between $A$ and $B$ at the onset. Since $\hat{A}^p$ and $\hat{A}^q$ generally do not commute when $p\neq q$, $\hat{F}_{\tlA}^{p,q}$ in Eq.~\ref{eq:FtlAB-tlA-matrix-summary} is non-trivial, and $\mathcal{N}^{(2)}_{\tlA;B}$ can be non-zero, so $\tlA$ and $B$ can become entangled immediately at the onset for a generic interaction Hamiltonian between $A$ and $B$. We confirm our perturbative results in Eq.~\ref{eq:negativity-tlA-B-1-summary}-\ref{eq:FtlAB-tlA-matrix-summary} with numerical calculations in Fig.~\ref{fig:pure_qutrit_tlAB_perturb_initial} in Appendix~\ref{sec:verification}.

We can in general maximize the value of negativity transmissibility $T(\rho_{\tlA A},\rho_B,\hat{H}_{{\rm tot}})$ in order to maximize the speed of entanglement transfer from $\tlA;A$ to $\tlA;B$. Using Eq.~\ref{eq:negativity-A-B-2-summary} and \ref{eq:negativity-tlA-B-2-summary}, we can compare the initial speed of entanglement generation within $A;B$ and $\tlA;B$ and probe how much entanglement within $\tlA;A$ get transferred to $\tlA;B$ or $A;B$ initially. We next study the properties of negativity transmissibility $\mathcal{N}_{\tlA;B}^{(2)}$.

\subsubsection{Free Hamiltonians do not contribute to the negativity transmissibility \texorpdfstring{$\mathcal{N}_{\tlA;B}^{(2)}$}{negatlAB}.} \label{sec:free-negativity-tlA-B}

Similar to Sec.~\ref{sec:free-negativity-A-B}, we now examine the effects of free Hamiltonians on the perturbation of negativity between $\tlA$ and $B$, and we show that free Hamiltonians also can not contribute to $\mathcal{N}_{\tlA;B}^{(2)}$.

We first consider the free Hamiltonian for $A$. $\hat{H}_{{\rm tot}}$ can be expressed in the form of Eq.~\ref{eq:H-int-A-free}. Using Eq.~\ref{eq:F-tlA-B-summary}, \ref{eq:FtlAB-tlA-matrix-summary} and \ref{eq:FAB-B-matrix-summary}, we can show that all terms containing $\hat{C}$ will vanish following exactly the same argument in Eq.~\ref{eq:FAB-C} and \ref{eq:FAB-C-B} for the case of the system $A;B$, since $\hat{F}_{B}^{p,q}$ shares exactly the same expression in both Eq.~\ref{eq:F-A-B-summary} and \ref{eq:F-tlA-B-summary}. 

Next we consider the free Hamiltonian for $B$ where $\hat{H}_{{\rm tot}}$ is expressed in the form of Eq.~\ref{eq:H-int-B-free}. We can express $\hat{F}_{\tlA;B}$ in Eq.~\ref{eq:F-tlA-B-summary} as:
\begin{align}
\hat{F}_{\tlA;B}&=\sum_{p\geq 2}\left\{\sum_{i,j}\left(\bra*{a_j}\hat{A}^{p}\mathbb{I}_A-\mathbb{I}_A\hat{A}^{p}\ket*{a_i}\right)\alpha_i\alpha_j\ket*{\tla_i}\bra*{\tla_j}\right\}^*\ot\hat{P}_B^D\hat{B}^p\rho_B\hat{D}\hat{P}_B^D\nonumber\\  
&\qquad+\sum_{q\geq 2}\left\{\sum_{i,j}\left(\bra*{a_j}\mathbb{I}_A\hat{A}^{q}-\hat{A}^{q}\mathbb{I}_A\ket*{a_i}\right)\alpha_i\alpha_j\ket*{\tla_i}\bra*{\tla_j}\right\}^*\ot\hat{P}_B^D\hat{D}\rho_B\hat{B}^q\hat{P}_B^D\nonumber\\
&\qquad+\sum_{p\geq 2,q\geq 2}\left\{\sum_{i,j}\left(\bra*{a_j}\hat{A}^{p}\hat{A}^{q}-\hat{A}^{q}\hat{A}^{p}\ket*{a_i}\right)\alpha_i\alpha_j\ket*{\tla_i}\bra*{\tla_j}\right\}^*\ot\hat{P}_B^D\hat{B}^p\rho_B\hat{B}^q\hat{P}_B^D
\label{eq:FtlAB-D}
\end{align}
We see that the first and second terms in Eq.~\ref{eq:FtlAB-D} vanish, since $\hat{A}^{p}\bIA-\bIA\hat{A}^p=0$. Therefore, the free Hamiltonian term $\bIA\ot\hat{D}$ can not contribute to $\mathcal{N}_{\tlA;B}^{(2)}$. 

Hence, we have shown that terms of the form $\hat{C}\ot\bIB$ and $\bIA\ot\hat{D}$ (i.e. free Hamiltonians of $A$ and $B$) do not contribute to the first and second-order time derivatives of the negativity for $\tlA ;B$. We can therefore ignore the free Hamiltonians of $A$ and $B$ to second-order when we study the dynamics of entanglement between $\tlA$ and $B$ at the onset of interaction. Free Hamiltonians are simply not useful for entanglement transfer from $\tlA;A$ to $\tlA;B$ at the onset. However, free Hamiltonian can in fact impact the entanglement within $\tlA;B$ beyond the perturbation regime, which will be discussed in Sec.~\ref{sec:pf-free-tlA-B-separable}.

\subsubsection{\texorpdfstring{$\hat{H}_{{\rm tot}}=\hat{C}\ot\bIB+\hat{A}\ot\hat{B}+\bIA\ot\hat{D}$}{product Hamiltonian with free Hamiltonians} can not immediately entangle \texorpdfstring{$\tlA$}{A} and \texorpdfstring{$B$}{B}.} \label{sec:pure-pf-tlAB-unentangled-perturbed}

When we consider a product Hamiltonian $\hat{H}_{{\rm tot}}=\hat{A}\ot \hat{B}$, Eq.~\ref{eq:FtlAB-tlA-matrix-summary} becomes
\begin{align}
\hat{F}_{\tlA}=\left\{\sum_{i,j}\left(\bra*{a_j}\hat{A}\hat{A}^-\hat{A}\hat{A}\ket*{a_i}\right)\alpha_i\alpha_j\ket*{\tla_i}\bra*{\tla_j}\right\}^*=0,
  \label{eq:FtlAB-tlA-product-form}
\end{align}
Therefore, $\hat{F}_{\tilde{A};B}=0$ with the product Hamiltonian, and consequently the negativity transmissibility vanishes $\mathcal{N}_{\tlA;B}^{(2)}=0$ according to Eq.~\ref{eq:negativity-tlA-B-2-summary} and \ref{eq:F-tlA-B-summary}. As we just established, the free Hamiltonians will not contribute to $\mathcal{N}_{\tlA;B}^{(2)}$. Therefore, a total Hamiltonian between $A$ and $B$ with the form $\hat{H}_{{\rm tot}}=\hat{C}\ot\bIB+\hat{A}\ot\hat{B}+\bIA\ot\hat{D}$ is not able to generate negativity between $\tilde{A}$ and $B$ upto second-order at the onset. We will explore the separability of $\tlA$ and $B$ beyond perturbation regime with a product Hamiltonian in Sec.~\ref{sec:pf-nf-tlA-B-separable}. 

Eq.~\ref{eq:S-AB-product} shows that $A$ and $B$ can become immediately entangled at the onset of interaction under a product Hamiltonian, while the negativity between $\tlA$ and $B$ remains zero upto the second order according to Eq.~\ref{eq:FtlAB-tlA-product-form}. This suggests that more entanglement within $\tlA;A$ gets transferred to $A;B$ compared to $\tlA;B$ at the onset.

\section{Delocalization of Entanglement}\label{sec:delocalization}

In the previous two sections, we analyzed the dynamics of negativity for systems $A;B$ and $\tlA;B$ at the onset of interaction in the cases of ${\rm det}(\rho_B)=0$ and ${\rm det}(\rho_B)\neq0$ respectively. We now perturbatively expand the negativity for the system $\tlA;A$ and synthesize these results to study the negativity for systems $\tlA;AB$ (the bipartite system between $\tlA$ and $AB$ where we treat $AB$ as a single subsystem) and $B;\tlA A$. We observe that the entanglement for systems $\tlA;AB$ and $B;\tlA A$ can become delocalized, though the degree of delocalization depends on the choice of the total Hamiltonian $\hat{H}_{{\rm tot}}$ between $A$ and $B$ and the initial state of $\rho_B$. Our analysis on the delocalization of the $\tlA;AB$ and $B;\tlA A$ entanglement provides specific examples on how to generate a delocalized state and how to access the delocalized part of the entanglement, which is crucial for leveraging the advantages of quantum computing and quantum information processing. 

\subsection{Perturbative Results for the Negativity of \texorpdfstring{$\tilde{A};A$}{AA}}
\label{sec:negativity-perturbed-tlA-A}

To complete the study of bipartite entanglement in our system using perturbation, we will calculate the first and second-order derivative of negativity for the system $\tlA;A$. The perturbation of $\mathcal{N}_{\tlA;A}(t)$ is of particular importance, since it can represent the loss of entanglement to the environment when the system of interest $\tlA;A$ starts at a pure state. When we desire to prevent the loss of entanglement, the interaction Hamiltonian $\hat{H}_{{\rm tot}}$ between $A$ and $B$ is usually small, which renders the perturbation approach more applicable compared to the previous studies of the $A;B$ and $\tlA;B$ systems. We can therefore benefit from perturbative results that quantify how much bipartite entanglement between $\tlA$ and $A$ is lost initially. 

The detailed calculation steps for the perturbations of $\mathcal{N}_{\tlA;A}(t)$ are shown in Appendix~\ref{sec:negativity-tlA-A-calculation}, and we summarize the results below:
\begin{align}
&\mathcal{N}_{\tlA;A}^{(1)}=0
\label{eq:negativity-tlA-A-1-summary}\\
&V(\rho_A,\rho_B,\hat{H}_{{\rm tot}}):=\mathcal{N}_{\tlA;A}^{(2)}\nonumber\\
&=-\frac{1}{2}\sum_{p,q}{\rm ucov}(\hat{B}^p,\hat{B}^q)\left(\Tr[\sqrt{\rho_A}]\Tr[\hat{A}^p\sqrt{\rho_A}\hat{A}^q]-\Tr[\sqrt{\rho_A}\hat{A}^p]\Tr[\sqrt{\rho_A}\hat{A}^q]\right),
\label{eq:negativity-tlA-A-2-summary}
\end{align}
where we defined the unsymmetrized covariance between two Hamiltonians (observables) $\hat{B}^p$ and $\hat{B}^q$ as
\begin{align}
{\rm ucov}(\hat{B}^p,\hat{B}^q):=\Tr[\hat{B}^p\rho_B\hat{B}^q]-\Tr[\hat{B}^p\rho_B]\Tr[\hat{B}^q\rho_B]\label{eq:ucov-Bpq-summary}.
\end{align}
This is in contrast to the symmetrized covariance (the quantum analogue of covariance) between two observables \cite{11quantumcovariance}:
\begin{equation}
    {\rm Cov}(\hat{B}^p,\hat{B}^q):=\frac{1}{2}\Tr[\hat{B}^p\rho_B\hat{B}^q+\hat{B}^q\rho_B\hat{B}^p]-\Tr[\hat{B}^p\rho_B]\Tr[\hat{B}^q\rho_B]\label{eq:Cov-Bpq}
\end{equation}
Both unsymmetrized and symmetrized definitions of covariance reduce to the familiar expression of variance when $p=q$, and ${\rm ucov}(\hat{B}^p,\hat{B}^q)={\rm Cov}(\hat{B}^p,\hat{B}^q)$ when $\hat{B}^{p}$ and $\hat{B}^q$ commute. 

As shown in Eq.~\ref{eq:negativity-tlA-A-1-summary}, the first-order derivative of the negativity for $\tlA;A$ also vanishes just as the cases for the systems $A;B$ and $\tlA;B$ in Sec.~\ref{sec:pure, perturbative}. We define the second-order time-derivative of the negativity for $\tlA;A$ in Eq.~\ref{eq:negativity-tlA-A-2-summary} as the negativity vulnerability $V(\rho_A,\rho_B,\hat{H}_{{\rm tot}})$, which indicates how fast the initial entanglement within $\tlA;A$ will be lost due to the interaction proceeding between $A$ and $B$. Notice that the negativity vulnerability $V(\rho_A,\rho_B,\hat{H}_{{\rm tot}})$ should generally have negative values to indicate the loss of entanglement. Therefore, in order to protect the entanglement between $\tlA$ and $A$ and minimize the loss of the entanglement to the environment, we need to maximize the value of negativity vulnerability in Eq.~\ref{eq:negativity-tlA-A-2-summary} by adjusting $\rho_A$, $\rho_B$, or $\hat{H}_{{\rm tot}}$. We will give an example of such maximization procedure in the special case of $A$ and $B$ as qubits and a product Hamiltonian in the following Sec.~\ref{sec:pf-tlAA-lost-entanglement}. We verify our perturbative results in Eq.~\ref{eq:negativity-tlA-A-1-summary} and \ref{eq:negativity-tlA-A-2-summary} with the numerical examples in Fig.~\ref{fig:pure_qutrit_tlAA_perturb_initial} of Appendix~\ref{sec:verification}, and we see that our perturbative calculations agree with the numerical results at the onset of interaction. We emphasize that the expression of negativity vulnerability in Eq.~\ref{eq:negativity-tlA-A-2-summary} applies to both situations where ${\rm det}(\rho_B)=0$ and ${\rm det}(\rho_B)\neq0$, which is different from the dynamics of the $A;B$ and $\tlA;B$ systems in the previous Sec.~\ref{sec:mixed, finite time} and \ref{sec:pure, perturbative} where the negativity susceptibility and transmissibility introduced in Eq.~\ref{eq:negativity-A-B-2-summary} and \ref{eq:negativity-tlA-B-2-summary} only apply in the case of ${\rm det}(\rho_B)=0$.

\subsubsection{Free Hamiltonians do not contribute to the negativity vulnerability \texorpdfstring{$\mathcal{N}_{\tlA; A}^{(2)}$}{negaAA}.}\label{sec:free-negativity-tlA-A}

Similar to Sec.~\ref{sec:free-negativity-A-B} and \ref{sec:free-negativity-tlA-B}, we can also show that free Hamiltonians on $A$ or $B$ do not contribute to the negativity vulnerability $\mathcal{N}_{\tlA;A}^{(2)}$.  

We first consider the free Hamiltonian $\hat{C}$ for the subsystem $A$ where $\hat{H}_{{\rm tot}}$ is given in Eq.~\ref{eq:H-int-A-free}. We see that all terms in Eq.~\ref{eq:negativity-tlA-A-2-summary} containing the free Hamiltonian $\hat{C}\ot\mathbb{I}_B$ will vanish, since all the unsymmetrized covariance terms involving the identity operator will vanish:  
\begin{equation}
    {\rm ucov}(\hat{B}^p,\mathbb{I}_B)=\Tr[\hat{B}^p\rho_B\mathbb{I}_B]-\Tr[\hat{B}^p\rho_B]\Tr[\mathbb{I}_B\rho_B]=0.
\end{equation}
Next we consider the impact of the free Hamiltonian $\hat{D}$ for $B$ where $\hat{H}_{{\rm tot}}$ is given in Eq.~\ref{eq:H-int-B-free}. All terms in Eq.~\ref{eq:negativity-tlA-A-2-summary} containing the free Hamiltonian $\mathbb{I}_A\ot\hat{D}$ will also vanish, since when $\hat{A}^q=\mathbb{I}_A$, the second part of Eq.~\ref{eq:negativity-tlA-A-2-summary} will become
\begin{equation}
\Tr[\sqrt{\rho_A}]\Tr[\hat{A}^p\sqrt{\rho_A}\,\mathbb{I}_A]-\Tr[\sqrt{\rho_A}\hat{A}^p]\Tr[\sqrt{\rho_A}\,\mathbb{I}_A]=0.
\end{equation}
Therefore, the free Hamiltonians on both $A$ and $B$ are initially negligible for the dynamics of negativity between $\tlA$ and $A$, and the two free Hamiltonian terms $\hat{C}\ot\bIB$ and $\bIA\ot\hat{D}$ can not contribute to the negativity vulnerability $\mathcal{N}_{\tlA ;A}^{(2)}$. 

Hence, we have shown that terms of the form $\hat{C}\ot\bIB$ and $\bIA\ot\hat{D}$ (i.e. free Hamiltonians of $A$ and $B$) do not contribute to the first and second-order time derivatives of the negativity for any of the bipartite systems in our system. We can therefore ignore the free Hamiltonians when we consider the dynamics of negativity for $A;B$ (Sec.~\ref{sec:free-negativity-A-B}), $\tlA;B$ (Sec.~\ref{sec:free-negativity-tlA-B}), and $\tlA;A$ (Sec.~\ref{sec:free-negativity-tlA-A}) at the onset of interaction upto the second order, and the phenomenon of resonance can only occur at most at the third order at the onset in our system.

\subsubsection{\texorpdfstring{$\tlA$}{A} and \texorpdfstring{$A$}{A} lose their entanglement under a generic \texorpdfstring{$\hat{H}_{{\rm tot}}=\hat{A}\ot\hat{B}$}{product Hamiltonian}.}
\label{sec:pf-tlAA-lost-entanglement}

When we consider a product Hamiltonian with $\hat{H}_{{\rm tot}}=\hat{A}\ot\hat{B}$, the negativity vulnerability in Eq.~\ref{eq:negativity-tlA-A-2-summary} reduces to
\begin{align}
V(\rho_A,\rho_B,\hat{A}\ot\hat{B}):=\mathcal{N}_{\tlA ;A}^{(2)}&=-\frac{1}{2}(\Delta B)^2\left(\Tr[\sqrt{\rho_A}]\Tr[\hat{A}\sqrt{\rho_A}\hat{A}]-\Tr[\sqrt{\rho_A}\hat{A}]^2\right)\label{eq:negativity-tlA-A-product-form}\\
&=-\frac{1}{2}(\Delta B)^2G_A\label{eq:negativity-tlA-A-product-form-simple}
\end{align}
We observe that the second term involving $\hat{A}$ and $\rho_A$ in Eq.~\ref{eq:negativity-tlA-A-product-form} will reduce to the variance $(\Delta A)^2$ if we replace all the appearances of $\sqrt{\rho_A}$ with $\rho_A$. Since $\sqrt{\rho_A}$ is the square root of the probability of each state in the ensemble, we define the notion of the probability amplitude-based variance $G_A$ with respect to the initial density matrix $\rho_A$ and the Hamiltonian (observable) $\hat{A}$ on the system $A$:
\begin{align}
G_{A}&:=\Tr[\sqrt{\rho_A}]\Tr[\hat{A}\sqrt{\rho_A}\hat{A}]-\Tr[\sqrt{\rho_A}\hat{A}]^2.
\label{eq:GA}
\end{align}
To show $\mathcal{N}_{\tlA;A}^{(2)}\leq 0$ under $\hat{H}_{{\rm tot}}=\hat{A}\ot\hat{B}$, we express Eq.~\ref{eq:GA} in the matrix component form similar to Eq.~\ref{eq:negativity-tlA-A-2-dirac-symmetric}:
\begin{align}
G_{A}&=\sum_{u,v,y}\alpha_u\alpha_vA_{yv}A_{vy}-\sum_{uv}\alpha_u\alpha_vA_{uu}A_{vv}\\
&=\frac{1}{2}\sum_{u,v}\alpha_u\alpha_v\left(\sum_{y}A_{yv}A_{vy}+\sum_{y}A_{yu}A_{uy}-2A_{uu}A_{vv}\right)\\
&=\frac{1}{2}\sum_{u,v}\alpha_u\alpha_v\left(\sum_{y\neq v}|A_{yv}|^2+\sum_{y\neq u}|A_{yu}|^2+(A_{uu}-A_{vv})^2\right)\geq 0
\label{eq:positivity-GA}
\end{align}
Therefore, we see that $\mathcal{N}_{\tlA;A}^{(2)}\leq 0$, and the negativity for $\tlA;A$ is guaranteed to be non-increasing at the onset of interaction, which is expected, since $\tlA$ and $A$ initially purify each other, and there is no interaction Hamiltonian between $\tlA$ and $A$. In particular, we see that the negativity vulnerability $\mathcal{N}_{\tlA; A}^{(2)}$ only vanishes when either $\Delta B$ or $G_A$ vanishes. The variance term $(\Delta B)^2=0$ only when $\rho_B$ is pure with an eigenstate of the observable $\hat{B}$, while $G_A=0$ is only achieved when $\hat{A}\propto\mathbb{I}_A$ under the assumption that ${\rm det}(\rho_A)\neq 0$. Therefore, under a generic product Hamiltonian $\mathcal{N}_{\tlA;A}^{(2)}<0$, which suggests that $\tlA$ and $A$ immediately lost their entanglement at the onset. 

We can further show that $G_A$ is bounded below by the variance $(\Delta A)^2$. We can write $(\Delta A)^2$ in a form similar to Eq.~\ref{eq:positivity-GA}: 
\begin{align}
(\Delta A)^2&=\Tr[\hat{A}^2\rho_A]-\Tr[\hat{A}\rho_A]^2=\sum_{u,v} \lambda_A^uA_{uv}A_{vu}-\sum_{u,v}\lambda_A^u\lambda_A^v A_{uu}A_{vv}\\
&=\left(\sum_v \lambda_A^v\right)\sum_{u,y} \lambda_A^uA_{uy}A_{yu}-\sum_{u,v}\lambda_A^u\lambda_A^v A_{uu}A_{vv}\\
&=\frac{1}{2}\sum_{u,v}\lambda_A^u\lambda_A^v\left(\sum_{y\neq v}|A_{yv}|^2+\sum_{y\neq u}|A_{yu}|^2+(A_{uu}-A_{vv})^2\right)
\label{eq:positivity-DA}
\end{align}
Combining Eq.~\ref{eq:positivity-GA} and \ref{eq:positivity-DA}, we have
\begin{align}
    G_A-(\Delta A)^2&=\frac{1}{2}\sum_{u,v}(\alpha_u\alpha_v-\lambda_A^u\lambda_A^v)\left(\sum_{y\neq v}|A_{yv}|^2+\sum_{y\neq u}|A_{yu}|^2+(A_{uu}-A_{vv})^2\right)\\
    &\geq 0,
\end{align}
since $\alpha_u\alpha_v-\lambda_A^u\lambda_A^v\geq 0$ with $0\leq\alpha_i=\sqrt{\lambda_A^i}\leq\lambda_A^i\leq 1$ for all $i$. Therefore, the amplitude-based variance $G_A$ is bounded below by the variance $(\Delta A)^2$. When $\rho_A$ is in a pure state, $\sqrt{\rho_A}=\rho_A$, so $G_A=(\Delta A)^2$ in Eq.~\ref{eq:GA}. $G_A$ reduces to normal variance for a pure state.   

In the previous work of \cite{20EmilyFirstPaper}, the notion of the 2-fragility $f_{2,A}$ was introduced:
\begin{equation}
    f_{2,A}:=-\frac{1}{2}\Tr[[\hat{A},\rho_A]^2],
    \label{eq:2-fragility}
\end{equation}
which determines the proclivity
of the system $A$ to lose its own purity at the onset of the interaction with $B$. The probability amplitude-based variance $G_A$ in Eq.~\ref{eq:GA} can be interpreted as a measurement for the tendency of $A$ to lose its entanglement with $\tilde{A}$ under the interaction with $B$. Interestingly, combining the result in \cite{20EmilyFirstPaper}, we have
\begin{equation}
    G_A\geq(\Delta A)^2\geq f_{2,A}.
\end{equation}
Therefore, the tendency of the system $A$ to lose its own purity is different from the tendency of the system $A$ to lose its entanglement with $\tlA$ (which initially purifies $A$). Both $G_A$ and $f_{2,A}$ are bounded by the variance $(\Delta A)^2$ and reduce to the variance when the system $A$ starts at a pure state. However, when $A$ is pure, $G_A$ loses its interpretation as a term in the negativity vulnerability $\mathcal{N}_{\tlA;A}^{(2)}$ in Eq.~\ref{eq:negativity-tlA-A-product-form}, since our calculation of the negativity perturbation in Appendix.~\ref{sec:negativity-tlA-A-calculation} breaks down due to the degenerate eigenvalue $0$. In fact, when $A$ is pure, there will be no initial entanglement between $\tlA$ and $A$. Without an interaction Hamiltonian between $\tlA$ and $AB$, no entanglement between $\tlA$ and $A$ can be generated. However, taking $\rho_A$ to be pure in the probability amplitude-based variance $G_A$ can still be reasonable: we can let a generically mixed $\rho_A$ to approach the pure state asymptotically. The probability amplitude-based variance $G_A$ might also have other significance beyond contributing to the negativity vulnerability.

\subsubsection{Maximize the negativity vulnerability \texorpdfstring{$\mathcal{N}_{\tlA;A}^{(2)}$}{negaAA} for qubits under \texorpdfstring{$\hat{H}_{{\rm tot}}=\hat{A}\ot\hat{B}$}{product Hamiltonian}.}
We next show how we can minimize the amplitude-based variance $G_A$, therefore maximizing the negativity vulnerability and minimizing the loss of entanglement within $\tlA;A$ in the case of $\hat{H}_{{\rm tot}}=\hat{A}\ot\hat{B}$. For the example, we consider $A$ as a qubit. In the usual eigenbasis of $\rho_A$ with the two eigenvalues $\lambda_A^1=\lambda$ and $\lambda_A^2=1-\lambda$, Eq.~\ref{eq:positivity-GA} becomes
\begin{align}
    G_A = \Big[2|A_{12}|^2+(A_{11}-A_{22})^2\Big]\sqrt{\lambda}\sqrt{1-\lambda}+|A_{12}|^2
\end{align}

With a fixed $\hat{A}$, we see that $G_A$ is the largest when $\lambda=\frac{1}{2}$, that is $\tlA$ and $A$ are maximally entangled. The more entangled $\tlA$ and $A$ are, the faster $\tlA$ and $A$ lose their entanglement at the onset of interaction. When $\lambda=0$ or $1$, that is $\rho_A$ being pure, $G_A$ vanishes, since $A$ and $\tlA$ become initially unentangled. With fixed $\rho_A$, we see that $G_A$ vanishes when $\hat{A}=\mathbb{I}_A$ (only free Hamiltonian $\hat{B}$ on B). Under fixed $\rho_A$, we can certainly minimize $G_A$ by adjusting $A_{12}$, $A_{11}$ and $A_{12}$ under certain constraints on the Hamiltonian $\hat{A}$.  

When we want to protect the initial entanglement in $\tlA;A$ from the environment $B$, we usually can not choose or have no control over the interaction Hamiltonian $\hat{A}\ot\hat{B}$. For a more realistic procedure of minimizing $G_A$, $\hat{A}$ should be considered as fixed. We also usually start with a given or fixed amount of the initial entanglement between $\tlA$ and $A$ with some control over the initial state $\rho_A$. Therefore, we can minimize $G_A$ by changing $\rho_A$ under the constraint of the eigenvalues of $\rho_A$ staying the same. Different from the rest of the paper, we will work in the eigenbasis of the Hamiltonian $\hat{A}$ where $\hat{A}=[[A_{1},0],[0,A_{2}]]$ is characterized by the two measurement values of the observable, and the initial density matrix $\rho_A$ for the qubit $A$ can be written as $\rho_A=\frac{1}{2}(\mathbb{I}_A+a_x\sigma_x+a_y\sigma_y+a_z\sigma_z)$ in the eigenbasis of $\hat{A}$ under the Bloch sphere representation. Let the radius of the Bloch vector corresponding to $\rho_A$ be $r=\sqrt{a_x^2+a_y^2+a_z^2}$, so $0\leq r\leq 1$. We know that the eigenvalues of $\rho_A$ can be expressed as $\lambda_A=\frac{1\pm r}{2}$, and the initial amount of entanglement between $\tlA$ and $A$ can be described with the negativity $\mathcal{N}_{\tlA ;A}(t_0)=\sqrt{1-r^2}/2$ according to Eq.~\ref{eq:tlA-A-negativity-0}. With $\mathcal{N}_{\tlA ;A}(t_0)$ considered fixed, $r$ will be considered as a constant, so we can rotate the Bloch vector $(a_x,a_y,a_z)$ under the constraint $a_x^2+a_y^2+a_z^2=r^2$ to minimize the amplitude-based variance $G_A$. Substituting our $\hat{A}$ and $\rho_A$ expressions in the eigenbasis of $\hat{A}$ into Eq.~\ref{eq:GA}, we can eventually find 
\begin{align}
  G_A = \frac{(A_1-A_2)^2}{4}\Big[\sqrt{1-r^2}+1-(1-\sqrt{1-r^2})\frac{a_z^2}{r^2}\Big].
\label{eq:GA-qubit-Bloch}
\end{align}
With $0\leq a_z\leq r$, we have 
\begin{align}
   \frac{(A_1-A_2)^2}{4}(2\sqrt{1-r^2})\leq G_A\leq \frac{(A_1-A_2)^2}{4}(\sqrt{1-r^2}+1),
\end{align}
where $G_A$ achieves its minimum when $a_z=\pm r$, while $G_A$ achieves its maximum when $a_z=0$. Interestingly, the minimization of $G_A$ for the case of qubits does not depend on the exact measurement values $A_1$ and $A_2$ of the Hamiltonian $\hat{A}$ and only depends on $a_z$. In order to minimize the loss of the entanglement within $\tlA;A$, we need the negativity vulnerability $V(\rho_A,\rho_B,\hat{A}\ot\hat{B})$ to be large and $G_A$ to be small, which corresponds to $a_z=r$, that is $\rho_A$ being diagonal in the eigenbasis of $\hat{A}$. When $\rho_A$ is maximally mixed (proportional to the identity matrix), we know $r=0$ and $G_A=(A_1-A_2)^2/2$, so there is no place to hide the entanglement loss as expected. When $\rho_A$ is neither completely mixed nor pure with $0<r<1$, we can always minimize the initial speed of the entanglement loss to the environment by ensuring $\rho_A$ is diagonal in the eigenbasis of $\hat{A}$. The entanglement within the $2\times 2$ dimensional $\tlA;A$ system is the most protected when $\rho_A$ is essentially an ensemble of the eigenstate of $\hat{A}$.

We have thus demonstrated the procedure of analytically minimizing the probability amplitude-based variance $G_A$ in order to maximize the negativity vulnerability $\mathcal{N}_{\tlA; A}^{(2)}$ in the case of a qubit under a product Hamiltonian. We can easily generalize the above procedure of minimizing $G_A$ to higher dimensional systems such as qutrit under a product Hamiltonian, which we expect to offer more degrees of freedom compared to the qubit to hide the entanglement within $\tlA;A$ away from the interaction Hamiltonian $H_{\rm int}=\hat{A}\ot\hat{B}$. We can also generalize the above procedure to a generic interaction Hamiltonian where we have to directly maximize the negativity vulnerability $\mathcal{N}_{\tlA;A}^{(2)}$ in Eq.~\ref{eq:negativity-tlA-A-2-summary}. In the case of a generic Hamiltonian, we can no longer separately consider systems $A$ and $B$ (minimizing $G_A$ and $(\Delta B)^2$ individually), and we expect the maximization of $\mathcal{N}_{\tlA;A}^{(2)}$ to be performed numerically under given constraints. We leave the study of these more general cases to future work.

\subsection{Delocalization in System \texorpdfstring{$\tlA;AB$}{tlA;AB}}
\label{sec:tlA-AB-delocalization}

\begin{figure}
\centering
\begin{subfigure}{.47\textwidth}
  \centering
  \includegraphics[width=0.95\linewidth]{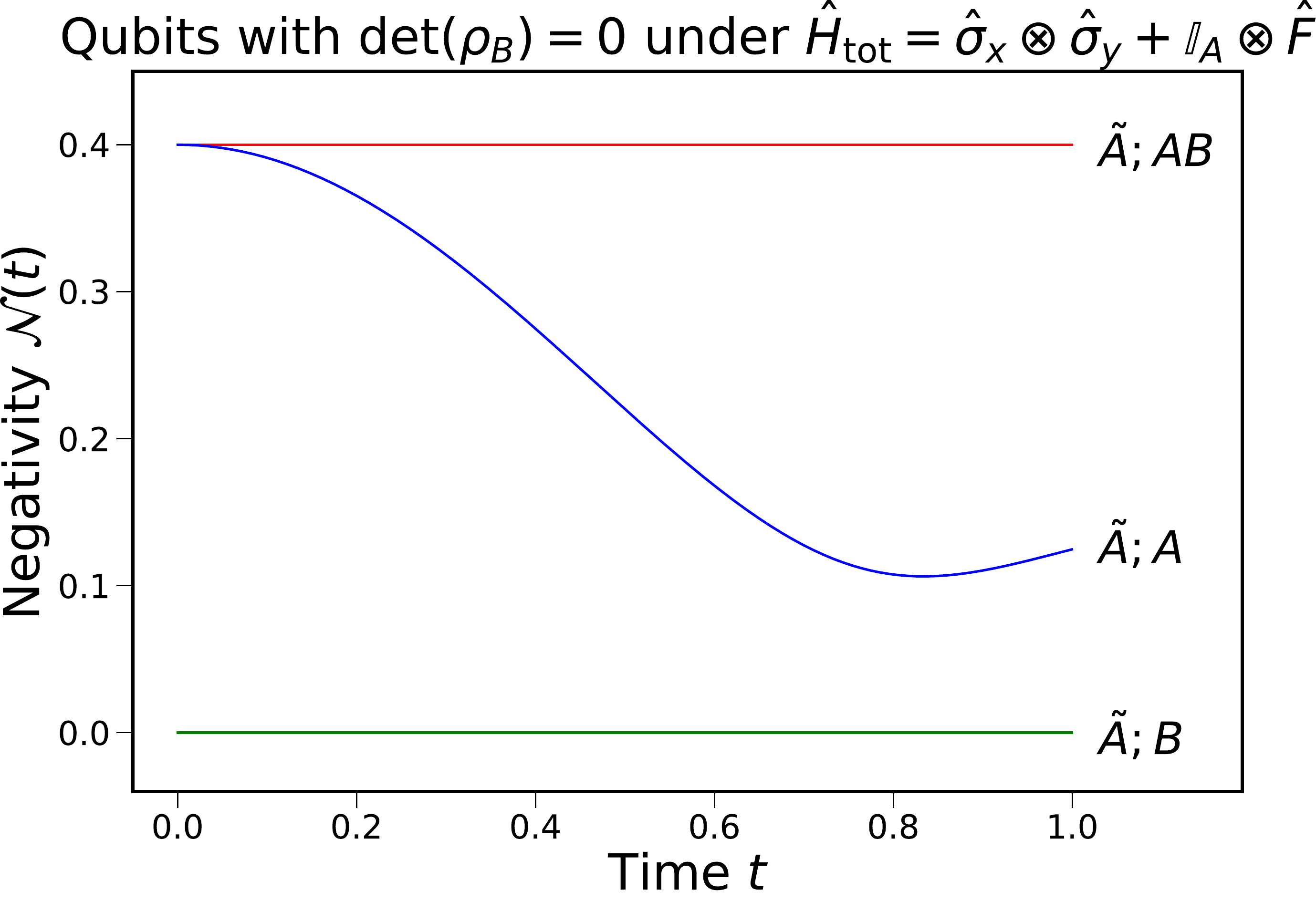}
  \caption{Delocalization of the $\tlA;AB$ entanglement in the case that $\tlA;B$ are unentangled throughout the time evolution under $\hat{H}_{{\rm tot}}=\hat{\sigma}_x\ot\hat{\sigma}_y+\mathbb{I}_A\ot\hat{F}$.}
  \label{fig:delocal-tlA-Bfree}
\end{subfigure}
\hfill
\begin{subfigure}{.47\textwidth}
  \centering
  \includegraphics[width=0.9\linewidth]{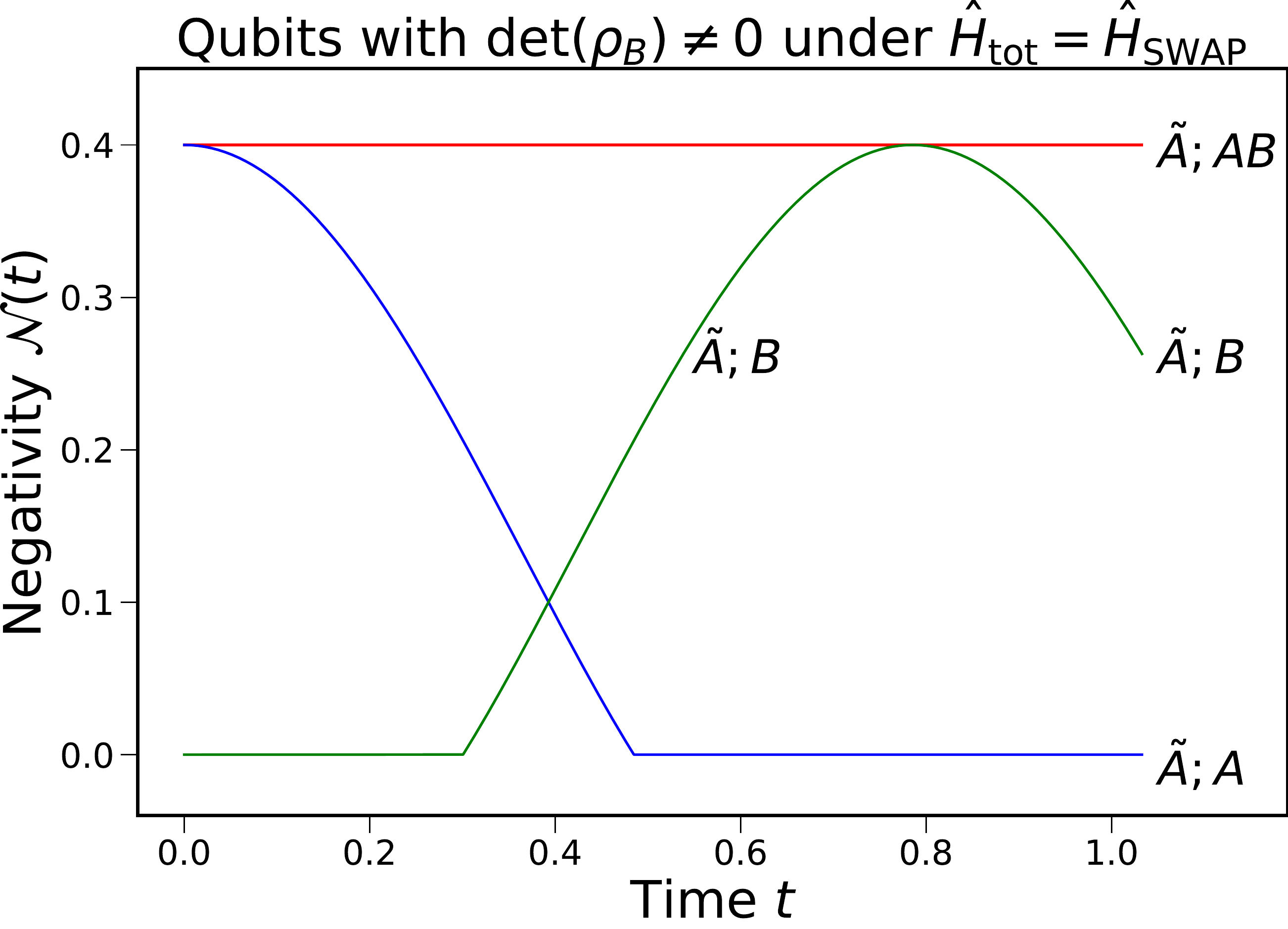}
  \caption{Delocalization of the $\tlA;AB$ entanglement in the case that $\tlA;B$ are initially unentangled for a finite amount of time under ${\rm det}(\rho_B)\neq 0$.}
  \label{fig:delocal-tlA-SWAP}
\end{subfigure}
\caption{Delocalization of the $\tlA;AB$ entanglement in two cases. We choose $\tlA$, $A$, and $B$ as qubits, so the PPT criterion applies. The initial state of $A$ in both plots is given by Eq.~\ref{eq:A-qubit} in Appendix~\ref{sec:qubit}. In Fig.~\ref{fig:delocal-tlA-Bfree}, $B$ is chosen to be pure, and $\hat{H}_{{\rm tot}}=\hat{\sigma}_x\ot\hat{\sigma}_y+\mathbb{I}_A\ot\hat{F}$, where $\hat{\sigma}_{x,y,z}$ are Pauli matrices and $\hat{F}$ (the free Hamiltonian on $B$) is given in Eq.~\ref{eq:free-B-qubit}. In Fig.~\ref{fig:delocal-tlA-SWAP}, $\rho_B$ is given by Eq.~\ref{eq:qubit-B-mixed}, and $\hat{H}_{{\rm tot}}=\hat{H}_{\rm SWAP}$ is given in Eq.~\ref{eq:H-SWAP-qubit}.}
\label{fig:delocal-tlA-numerical}
\end{figure}

In the previous sections, we calculated the second-order perturbation of negativity for the bipartite systems $A;B$, $\tlA;B$, and $\tlA;A$ at the onset. We now treat $AB$ as a single subsystem and consider the bipartite entanglement between $\tlA$ and $AB$. Under our system setup, there is no interaction between $\tlA$ and $AB$, so the entanglement, thereby the negativity, of the system $\tlA;AB$ will remain unchanged, which can be directly observed from the constant negativity between $\tlA$ and $AB$ in Fig.~\ref{fig:delocal-tlA-numerical}. We now proceed to analyze how the entanglement between $\tlA$ and $AB$ is distributed within the $AB$ system.

From Sec.~\ref{sec:negativity-perturbed-tlA-A}, we know that the initial change of negativity for $\tlA;A$ is characterized by the negativity vulnerability $\mathcal{N}_{\tlA; A}^{(2)}$ in Eq.~\ref{eq:negativity-tlA-A-2-summary}. Since all of the entanglement between $\tlA$ and $AB$ is initially localized to $A$, and there is no interaction Hamiltonian between $\tlA$ and any other subsystems, the entanglement between $\tlA$ and $A$ can not increase at the onset. In fact, we have shown in Sec.~\ref{sec:free-negativity-tlA-A} and \ref{sec:pf-tlAA-lost-entanglement} that for a total Hamiltonian between $A$ and $B$ with the form $\hat{H}_{{\rm tot}}=\hat{C}\ot\mathbb{I}_B+\hat{A}\ot\hat{B}+\mathbb{I}_A\ot\hat{D}$, the negativity vulnerability $\mathcal{N}_{\tlA; A}^{(2)}$ is guaranteed to be negative as long as $\rho_B$ is not a pure eigenstate of $\hat{B}$ and $\hat{A}\not\propto\mathbb{I}_A$, which suggests that $\tlA$ and $A$ will immediately lose their entanglement at the onset.

As the entanglement between $\tlA$ and $AB$ remains constant throughout the interaction, the loss of the entanglement between $\tlA$ and $A$ indicates that the bipartite entanglement of $\tlA;AB$ must be redistributed within the $AB$ subsystem: either $\tlA$ becomes entangled with $B$ or the entanglement for $\tlA;AB$ becomes delocalized within the $AB$ subsystem. The delocalization of the bipartite $\tlA;AB$ entanglement is a manifestation of the tripartite entanglement among $\tlA A B$, since the delocalized part of $\tlA;AB$ entanglement can neither be accounted by the bipartite $\tlA;A$ entanglement or $\tlA;B$ entanglement. The degree of delocalization of the $\tlA;AB$ entanglement depends on whether $\tlA$ and $B$ can become entangled and how much they become entangled. 

\begin{figure}
\centering
\begin{subfigure}{.45\textwidth}
  \centering
  \includegraphics[width=0.87\linewidth]{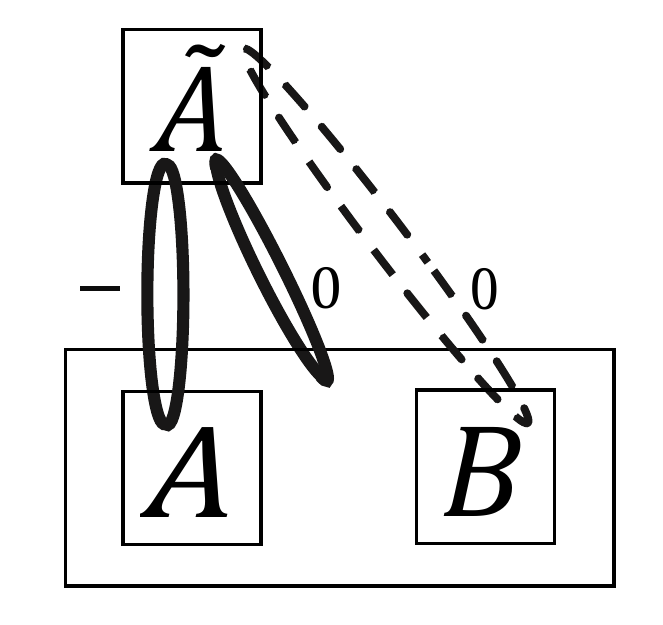}
  \caption{Delocalization of the $\tlA;AB$ entanglement within the $AB$ subsystem.}
  \label{fig:delocal-tlA}
\end{subfigure}
\hfill
\begin{subfigure}{.45\textwidth}
  \centering
  \includegraphics[width=.74\linewidth]{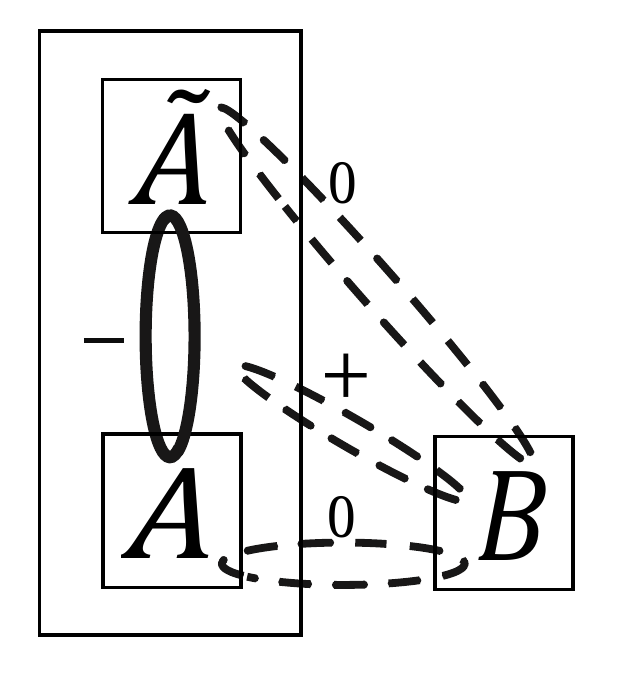}
  \caption{Delocalization of the $B;\tlA A$ entanglement within the $\tlA A$ subsystem.}
  \label{fig:delocal-B}
\end{subfigure}
\caption{Boxes represent systems (which can be composite) under consideration, and we adopt the convention that systems whose boxes are linked by a solid ellipse are entangled with another. A dashed ellipse indicates that the two systems are unentangled. The $+,-,0$ signs indicate whether the entanglement (or negativity) of the two systems increases, decreases, or remains constant at the onset of interaction. The absence of any form of ellipse between two systems suggests that we do not wish to specify or discuss the status of the corresponding entanglement.}
\label{fig:delocal-demo}
\end{figure}

When $\tlA$ and $B$ are unentangled, the $\tlA;AB$ entanglement certainly becomes delocalized within $AB$. The loss of the localized entanglement between $\tlA$ and $A$ is transferred to the entanglement between $\tlA$ and the delocalized part of $AB$, which is illustrated schematically in Fig.~\ref{fig:delocal-tlA}. The key features of the entanglement dynamics between $\tlA$ and $B$ are summarized in Table~\ref{tab:tlA-B-summary} in Sec.~\ref{sec:system setup}, and we see that the $\tlA;B$ entanglement dynamics, either at the onset or throughout the course of interaction, depend on the state of the initial density matrix $\rho_B$ and the form of total Hamiltonian $\hat{H}_{\rm tot}$ between $A$ and $B$. Therefore, the degree of delocalization of the $\tlA;AB$ entanglement is also determined by the forms of $\rho_B$ and $\hat{H}_{{\rm tot}}$. The less entangled $\tlA$ and $B$ are, the more delocalized the $\tlA;AB$ entanglement will be.

When $\tlA$ and $B$ remain unentangled throughout the course of interaction (the case of $\hat{H}_{{\rm tot}}=\hat{A}\ot\hat{B}+\bIA\ot\hat{D}$ summarized in Table~\ref{tab:tlA-B-summary}), all the loss of the $\tlA;A$ entanglement is transferred to the entanglement between $\tlA$ and the delocalized part of $AB$ throughout the time evolution, which is illustrated by the negativity dynamics depicted in Fig.~\ref{fig:delocal-tlA-Bfree}. When $\tlA$ and $B$ remain unentangled (or weakly entangled with vanishing negativity) only for a finite amount of time in the case of ${\rm det}(\rho_B)\neq 0$ summarized in Table~\ref{tab:tlA-B-summary}, the delocalization phenomenon can be just temporary at the initial time, which can be see in Fig.~\ref{fig:delocal-tlA-SWAP}. Under the conditions set in Fig.~\ref{fig:delocal-tlA-SWAP}, $\tlA;B$ remain unentangled before $t\approx0.3$, which suggests that all the loss of the $\tlA;A$ entanglement is transferred to the entanglement between $\tlA$ and the delocalized part of $AB$ during this period. After $t\approx0.3$, $\tlA$ and $B$ gain entanglement first at the expense of the $\tlA;A$ entanglement (during $0.3 \lessapprox t\lessapprox 0.5$) and then at the expense of the delocalized $\tlA;AB$ entanglement (during $0.5 \lessapprox t\lessapprox 0.8$). After $t\approx 0.5$, we see that the $\tlA;AB$ entanglement is becoming more localized to $B$, and the delocalized $\tlA;AB$ entanglement is converted to the localized $\tlA;B$ entanglement. At $t\approx 0.8$, the $\tlA;AB$ entanglement is completely localized to $B$ with no delocalization phenomenon. The $\tlA;A$ entanglement is completely transferred to $\tlA;B$ at this point. 

Generally, this behavior of delocalization of the $\tlA;AB$ entanglement decreases the quantum channel capacity and presents a challenge to our goal of transferring entanglement from $\tlA;A$ to $\tlA;B$ in the localized form. However, the existence of delocalized entanglement also provides one of the key advantages of quantum computing, since the combined system $\rho_{AB}$ acting on the tensor product space $\hA\ot\hB$ can store much more information than the sum of the information stored by $\rho_A$ and $\rho_B$ individually. The additional degrees of freedom in $\rho_{AB}$ exists in the delocalized form. We can store localized information into the delocalized part of the Hilbert space similar to the example in Fig.~\ref{fig:delocal-tlA-Bfree}, and we can access the delocalized part of the Hilbert space by localizing the delocalized entanglement similar to the $0.5 \lessapprox t\lessapprox 0.8$ period in Fig.~\ref{fig:delocal-tlA-SWAP}.

\subsection{Delocalization in System \texorpdfstring{$B;\tlA A$}{B;tlAA}}
\label{sec:B-tlAA-delocalization}

Similar to the previous Sec.~\ref{sec:tlA-AB-delocalization} where we considered $AB$ as a single subsystem, we can also treat $\tlA A$ as a single subsystem and consider the dynamics of the bipartite entanglement between $B$ and $\tlA A$. The total Hamiltonian between $B$ and $\tlA A$ can then be written as $\hat{H}_{{\rm tot}}^{B;\tlA A}=\sum_{p}\hat{B}^p\ot(\mathbb{I}_{\tlA}\ot\hat{A}^p)$ based on Eq.~\ref{eq:tri-Hamiltonian} and \ref{eq:total-Hamiltonian}. Under our system setup, $\rho_{\tlA A}$ starts in a pure state. 

\begin{figure}[htbp!]
\centerline{\includegraphics[width=0.5\hsize]{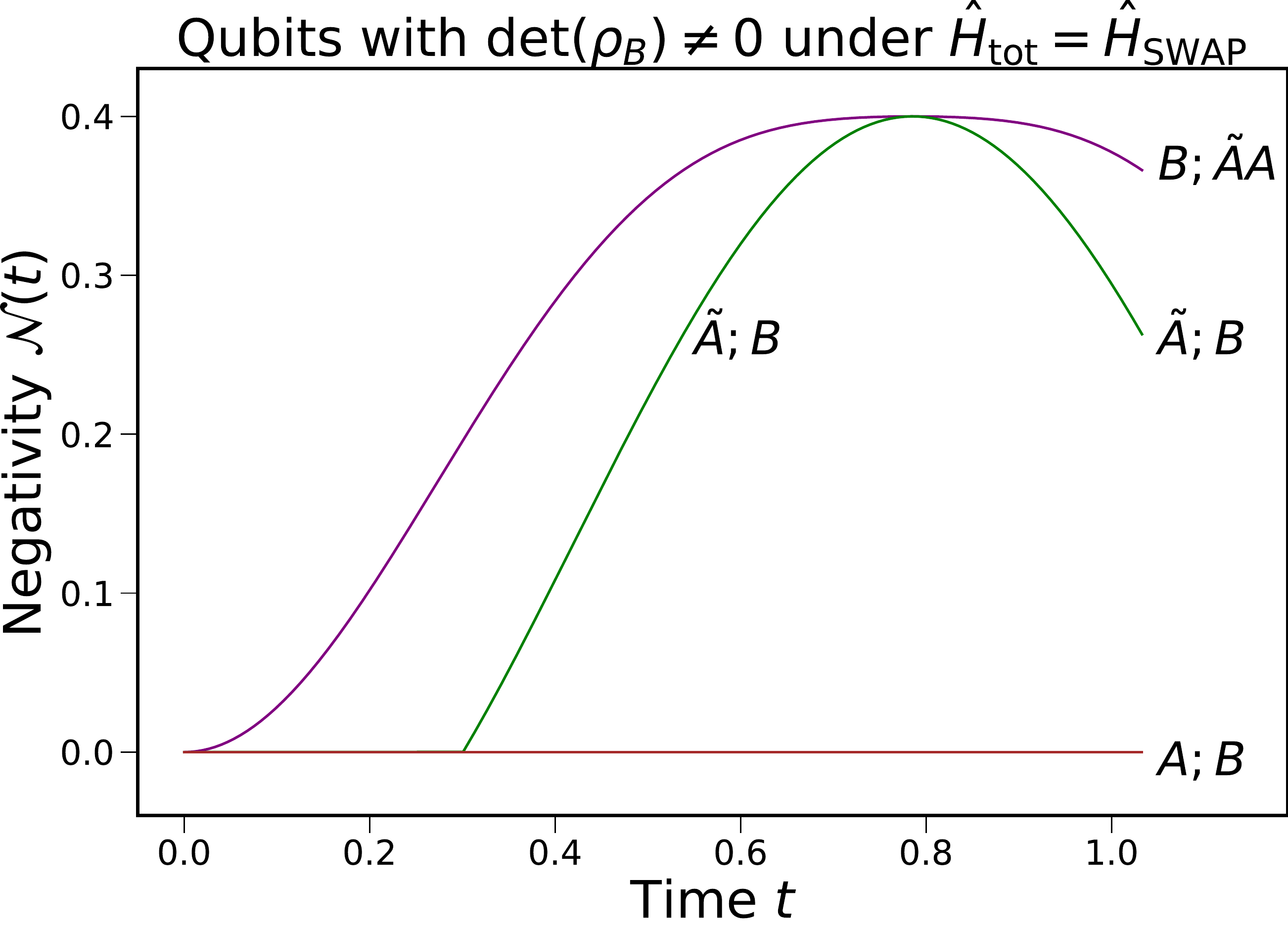}}
\caption{Delocalization of the $B;\tlA A$ entanglement under ${\rm det}(\rho_B)\neq 0$. For $t\lessapprox 0.3$, the newly generated $B;\tlA A$ entanglement (transferred from the initial $\tlA;A$ entanglement) is completely delocalized as $B$ remain unentangled with both $A$ and $\tlA$. We here choose $\tlA$, $A$, and $B$ as qubits, and the initial states and $\hat{H}_{{\rm tot}}$ used here are the same at those for Fig.~\ref{fig:delocal-tlA-SWAP} listed in Appendix.~\ref{sec:qubit}.}
\label{fig:mixed_qubit_B-tlAA_delocal_SWAP}
\end{figure}

When ${\rm det}(\rho_B)\neq 0$, the $B;\tlA A$ system will then reduce to the case studied in Sec.~\ref{sec:negativity-perturbed-A-B} with $A$ and $B$ replaced by $B$ and $\tlA A$ respectively. Applying results from Eq.~\ref{eq:negativity-A-B-1-summary} to \ref{eq:FAB-B-matrix-summary}, we can find the expressions for the first and second-order perturbation of the negativity between $B$ and $\tlA A$ at the onset of interaction:
\begin{align}
    \mathcal{N}_{B;\tlA A}^{(1)}&=0
    \label{eq:negativity-B-tlAA-1}\\
    \mathcal{N}_{B;\tlA A}^{(2)}&=\frac{1}{2}\left(\Tr\left[\sqrt{\hat{F}_{B;\tlA A}^{\dagger} \hat{F}_{B;\tlA A}}\right]-\Tr\left[ \hat{F}_{B;\tlA A}\right]\right)\label{eq:negativity-B-tlAA-2}\\
    \hat{F}_{B;\tlA A}:&=\sum_{p,q}\Big(\hat{B}^{q}\rho_B\hat{B}^{p}-\rho_B \hat{B}^{p}\rho_B^{-1}\hat{B}^{q}\rho_B\Big)^*\ot \Big(\hat{P}_{\tlA A}^D(\mathbb{I}_{\tlA}\ot\hat{A}^p)\rho_{\tlA A}(\mathbb{I}_{\tlA}\ot\hat{A}^q)\hat{P}_{\tlA A}^D\Big)\\
    \hat{P}_{\tlA A}^D&=\mathbb{I}_{\tlA A}-\rho_{\tlA A},
\end{align}
where $\rho_{\tlA A}$ is given in Eq.~\ref{eq:AtlA-rho0}. When the total Hamiltonian between $B$ and $\tlA A$ takes the product form $\hat{H}_{{\rm tot}}^{B;\tlA A}=\hat{B}\ot(\mathbb{I}_{\tlA}\ot\hat{A})$, we know that $\mathcal{N}_{B;\tlA A}^{(2)}>0$ when $[\hat{B},\rho_B]\neq 0$ and $(\mathbb{I}_{\tlA A}-\rho_{\tlA A})(\mathbb{I}_{\tlA}\ot\hat{A})\rho_{\tlA A}\neq 0$ based on the analysis in Sec.~\ref{sec:pure-pf-AB-entangled}, so a generic product Hamiltonian is sufficient to immediately generate entanglement between $B$ an $\tlA A$ at the onset when $\rho_B$ is not maximally mixed. We also expect that a generic interaction Hamiltonian containing multiple terms can also immediately entangle $B$ with $\tlA A$ at the onset, which can be seen from the topmost curve in Fig.~\ref{fig:mixed_qubit_B-tlAA_delocal_SWAP} as an example. 

When ${\rm det}(\rho_B)=0$, our perturbative calculations in Sec.~\ref{sec:pure, perturbative} no longer apply. With the initial density matrices of both $B$ and $\tlA A$ subsystems containing vanishing eigenvalues, the perturbation of negativity becomes more complicated due to the additional structures present in the degenerate eigenspace for the vanishing eigenvalue. However, we still expect $B$ and $\tlA A$ to gain entanglement at the onset of interactions under a generic total Hamiltonian between $A$ and $B$, since zero eigenvalues are generally prone to entanglement generation. In the special case when $B$ is pure, the total $\tlA A B$ system is pure, so we can use n-R\'enyi entropy as an entanglement measure for $B;\tlA A$. The perturbative result for the n-R\'enyi entropy in Eq.~\ref{eq:renyi-derivative} from \cite{20EmilyFirstPaper} shows that $B$ and $\tlA A$ will become immediately entangled under a generic product Hamiltonian on $AB$. Therefore, we expect $B$ and $\tlA A$ to become entangled under a generic $\rho_B$ and $\hat{H}_{{\rm tot}}$. 

The generation of entanglement within the bipartition $B;\tlA A$ is in fact an indication for the establishment of genuine tripartite entanglement in the total system $\tlA A B$, which is equivalent to all three bipartitions $\tlA;AB$, $A;\tlA B$, and $B;\tlA A$ being bipartitely entangled \cite{19CunhaTripartiteEntanglement}. The entanglement (negativity) between $\tlA;AB$ will remain constant as established in Sec.~\ref{sec:tlA-AB-delocalization}, and $B;\tlA A$ (with $\rho_B$ pure) will become immediately entangled at the onset under a generic $\hat{H}_{{\rm tot}}$ as discussed in this section. The bipartition $A;\tlA B$ is not studied in this work, but since $A$ is initially entangled with $\tlA$ under our setup, we know $A$ and $\tlA B$ are initially entangled. Therefore, the total system $\tlA A B$ will become tripartitely entangled as $B$ become entangled with $\tlA A$ at the onset.

We next proceed to analyze how the entanglement between $B$ and $\tlA A$ is distributed within the $\tlA A$ system. In the case of ${\rm det}(\rho_B)\neq 0$, we know $\mathcal{N}_{A;B}(t)=\mathcal{N}_{\tlA ;B}(t)=0$ for a finite amount of time after the start of interaction between $A$ and $B$ based on Sec.~\ref{sec:mixed, finite time}. However, we know the negativity for $B;\tlA A$ increases immediately from Eq.~\ref{eq:negativity-B-tlAA-2} under a generic $\hat{H}_{\rm tot}$, which indicates that most, if not all, of the entanglement generated between $B$ and $\tlA A$ exists between $B$ and the delocalized part of $\tlA A$. In the case of $A$ and $B$ as qubits (dim$(\hA)$dim$(\hB)\leq 6$), $B$ remain unentangled with both $A$ and $\tlA$ for a finite amount of time while $B$ becomes entangled with $\tlA A$ at the same time, which indicates that the generated $B;\tlA A$ entanglement is completely delocalized, which is demonstrated in the region with $t\lessapprox 0.3$ in Fig.~\ref{fig:mixed_qubit_B-tlAA_delocal_SWAP}. The existing $\tlA A$ entanglement is transferred to the entanglement between $B$ and the delocalized part of $\tlA A$. However, such complete delocalization of the $B;\tlA A$ entanglement is only temporary. We expect $B$ and $A$ to become entangled after a finite amount time ($\tlA$ and $B$ can also become entangled depending on the forms of $\hat{H}_{{\rm tot}}$ and $\rho_B$ shown in Table.~\ref{tab:tlA-B-summary}), so part of the $B;\tlA A$ entanglement become localized again after this initial period as seen in the $t\gtrapprox 0.3$ region of Fig.~\ref{fig:mixed_qubit_B-tlAA_delocal_SWAP}. The $t\lessapprox 0.3$ period in Fig.~\ref{fig:mixed_qubit_B-tlAA_delocal_SWAP} provides an example for generating purely delocalized entanglement. 

When ${\rm det}(\rho_B)=0$, we have shown in Sec.~\ref{sec:negativity-perturbed-A-B} that $B$ will become immediately entangled with $A$ under a generic total Hamiltonian for $AB$ and a generic state $\rho_B$, which prevents the existence of the complete delocalization of the $B;\tlA A$ entanglement under generic cases. However, a complete delocalization of the $\tlA;AB$ entanglement is still possible for ${\rm det}(\rho_B)=0$ under some special circumstances: we can simply add an ancilla $\tilde{B}$ to purify $B$ in the system described in Fig.~\ref{fig:mixed_qubit_B-tlAA_delocal_SWAP} such that ${\rm det}(\rho_{\tilde{B}B})=0$ while the $\tilde{B}B;\tlA A$ entanglement is still completely delocalized within $\tilde{A}A$ for an initial period. 

Combining the analysis in this section and the previous Sec.~\ref{sec:tlA-AB-delocalization}, we see that when ${\rm det}(\rho_B)\neq 0$, both delocalization of the $\tlA;AB$ entanglement and delocalization of the $B;\tlA A$ entanglement can occur simultaneously for a finite amount of time after the start of interaction between $A$ and $B$, which is demonstrated by the $t\lessapprox 0.3$ region in Fig.~\ref{fig:delocal-tlA-SWAP} and \ref{fig:mixed_qubit_B-tlAA_delocal_SWAP} with the same initial states and $\hat{H}_{{\rm tot}}$. The original $\tlA;A$ entanglement is transferred to the entanglement between $\tlA$ and the delocalized part of $AB$ as well as the entanglement between $B$ and the delocalized part of $\tlA A$. In the case of $A$ and $B$ as qubits, we see that the localized $\tlA;A$ entanglement is only transferred to the entanglement in the delocalized form, which provides a mechanism to store information in the delocalized part of the Hilbert space.

\section{When can an interaction \texorpdfstring{$\hat{H}_{{\rm int}}=\hat{A}\ot \hat{B}$}{product Hamiltonian} transmit entanglement?} \label{sec:tlA-B-separable}

In Sec.~\ref{sec:mixed, finite time}, we see that the negativity for $\tlA;B$ will vanish for a finite amount of time when ${\rm det}(\rho_B)\neq 0$ regardless of the total Hamiltonian $\hat{H}_{{\rm tot}}$ in $AB$. In Sec.~\ref{sec:negativity-perturbed-tlA-B}, we have established that the negativity transmissibility $\mathcal{N}_{\tlA;B}^{(2)}$ given in Eq.~\ref{eq:negativity-tlA-B-2-summary} vanishes at the onset under a total Hamiltonian of the form $\hat{H}_{{\rm tot}}=\hat{C}\ot\bIB+\hat{A}\ot\hat{B}+\bIA\ot\hat{D}$ between $A$ and $B$ when ${\rm det}(\rho_B)=0$. In this section, we go beyond the perturbative calculation of the $\tlA;B$ negativity and study under what forms of $\hat{H}_{{\rm tot}}$ can the the entanglement within $\tlA;A$ be transmitted to $\tlA;B$ during the interaction between $A$ and $B$.

\subsection{Without Free Hamiltonians: no entanglement transmission}\label{sec:pf-nf-tlA-B-separable}
We first consider a product Hamiltonian $\hat{H}_{{\rm tot}}=\hat{H}_{{\rm int}}=\hat{A}\ot \hat{B}$ where free Hamiltonians on $A$ and $B$ are ignored. According to Eq.~\ref{eq:time-evolution-exact}, the time evolution is given by
\begin{equation}
\rho(t)=U(t)\rho_{\rm tri}U^{\dag}(t)=e^{-it(\mathbb{I}_{\tlA}\ot\hat{A}\ot\hat{B})}\rho_{\tlA A}\ot\rho_{B}e^{it(\mathbb{I}_{\tlA}\ot\hat{A}\ot\hat{B})}
\label{eq:pf-nf-rho-exact}
\end{equation}
We express $U(t)$ in the eigenbasis of $\hat{A}$ and $\hat{B}$. Assume $\hat{A}=\sum_{i}h_A^i\ket*{h_A^i}\bra*{h_A^i}$ and $\hat{B}=\sum_{j} h_B^j\ket*{h_B^j}\bra*{h_B^j}$, then $\hat{A}\ot\hat{B} \ket*{h_A^i}\ot \ket*{h_B^j}=h_A^{i}h_{B}^j\ket*{h_A^i}\ot\ket*{h_B^j}$, which are the eigenvalues and eigenvectors of $\hat{A}\ot \hat{B}$. It is important for our proof that the eigenvectors of $\hat{A}\ot \hat{B}$ are in the product form themselves. The unitary operator can then be decomposed in the following form:
\begin{align}
    U(t)&=\mathbb{I}_{\tlA}\ot\exp\{-it\hat{A}\ot \hat{B}\}=\mathbb{I}_{\tlA}\ot\sum_{i,j}\exp\{-it h_A^i h_B^j\}\ket*{h_A^i}\bra*{h_A^i}\ot\ket*{h_B^j}\bra*{h_B^j}\\
    &=\mathbb{I}_{\tlA}\ot\sum_{j}\exp\{-it h_B^j \hat{A}\}\ot\ket*{h_B^j}\bra*{h_B^j}\label{eq:pf-nf-time-evolution-A}\\
    &=\mathbb{I}_{\tlA}\ot\sum_{i}\ket*{h_A^i}\bra*{h_A^i}\ot\exp\{-it h_A^i\hat{B}\} \label{eq:pf-nf-time-evolution-B}
\end{align}

Using Eq.~\ref{eq:pf-nf-rho-exact} and \ref{eq:pf-nf-time-evolution-B}, we proceed to show $\tlA$ and $B$ remain unentangled throughout the time evolution:
\begin{align}
&\rho_{\tlA B}(t)=\Tr_A[\rho(t)]=\Tr_A[U(t)\rho_{\tlA A}\ot \rho_B U^{\dag}(t)]\label{eq:pf-nf-tla-b-separable-0}\\
&=\Tr_A[\sum_{i}\ket*{h_A^i}\bra*{h_A^i}\ot\exp\{-it h_A^i\hat{B}\}\rho_{\tlA A}\ot\rho_B \sum_{j}\ket*{h_A^j}\bra*{h_A^j}\ot\exp\{it h_A^j\hat{B}\}]\label{eq:pf-nf-tla-b-separable-1}\\
&=\sum_{i,j}\Tr_A[\ket*{h_A^i}\bra*{h_A^i}\rho_{\tlA A}\ket*{h_A^j}\bra*{h_A^j}]\ot\exp\{-ith_A^i\hat{B}\}\rho_B\exp\{ith_A^j\hat{B}\}\label{eq:pf-nf-tla-b-separable-2}\\
&=\sum_{e}\bra*{h_A^e}\rho_{\tlA A}\ket*{h_A^e}\ot\exp\{-ith_A^e\hat{B}\}\rho_B\exp\{it h_A^e\hat{B}\}\\
&=\sum_{e}\Tr_{\tlA}[\bra*{h_A^e}\rho_{\tlA A}\ket*{h_A^e}]\frac{\bra*{h_A^e}\rho_{\tlA A}\ket*{h_A^e}}{\Tr_{\tlA}[\bra*{h_A^e}\rho_{\tlA A}\ket*{h_A^e}]}\ot\exp\{-ith_A^e\hat{B}\}\rho_B\exp\{it h_A^e\hat{B}\}\\
&=\sum_{e}\bra*{h_A^e}\rho_{A}\ket*{h_A^e}\frac{\bra*{h_A^e}\rho_{\tlA A}\ket*{h_A^e}}{\Tr[\bra*{h_A^e}\rho_{\tlA A}\ket*{h_A^e}]}\ot\exp\{-ith_A^e\hat{B}\}\rho_B\exp\{it h_A^e\hat{B}\}
\label{eq:pf-nf-tla-b-separable}
\end{align}

We analyze each term in Eq.~\ref{eq:pf-nf-tla-b-separable}. Let $\rho_{\tlA}^e:=\bra*{h_A^e}\rho_{\tlA A}\ket*{h_A^e}/\Tr[\bra*{h_A^e}\rho_{\tlA A}\ket*{h_A^e}]$. $\rho_{\tlA}^e$ is clearly a density matrix acting on $\hlA$, since $\rho_{\tlA}^e$ is Hermitian and positive-definite with $\Tr[\rho_{\tlA}^e]=1$ according to its definition. Let $\rho_B^e(t):=\exp\{-ith_A^e\hat{B}\}\rho_B\exp\{it h_A^e\hat{B}\}$, which is simply the time evolution of the system $B$ under the new free Hamiltonian $h_A^e\hat{B}$, so $\rho_B^e(t)$ is a density matrix acting on $\hB$. Last, we define $p_e:=\bra*{h_A^e}\rho_{A}\ket*{h_A^e}$, We have $p_e\geq 0$ and $\sum_e p_e=\Tr[\rho_A]=1$ where $\rho_A$ is the initial density matrix for the system A. We can now write 
\begin{equation}
\rho_{\tlA B}(t)=\sum_{e}p_e\rho_{\tlA}^e\ot\rho_B^e(t),
\label{eq:pf-nf-tla-b-final}
\end{equation}
and according to Eq.~\ref{eq:separable} in Sec.~\ref{sec:entanglement}, we see that $\rho_{\tlA B}(t)$ is separable for any time $t$, so $\tlA$ and $B$ remain unentangled during the entire time evolution. Therefore, it is impossible to transfer any entanglement from $\tlA;A$ to $\tlA;B$ with a product Hamiltonian between $A$ and $B$. 

Notice that the above separability proof from Eq.~\ref{eq:pf-nf-tla-b-separable-0} to \ref{eq:pf-nf-tla-b-final} is independent of the exact form of the initial density matrices $\rho_{\tlA A}$ and $\rho_B$. Though we assumed that $\tlA$ and $A$ purify each other and ${\rm det}(\rho_A)\neq 0$ for our system setup, $\tlA$ and $B$ will remain unentangled under the product Hamiltonian regardless of the initial relation between systems $\tlA$ and $A$ as long as the the total system is initially in a product state $\rho_{\tlA A}\ot\rho_{B}$. In fact, the above proof for the separability of $\tlA;B$ under a product Hamiltonian between $A$ and $B$ can be generalized to the situation where $\tlA A$ and $B$ are initially separable (that is $\rho_{\tlA A B}=\sum_k q_k\rho_{\tlA A}^k\ot\rho_B^k$ where $\sum_k q_k=1$) by adding an additional summation over all expressions of the proof. The above proof also works when $\tlA$, $A$, and $B$ are infinite dimensional systems.

If $B$ is initially entangled with another system $\tlB$ (when $B$ is mixed we can purify $B$ with an ancillary $\tlB$ similar to what we did to $A$), then by the symmetry of our system setup, we know that $A$ and $\tlB$ will also remain unentangled during the course of their time evolution if we choose $\hat{H}_{{\rm tot}}=\hat{A}\ot \hat{B}$.

\subsubsection{Non-zero quantum discord for qubits with \texorpdfstring{$\rho_B$}{rhoB} pure under \texorpdfstring{$\hat{H}_{{\rm tot}}=\hat{A}\ot\hat{B}$}{product Hamiltonian}.}\label{sec:quantum-discord} 

Even though $\tlA$ and $B$ remain unentangled throughout the time evolution under $\hat{H}_{{\rm tot}}=\hat{A}\ot\hat{B}$ for $AB$, $\tlA$ and $B$ can still have quantum correlation in the form of discord \cite{01OllivierDiscord}. The condition for $D(A,B)=0$, a vanishing discord between system A and B, is the more restrictive condition (compared to the separability criterion in Eq.~\ref{eq:separable}) that the bipartite state can be written in the form $\rho=\sum_k\rho_1^k\ot\ket{k}_2\bra{k}_2$, where $\{\ket{k}_2\}$ is an orthonormal basis for the Hilbert space $\mathcal{H}_2$. Clearly, discord is asymmetric, that is generically $D(A,B)\neq D(B,A)$. Discord is also a resource that could provide a quantum advantage in some cases, albeit smaller than the advantage provided by entanglement \cite{12PianiDiscordProblem}. Using results in \cite{13BrownDiscord}, we will show that $\tlA$ and $B$ can possess quantum discord when $\tlA, A$ and $B$ are qubits and the initial state $\rho_B$ is pure under a product Hamiltonian. 

As discussed in Sec.~\ref{sec:B-tlAA-delocalization}, the total system $\tlA A B$ will become genuinely tripartitely entangled as $B$ and $\tlA A$ become entangled at the onset under a generic product Hamiltonian on $AB$. When the initial state $\rho_B$ is pure, the total system $\tlA A B$ will remain in a pure state. $\tlA;A$ is initially entangled in our system setup, and $A;B$ (with $\rho_B$ pure) will become immediately entangled at the onset under a generic $\hat{H}_{{\rm tot}}=\hat{A}\ot\hat{B}$ as shown in Sec.~\ref{sec:pure-pf-AB-entangled}. For three qubits in a total pure state, \cite{13BrownDiscord} showed that the presence of both bipartite and genuine tripartite entanglement in the total system require the presence of discord between $\tlA$ and $B$. In particular, with $A$ entangled with $\tlA$, $D(\tlA, B)>0$; and with $A$ become entangled with $B$, we will have $D(B,\tlA)$ become positive.

Therefore, the system $\tlA;B$ can have quantum discord even though they remain unentangled under $\hat{H}_{{\rm tot}}=\hat{A}\ot\hat{B}$ in the case of qubits with $\rho_B$ being pure. Quantum discord is indeed easier to generate and transfer compared to entanglement. We can in fact perform analytical calculation to quantify the initial change of discord using perturbation theory or plot the dynamics of discord numerically in the case of three qubits at a total pure state, since the entanglement of formation, which enters the expression of discord, simplifies to concurrence in the $2\times 2$ dimensional system \cite{98EFQubit}. We can no longer conclude the presence of discord between $\tlA$ and $B$ when systems become larger than qubits or the initial state $\rho_B$ becomes mixed, which is no longer covered by \cite{13BrownDiscord} as discord can no longer be easily calculated analytically. However, we still suspect the presence of quantum discord between $\tlA$ and $B$ in the generic case. 

\subsection{With Free Hamiltonians}\label{sec:pf-free-tlA-B-separable}

We next consider the cases where the free Hamiltonian for system $A$ or $B$ becomes non-negligible. Let the free Hamiltonian of $A$ be $\hat{C}$ and the free Hamiltonian of $B$ be $\hat{D}$. We examine the two cases where $\hat{H}_{{\rm tot}}=\hat{A}\ot \hat{B}+\mathbb{I}_A\ot \hat{D}$ or  $\hat{A}\ot \hat{B}+\hat{C}\ot\mathbb{I}_{B}$ respectively, that is we include either one of the free Hamiltonians for $A$ and $B$. We show that $\tlA$ and $B$ remain unentangled with the inclusion of only free Hamiltonian $\hat{D}$ for system $B$, while $\tlA$ and $B$ can become entangled when the free Hamiltonian $\hat{C}$ for system $A$ is included, yet inefficiently at the onset.

\subsubsection{The case \texorpdfstring{$\hat{H}_{{\rm tot}}=\hat{A}\ot \hat{B}+\mathbb{I}_A\ot \hat{D}$}{product Hamiltonian with free B}: no entanglement transmission.}
\label{sec:pf-fb-tla-b-separable}

\begin{figure}[htbp]
\centerline{\includegraphics[width=0.65\hsize]{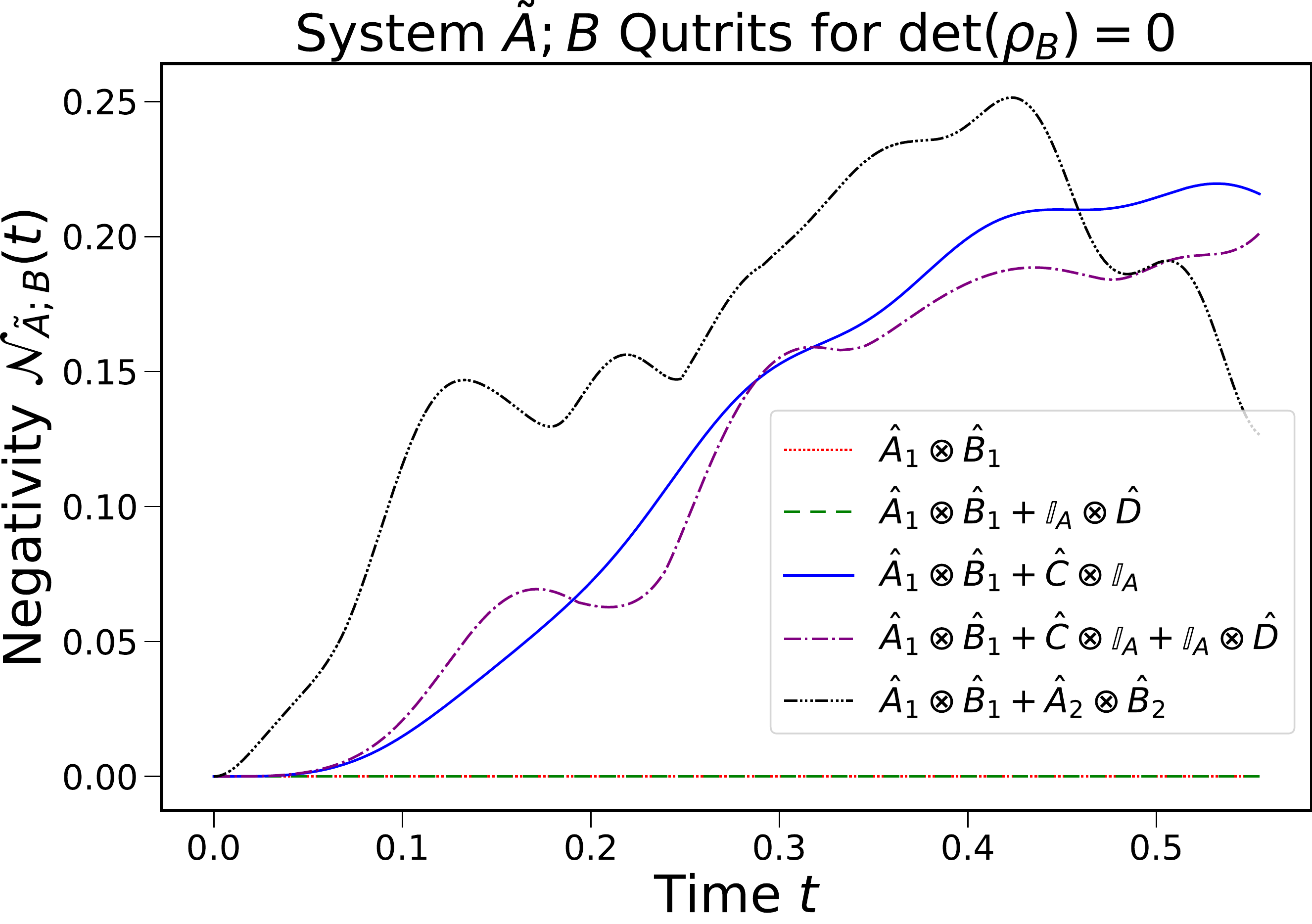}}
\caption{The negativity dynamics for $\tlA;B$ under different forms of total Hamiltonian $\hat{H}_{\rm tot}$ between $A$ and $B$ in our system setup. We choose both $A$ and $B$ as qutrits with $\rho_A$ given in Eq.~\ref{eq:A-qutrit} and $\rho_B$ assumed to be pure. The Hamiltonians listed in the legend are chosen as Eq.~\ref{eq:A_1-qutrit}-\ref{eq:D-qutrit} in Appendix~\ref{sec:qutrit}. We see that the negativity for $\tlA;B$ vanish for a product-form interaction Hamiltonian $\hat{A}\ot\hat{B}$ with or without a free Hamiltonian $\hat{D}$ on $B$, while the addition of a free Hamiltonian $\hat{C}$ on $A$ can entangle $\tlA$ and $B$. A generic interaction Hamiltonian of multiple terms can also entangle $\tlA$ and $B$.}
\label{fig:pure_qutrit_tlAB}
\end{figure}

We first consider the case of a product interaction Hamiltonian with a free Hamiltonian for $B$. Assuming $\hat{A}=\sum_{i}h_A^i\ket*{h_A^i}\bra*{h_A^i}$ and $\hat{B}=\sum_{j} h_B^j\ket*{h_B^j}\bra*{h_B^j}$, $\hat{A}\ot \hat{B}+\mathbb{I}_A\ot\hat{D}$ have eigenvalues $v_{li}$ and eigenvectors of the product form $\ket*{h_A^i}\ot\ket*{v_{li}}$, where $\{v_{li}|l\in\mathbb{N}\}$ and $\{\ket*{v_{li}}|l\in\mathbb{N}\}$ are eigenvalues and eigenvectors of the self-adjoint operator $h_A^i\hat{B}+\hat{D}$ on $\hB$. Indeed,
\begin{align}
(\hat{A}\ot \hat{B}+\mathbb{I}_A\ot \hat{D})\ket*{h_A^i}\ot \ket*{v_{li}}&= h_A^i\ket*{h_A^i}\ot \hat{B}\ket*{v_{li}}+\ket*{h_A^i}\ot \hat{D}\ket*{v_{li}}\\
&=\ket*{h_A^i}\ot(h_A^i\hat{B}+\hat{D})\ket*{v_{li}}=v_{li}\ket*{h_A^i}\ot\ket*{v_{li}}
\label{eq:AB+1D eigendecomp}
\end{align}
Eq.~\ref{eq:AB+1D eigendecomp} yields
\begin{align}
    U^{{\rm with }\; \hat{D}}(t)
    &=\mathbb{I}_{\tlA}\ot\sum_{i,l}\exp\{-it v_{li}\}\ket*{h_A^i}\bra*{h_A^i}\ot\ket*{v_{li}}\bra*{v_{li}}\\
    &=\mathbb{I}_{\tlA}\ot\sum_{i}\ket*{h_A^i}\bra*{h_A^i}\ot\exp\{-it(h_A^i\hat{B}+\hat{D})\} \label{eq:pf-fB-time-evolution-B}
\end{align}
The proof for $\tlA$ and $B$ remaining unentangled with the inclusion of $\hat{D}$ for system $B$ is then similar to the case in Eq.~\ref{eq:pf-nf-tla-b-separable-1}-\ref{eq:pf-nf-tla-b-separable} under Sec.~\ref{sec:pf-nf-tlA-B-separable} by replacing $h_A^i\hat{B}$ with $h_A^i\hat{B}+\hat{D}$:
\begin{align}
\rho_{\tlA B}^{{\rm with }\; \hat{D}}(t)
&=\sum_{i,j}\Tr_A[\ket*{h_A^i}\bra*{h_A^i}\rho_{\tlA A}\ket*{h_A^j}\bra*{h_A^j}]\ot\nonumber\\ 
&\hspace{2cm}\exp\{-it(h_A^i\hat{B}+\hat{D})\}\rho_B\exp\{it(h_A^j\hat{B}+\hat{D})\}\label{eq:pf-fB-tla-b-separable-2}\\
&=\sum_{e}\bra*{h_A^e}\rho_{\tlA A}\ket*{h_A^e}\ot\exp\{-it(h_A^e\hat{B}+\hat{D})\}\rho_B\exp\{it(h_A^e\hat{B}+\hat{D})\}\\
&=\sum_{e}p_e\rho_{\tlA}^e\ot\rho_B^e(t),
\label{eq:pf-fB-tla-b-separable}
\end{align}
where we defined $\rho_{\tlA}^e:=\bra*{h_A^e}\rho_{\tlA A}\ket*{h_A^e}/\Tr[\bra*{h_A^e}\rho_{\tlA A}\ket*{h_A^e}]$, $\rho_B^e(t):=\exp\{-it(h_A^e\hat{B}+\hat{D})\}\rho_B\exp\{it (h_A^e\hat{B}+\hat{D})\}$, and $p_e:=\bra*{h_A^e}\rho_{A}\ket*{h_A^e}$ with $\sum_e p_e=\Tr[\rho_A]=1$. Therefore, we know $\rho_{\tlA B}^{{\rm with }\;\hat{D}}(t)$ remains separable according to Eq.~\ref{eq:separable}. The inclusion of a free Hamiltonian $\hat{D}$ for system $B$ on the product interaction Hamiltonian $\hat{A}\ot\hat{B}$ can not entangle $\tlA$ and $B$. Similar to the previous Sec.~\ref{sec:pf-nf-tlA-B-separable}, the above separability proof is also independent of the exact form of the initial density matrices $\rho_{\tlA A}$ and $\rho_B$, and it works for both finite and infinite dimensional systems. Since $\tlA$ and $B$ remain unentangled, the negativity for $\tlA;B$ vanishes, which is numerically verified in Fig.~\ref{fig:pure_qutrit_tlAB}.

\subsubsection{The case \texorpdfstring{$\hat{H}_{{\rm tot}}=\hat{A}\ot \hat{B}+\hat{C}\ot\mathbb{I}_B$}{product Hamiltonian with free A}: entanglement can be transmitted.} \label{sec:pf-fa-tla-b-entangled}
In the special case when $\hat{C}$ commutes with $\hat{A}$, we can factor the unitary operator $\hat{U}(t)$ into products, and $\tlA$ and $B$ remain unentangled:
\begin{align}
&\rho_{\tlA B}(t)=\Tr_A[\exp\{-it (\hat{A}\ot \hat{B}+\hat{C}\ot\mathbb{I}_B)\} \rho_{\tlA A}\ot\rho_B \exp\{it (\hat{A}\ot \hat{B}+\hat{C}\ot \mathbb{I}_B)\}] \label{eq:pf-fA-commute-1}\\
&=\Tr_A\left[\exp\{-it \hat{A}\ot \hat{B}\}\Big(\exp\{-it\hat{C}\}\rho_{\tlA A}\exp\{it\hat{C}\}\Big) \ot \rho_B  \exp{it\hat{A}\ot \hat{B}}\right]\\
&=\Tr_A\left[\exp\{-it \hat{A}\ot \hat{B}\}\rho_{\tlA A}^{\hat{C}}(t)\ot \rho_B  \exp{it\hat{A}\ot \hat{B}}\right],\label{eq:pf-fA-commute}
\end{align}
where we defined $\rho_{\tlA A}^{\hat{C}}(t)=\exp\{-it\hat{C}\}\rho_{\tlA A}\exp\{it\hat{C}\}$. We can then show that $\tlA$ and $B$ are separable by replacing $\rho_{\tlA A}$ with $\rho_{\tlA A}^{\hat{C}}(t)$ in the proof of Eq.~\ref{eq:pf-nf-tla-b-separable-0}-\ref{eq:pf-nf-tla-b-separable} in Sec.~\ref{sec:pf-nf-tlA-B-separable}. 

With a generic free Hamiltonian $\hat{C}$ which does not commute with $\hat{A}$, the proof strategy for separability used in Sec.~\ref{sec:pf-nf-tlA-B-separable} and \ref{sec:pf-fb-tla-b-separable} break down. $\tlA$ and $B$ can indeed become entangled, which is proved by the numerical example illustrated through the blue (solid) curve of Fig.~\ref{fig:pure_qutrit_tlAB}. Therefore, the total Hamiltonian of the form $\hat{H}_{\rm tot}=\hat{A}\ot\hat{B}+\hat{C}\ot\mathbb{I}_B$ can entangle $\tlA$ and $B$, while $\hat{H}_{\rm tot}=\hat{A}\ot\hat{B}+\mathbb{I}_A\ot\hat{D}$ can not. In this sense, the free Hamiltonian on $A$ is more useful than the free Hamiltonian on $B$ for the entanglement transfer from $\tlA;A$ to $\tlA;B$. The case for $\hat{H}_{\rm tot}=\hat{A}\ot \hat{B}+\hat{C}\ot\mathbb{I}_{B}+\mathbb{I}_A\ot \hat{D}$ with the inclusion of both free Hamiltonians is similar to the case of $\hat{H}_{\rm tot}=\hat{A}\ot\hat{B}+\hat{C}\ot\mathbb{I}_B$: $\tlA$ and $B$ remain separable when $[\hat{A},\hat{C}]=0$, while they can become entangled when $[\hat{A},\hat{C}]\neq 0$, which is demonstrated by the numerical example corresponding to the purple (dash-dot) curve in Fig.~\ref{fig:pure_qutrit_tlAB}. 

Even though the inclusion of the free Hamiltonian $\hat{C}$ on $A$ to the product interaction Hamiltonian $\hat{A}\ot\hat{B}$ can entangle $\tlA$ and $B$ during the time evolution, such entanglement transfer from $\tlA;A$ to $\tlA;B$ will be slow initially, characterized by the vanishing first and second derivatives of the negativity between $\tlA$ and $B$ at the onset. Both blue (solid) and purple (dash-dot) curves in Fig.~\ref{fig:pure_qutrit_tlAB} show this feature of slow negativity generation at the onset. When ${\rm det}(\rho_B)\neq 0$, the negativity for $\tlA;B$ vanishes for a finite amount of time as discussed in Sec.~\ref{sec:mixed, finite time}, which guarantees that all orders of derivatives of $\mathcal{N}_{\tlA;B}(t)$ will vanish at the onset. When ${\rm det}(\rho_B)=0$, we showed in Sec.~\ref{sec:negativity-perturbed-tlA-B} that the first and the second order perturbations $\mathcal{N}_{\tlA;B}^{(1)}$ and $\mathcal{N}_{\tlA;B}^{(2)}$ (corresponding to the first and second derivatives of $\mathcal{N}_{\tlA;B}(t)$ at the onset) vanish when the total Hamiltonian can be written in the form of $\hat{H}_{\rm tot}=\hat{A}\ot{B}+\hat{C}\ot\mathbb{I}_B+\mathbb{I}_A\ot\hat{D}$. Therefore, the transfer of entanglement from $\tlA;A$ to $\tlA;B$ will be slow at the onset even when we consider the free Hamiltonian $\hat{C}$ for $A$ in addition to the product interaction Hamiltonian $\hat{H}_{{\rm int}}=\hat{A}\ot\hat{B}$.

\subsection{Interactions with multiple terms: efficient transmission of entanglement.}\label{sec:m-tla-b-entangled}

When the interaction Hamiltonian between $A$ and $B$ contains multiple terms where $\hat{H}_{{\rm int}}=\sum_{p}\hat{A}^p\ot \hat{B}^p$, we showed in Sec.~\ref{sec:pure, perturbative} that $\tlA$ and $B$ will generically become entangled to the second-order at the onset when ${\rm det}(\rho_B)=0$, and we can characterize the rate of the entanglement generation with the negativity transmissibility introduced in Eq.~\ref{eq:negativity-tlA-B-2-summary}. We therefore expect a generic interaction Hamiltonian with multiple terms to be able to transfer the entanglement from $\tlA;A$ to $\tlA;B$ during the time evolution. 

The green (middle) curve in Fig.~\ref{fig:mixed_qubit_B-tlAA_delocal_SWAP} and the black (topmost) curve in Fig.~\ref{fig:pure_qutrit_tlAB} give examples of $\hat{H}_{{\rm int}}=\sum_{p}\hat{A}^p\ot \hat{B}^p$ entangling $\tlA;B$ in the cases of ${\rm det}(\rho_B)\neq 0$ and ${\rm det}(\rho_B)=0$ respectively. As discussed in Sec.~\ref{sec:mixed, finite time} and \ref{sec:pure, perturbative}, ${\rm det}(\rho_B)=0$ allows fast entanglement generation at the onset, while ${\rm det}(\rho_B)\neq 0$ forces $\mathcal{N}_{\tlA;B}(t)$ to vanish for a finite amount of time initially. Notice that in the case of Fig.~\ref{fig:mixed_qubit_B-tlAA_delocal_SWAP} where $\hat{H}_{\rm tot}=\hat{H}_{\rm SWAP}=\hat{\sigma_x}\ot\hat{\sigma_x}+\hat{\sigma_y}\ot\hat{\sigma_y}+\hat{\sigma_z}\ot\hat{\sigma_z}$, even though $\hat{\sigma_x}\ot\hat{\sigma_x}$, $\hat{\sigma_y}\ot\hat{\sigma_y}$, and $\hat{\sigma_z}\ot\hat{\sigma_z}$ commute with each other, that is the individual terms of $\hat{H}_{\rm tot}$ commute with each other, $\hat{H}_{\rm SWAP}$ can still entangle $\tlA$ and $B$. This suggests $\hat{H}_{\rm tot}$ can entangle $\tlA$ and $B$ even if all the individual Hamiltonian terms commute with each other.

From the previous \ref{sec:pf-fa-tla-b-entangled}, $\hat{H}_{{\rm tot}}=\hat{A}\ot\hat{B}+\hat{C}\ot\mathbb{I}_B$ allows entanglement transfer from $\tlA;A$ to $\tlA;B$, but the negativity for $\tlA;B$ initially vanishes upto the second order, thereby rendering the transmission slow. In comparison, the interaction Hamiltonian of multiple terms can transmit entanglement immediately at the second order when ${\rm det}(\rho_B)=0$. Therefore, when we want fast entanglement transfer from $\tlA;A$ to $\tlA;B$, we should always use a generic interaction Hamiltonian containing multiple terms, and the negativity transmissibility in Eq.~\ref{eq:negativity-tlA-B-2-summary} provides guidance on how to maximize the speed of the entanglement transfer. When we want to minimize the initial speed of the entanglement transmission, we should adopt a product-form interaction Hamiltonian. If we want to avoid the entanglement between $\tlA$ and $B$ altogether, we should choose specific interaction Hamiltonians $\hat{H}_{{\rm tot}}=\hat{A}\ot\hat{B}$ where $\hat{A}$ dominates the free Hamiltonian $\hat{C}$ of $A$ such that $\hat{C}$ can be neglected.

\section{Conclusions and Outlook} \label{sec:conclusion}

Due to entanglement, the state of a composite quantum system $\tilde{A}A$ can and generally does contain delocalized quantum information. The question arises, therefore, how 
the efficiency with which entanglement can be transmitted from system to system in an interaction depends on the Hamiltonians and states involved. This question is of foundational interest when applied to the fundamental interactions in nature. At the same time, this question is of practical interest for quantum technologies, for example, for the purpose of constructing large quantum processors from smaller modules. To this end, in order to access the full tensor product Hilbert space of the collection of modules, entanglement will need to be efficiently transmitted and spread across the modules. 

To study a prototypical scenario for the transmission of entanglement, we considered the situation in which a system $A$ is initially entangled with and purified by an ancilla $\tilde{A}$, i.e., where $A$ and $\tilde{A}$ initially share delocalized quantum information. 
We then let $A$ interact with a system $B$. During the interaction, system $\tilde{A}$ keeps constant entanglement with the combined system $AB$. At the same time, system $\tilde{A}$ can lose entanglement with $A$ and gain entanglement with $B$, i.e., entanglement with $\tilde{A}$ can be transmitted from $A$ to $B$. We analyzed the dynamics of this entanglement transfer and we also found that systems $\tilde{A}$ and $B$ can become discorded before they become entangled. This behavior is related to the fact that discord is a remnant of tri-partite entanglement after one of three subsystems, here $A$, is traced over, see \cite{13BrownDiscord}. 

Our focus, however, has been on the dynamics of entanglement transfer. To this end, we quantified magnitudes of entanglement by using the entanglement monotone negativity. We worked with finite-dimensional Hilbert spaces and we obtained both nonperturbative and perturbative results, as summarized in Table~\ref{tab:dynamics-entanglement-summary} and \ref{tab:tlA-B-summary}. 

For the perturbative results, we expanded to the second order in time, i.e., we studied the dynamics of the transfer of entanglement at the onset of the interaction between systems $A$ and $B$. We did not expand in coupling constants, i.e., we do allow strong couplings, which is useful since strong couplings can occur in quantum technologies \cite{14WeberMicrowaveCavity,15HeeresCavity}. We found, for example, the perturbative result that free Hamiltonians do not affect the change of negativity at the onset of an interaction, up to the second-order in time, which is consistent with our prior findings regarding the evolution of coherent information \cite{20EmilyFirstPaper,21EmilySecondPaper}. Further, we saw that the free Hamiltonian of $\tlA$ never impacts the entanglement transfer (see Eq.~\ref{eq:ftlA-no-impact-on-tlAB} in Sec.~\ref{sec:entanglement}). 

Considering possible interaction Hamiltonians, we saw, for example, that, to second order in time, interaction Hamiltonians of the simple form $H_{\rm int}=\hat{A}\ot \hat{B}$ cannot transfer entanglement regardless of the free Hamiltonians of the systems $\tlA$, $A$ and $B$.
In fact, such interaction Hamiltonians will leave $\tlA$ and $B$ unentangled throughout the entire time evolution - if or as long as the free Hamiltonian of $A$ can be neglected. 

More generally, we analyzed how arbitrary interaction Hamiltonians and arbitrary initial states  influence the proclivity of an interaction to either generate, transfer or lose entanglement. We found that these three proclivities are captured by three quantities that are Hamiltonian and state dependent and that we call negativity susceptibility (Eq.~\ref{eq:negativity-A-B-2-summary}), negativity transmissibility (Eq.~\ref{eq:negativity-tlA-B-2-summary}), and negativity vulnerability (Eq.~\ref{eq:negativity-tlA-A-2-summary}), respectively. 

For example, the notion of negativity transmissibility quantifies the efficiency of the transfer of entanglement at the onset of an interaction. The use of this measure may help experimental efforts to either prevent entanglement transfer (e.g., to prevent decoherence), or to optimize entanglement transfer, e.g., for quantum communication. For example, interaction Hamiltonians and/or the use of suitable regions in state space could be engineered so as to extremize the negativity transmissibility. The negativity transmissibility could also serve as all or part of the cost function for the training of a neural network, e.g., for the purpose of machine-learned quantum error correction or error prevention \cite{18NautrupRL-QEC,22ConvyML-CQEC-SQ,19NiemiecQCrypto-ANN}. 

For another example, the negativity vulnerability can be useful to identify those subsystems of the Hilbert space of $A$, in which entanglement with $\tilde{A}$ is relatively safe from decoherence and, conversely, to identify subsystems in which entanglement with $\tilde{A}$ is at an elevated risk of decohering. We note here that since a subsystem can be specified by giving the observables of the subsystem, it should be possible to describe entanglement vulnerability not only in terms of subsystems but also in terms of observables. This then suggests a connection to the study of entanglement dynamics in the Heisenberg picture \cite{07BenyHeisenbergQEC1,07BenyHeisenbergQEC2,07BenyHeisenbergQEC3,07HewittHeisenbergEntanglement}, which should depend on both the initial states and the observables (interaction Hamiltonians), similar to the notion of the \negaAB $S(\rho_A,\rho_B,\hat{H}_{{\rm tot}})$ in Eq.~\ref{eq:negativity-A-B-2-summary}. 

While our results here have been derived under the assumption that the Hilbert spaces involved are finite dimensional, it will also be very interesting to generalize these results to infinite-dimensional Hilbert spaces. What we can say so far, 
is, for example, that our calculations for the negativity susceptibility and negativity transmissibility for the $A;B$ and $\tlA;B$ systems in Sec.~\ref{sec:pure, perturbative} still hold if $B$ is infinite dimensional, under the condition that $\rho_B$ possesses only a finite number of non-zero eigenvalues and $\tlA$ and $A$ remain finite dimensional. 

This means that, in principle, it is possible to apply our results to a system consisting of two atoms, $A$ and $\tilde{A}$, and a quantum field, $B$, as long as the field's initial density matrix is of finite rank and as long as we take into account only finitely many energy levels of the atoms. For practical applications, the situation is somewhat subtle because a careful UV cutoff would generally be necessary. This is because, in practice one might expect the field to be in a state with a thermal background. On one hand, the condition that $\rho_B$ possesses finite rank would then be violated. On the other hand, it could be argued that for the field to violate the finite rank condition, the field would, unrealistically, have to thermalize infinitely far into the ultraviolet (i.e., e.g., even beyond the Planck scale). 

Our perturbative calculation for the negativity of the $\tlA;A$ system in Sec.~\ref{sec:negativity-perturbed-tlA-A} is more general. The notion of negativity vulnerability (Eq.~\ref{eq:negativity-tlA-A-2-summary}) applies to any situation where $A$ and $\tlA$ are finite dimensional while $B$ can be infinite dimensional without restrictions. In particular, the notion of negativity vulnerability for the $\tlA;A$ system can also be applied if $B$ is in a thermal state (while negativity susceptibility (Eq.~\ref{eq:negativity-A-B-2-summary}) and negativity transmissibility (Eq.~\ref{eq:negativity-tlA-B-2-summary}) for the $A;B$ and $\tlA;B$ systems do not straightforwardly apply if $B$ is in a thermal state). 

Also, our calculations have assumed that $A$ is finite dimensional and that ${\rm det}(\rho_A)\neq 0$. It will be very interesting to generalize our perturbative results to the case with ${\rm det}(\rho_A)=0$ and, finally, for the case in which $\tlA, A$ and $B$ are allowed to be infinite dimensional systems. For a related paper on the question what genuinely new phenomena may arise for infinite-dimensional systems, see, e.g., \cite{07BenyHeisenbergQEC3}. 

$$$$
\bf Acknowledgements. \rm AK acknowledges support through a Discovery Grant of the Natural Sciences and Engineering Research Council of Canada (NSERC) and a Discovery Project grant of the Australian Research Council (ARC). RW acknowledges support through the NSERC USRA program.

\section*{References}
\bibliographystyle{iopart-num}  
\bibliography{refs}

\providecommand{\newblock}{}
\begin{thebibliography}{10}
\expandafter\ifx\csname url\endcsname\relax
  \def\url#1{{\tt #1}}\fi
\expandafter\ifx\csname urlprefix\endcsname\relax\def\urlprefix{URL }\fi
\providecommand{\eprint}[2][]{\url{#2}}

\bibitem{tqs}
Hamma A, Santra S and Zanardi P 2012 {\em \prl\/} {\bf 109} 040502
  (\textit{Preprint} \eprint{1109.4391})

\bibitem{97LloydQCC}
{Lloyd} S 1997 {\em Phys. Rev. A\/} {\bf 55} 1613--1622 (\textit{Preprint}
  \eprint{quant-ph/9604015})

\bibitem{11Wilde}
{Wilde} M~M 2011 {\em arXiv e-prints\/} arXiv:1106.1445 (\textit{Preprint}
  \eprint{1106.1445})

\bibitem{18GyongyosiQCCReview}
{Gyongyosi} L, {Imre} S and {Viet Nguyen} H 2018 {\em arXiv e-prints\/}
  arXiv:1801.02019 (\textit{Preprint} \eprint{1801.02019})

\bibitem{17CuevasEDQCC}
{Cuevas} {\'A}, {Proietti} M, {Ciampini} M~A, {Duranti} S, {Mataloni} P,
  {Sacchi} M~F and {Macchiavello} C 2017 {\em \prl\/} {\bf 119} 100502
  (\textit{Preprint} \eprint{1612.07754})

\bibitem{1948ShannonCommunication}
Shannon C~E 2001 {\em {ACM} {SIGMOBILE} Mob. Comput. Commun. Rev.\/} {\bf 5}
  3--55 \urlprefix\url{https://doi.org/10.1145/584091.584093}

\bibitem{06CoverThomasInformationTheory}
Cover T~M and Thomas J~A 2006 {\em Elements of Information Theory (Wiley Series
  in Telecommunications and Signal Processing)\/} (USA: Wiley-Interscience)
  ISBN 0471241954

\bibitem{10NielsenChuangText}
{Nielsen} M~A and {Chuang} I~L 2010 {\em {Quantum Computation and Quantum
  Information}\/} (Cambridge University Press)

\bibitem{2010LaddQCReview}
{Ladd} T~D, {Jelezko} F, {Laflamme} R, {Nakamura} Y, {Monroe} C and {O'Brien}
  J~L 2010 {\em Nature\/} {\bf 464} 45--53 (\textit{Preprint}
  \eprint{1009.2267})

\bibitem{17BoyerDQC1}
{Boyer} M, {Brodutch} A and {Mor} T 2017 {\em \pra\/} {\bf 95} 022330
  (\textit{Preprint} \eprint{1606.05283})

\bibitem{18KlcoDigitizedSFQC}
Klco N and Savage M~J 2019 {\em Phys. Rev. A\/} {\bf 99} (\textit{Preprint}
  \eprint{1808.10378})
  \urlprefix\url{https://doi.org/10.1103%2Fphysreva.99.052335}

\bibitem{10YuanPhotonQCCommR}
Yuan Z~S, Bao X~H, Lu C~Y, Zhang J, Peng C~Z and Pan J~W 2010 {\em Physics
  Reports\/} {\bf 497} 1--40 ISSN 0370-1573
  \urlprefix\url{https://www.sciencedirect.com/science/article/pii/S0370157310001833}

\bibitem{2021ChenQComm}
Chen Y~A, Zhang Q and {Chen T \etal} 2021 {\em Nature\/} {\bf 589} 214--219

\bibitem{18StephanieQI}
Wehner S, Elkouss D and Hanson R 2018 {\em Science\/} {\bf 362}
  \urlprefix\url{https://www.science.org/doi/abs/10.1126/science.aam9288}

\bibitem{17DegenQS}
Degen C~L, Reinhard F and Cappellaro P 2017 {\em Rev. Mod. Phys.\/} {\bf 89}(3)
  035002 \urlprefix\url{https://link.aps.org/doi/10.1103/RevModPhys.89.035002}

\bibitem{17RosskpfQS}
{Rosskopf} T, {Zopes} J, {Boss} J~M and {Degen} C~L 2017 {\em npj Quantum
  Information\/} {\bf 3} 33 (\textit{Preprint} \eprint{1610.03253})

\bibitem{16HuntemannIonClock}
{Huntemann} N, {Sanner} C, {Lipphardt} B, {Tamm} C and {Peik} E 2016 {\em
  \prl\/} {\bf 116} 063001 (\textit{Preprint} \eprint{1602.03908})

\bibitem{10DongQControll}
Dong D and Petersen I 2010 {\em {IET} Control Theory {\&} Applications\/} {\bf
  4} 2651--2671 \urlprefix\url{https://doi.org/10.1049%2Fiet-cta.2009.0508}

\bibitem{17Gonzalez-Henao}
Gonzalez-Henao J~C, Pugliese E, Euzzor S, Meucci R, Roversi J~A and Arecchi F~T
  2017 {\em Sci. Rep.\/} {\bf 7}
  \urlprefix\url{https://doi.org/10.1038/s41598-017-09989-2}

\bibitem{20EmilyFirstPaper}
{Kendall} E and {Kempf} A 2020 {\em Journal of Physics A: Mathematical and
  Theoretical\/} {\bf 53} 425303 (\textit{Preprint} \eprint{2004.02829})

\bibitem{21EmilySecondPaper}
{Kendall} E, {{\v{S}}oda} B and {Kempf} A 2022 {\em Journal of Physics A:
  Mathematical and Theoretical\/} {\bf 55} 255301 (\textit{Preprint}
  \eprint{2110.06499})

\bibitem{02EMNegativity}
{Vidal} G and {Werner} R~F 2002 {\em \pra\/} {\bf 65} 032314 (\textit{Preprint}
  \eprint{quant-ph/0102117})

\bibitem{09YuSuddenDeathEntanglement}
{Yu} T and {Eberly} J~H 2009 {\em Science\/} {\bf 323} 598 (\textit{Preprint}
  \eprint{0910.1396})

\bibitem{1961renyientropy}
R{\'e}nyi A 1961 On measures of entropy and information {\em Proceedings of the
  Fourth Berkeley Symposium on Mathematical Statistics and Probability, Volume
  1: Contributions to the Theory of Statistics\/} vol~4 (University of
  California Press) pp 547--562

\bibitem{00EMproposal}
{Vidal} G 2000 {\em Journal of Modern Optics\/} {\bf 47} 355--376
  (\textit{Preprint} \eprint{quant-ph/9807077})

\bibitem{05EMReview}
{Plenio} M~B and {Virmani} S 2005 {\em arXiv e-prints\/} quant-ph/0504163
  (\textit{Preprint} \eprint{quant-ph/0504163})

\bibitem{96PeresSeparability}
{Peres} A 1996 {\em \prl\/} {\bf 77} 1413--1415 (\textit{Preprint}
  \eprint{quant-ph/9604005})

\bibitem{96HorodeckiSeparability}
{Horodecki} M, {Horodecki} P and {Horodecki} R 1996 {\em Physics Letters A\/}
  {\bf 223} 1--8 (\textit{Preprint} \eprint{quant-ph/9605038})

\bibitem{18CresswellNegativityExpansion}
Cresswell J~C, Tzitrin I and Goldberg A~Z 2019 {\em Phys. Rev. A\/} {\bf 99}
  (\textit{Preprint} \eprint{1809.07772})

\bibitem{02GurvitsLSBaMMBS}
{Gurvits} L and {Barnum} H 2002 {\em \pra\/} {\bf 66} 062311 (\textit{Preprint}
  \eprint{quant-ph/0204159})

\bibitem{Shor1995}
Shor P~W 1995 {\em Phys. Rev. A\/} {\bf 52}(4) R2493--R2496
  \urlprefix\url{https://link.aps.org/doi/10.1103/PhysRevA.52.R2493}

\bibitem{Monz2009}
Monz T, Kim K and {Villar A S \etal} 2009 {\em Phys. Rev. Lett.\/} {\bf
  103}(20) 200503
  \urlprefix\url{https://link.aps.org/doi/10.1103/PhysRevLett.103.200503}

\bibitem{11quantumcovariance}
Gibilisco P and Isola T 2011 {\em Journal of Mathematical Analysis and
  Applications\/} {\bf 384} 670--676 ISSN 0022-247X
  \urlprefix\url{https://www.sciencedirect.com/science/article/pii/S0022247X11005646}

\bibitem{19CunhaTripartiteEntanglement}
{Cunha} M~M, {Fonseca} A and {Silva} E~O 2019 {\em Universe\/} {\bf 5} 209
  (\textit{Preprint} \eprint{1909.00862})

\bibitem{01OllivierDiscord}
{Ollivier} H and {Zurek} W~H 2001 {\em \prl\/} {\bf 88} 017901
  (\textit{Preprint} \eprint{quant-ph/0105072})

\bibitem{12PianiDiscordProblem}
{Piani} M 2012 {\em \pra\/} {\bf 86} 034101 (\textit{Preprint}
  \eprint{1206.0231})

\bibitem{13BrownDiscord}
{Brown} E~G, {Webster} E~J, {Mart{\'\i}n-Mart{\'\i}nez} E and {Kempf} A 2013
  {\em Annals of Physics\/} {\bf 337} 153--162 (\textit{Preprint}
  \eprint{1212.3275})

\bibitem{98EFQubit}
{Wootters} W~K 1998 {\em \prl\/} {\bf 80} 2245--2248 (\textit{Preprint}
  \eprint{quant-ph/9709029})

\bibitem{14WeberMicrowaveCavity}
{Weber} S~J, {Chantasri} A, {Dressel} J, {Jordan} A~N, {Murch} K~W and
  {Siddiqi} I 2014 {\em Nature\/} {\bf 511} 570--573 (\textit{Preprint}
  \eprint{1403.4992})

\bibitem{15HeeresCavity}
{Heeres} R~W, {Vlastakis} B, {Holland} E, {Krastanov} S, {Albert} V~V,
  {Frunzio} L, {Jiang} L and {Schoelkopf} R~J 2015 {\em \prl\/} {\bf 115}
  137002 (\textit{Preprint} \eprint{1503.01496})

\bibitem{18NautrupRL-QEC}
Nautrup H~P, Delfosse N, Dunjko V, Briegel H~J and Friis N 2019 {\em Quantum\/}
  {\bf 3} 215 (\textit{Preprint} \eprint{1812.08451})

\bibitem{22ConvyML-CQEC-SQ}
{Convy} I, {Liao} H, {Zhang} S, {Patel} S, {Livingston} W~P, {Nguyen} H~N,
  {Siddiqi} I and {Whaley} K~B 2022 {\em New Journal of Physics\/} {\bf 24}
  063019 (\textit{Preprint} \eprint{2110.10378})

\bibitem{19NiemiecQCrypto-ANN}
{Niemiec} M 2019 {\em Quantum Information Processing\/} {\bf 18} 174
  (\textit{Preprint} \eprint{1810.00957})

\bibitem{07BenyHeisenbergQEC1}
{B{\'e}ny} C, {Kempf} A and {Kribs} D~W 2007 {\em \pra\/} {\bf 76} 042303
  (\textit{Preprint} \eprint{0705.1574})

\bibitem{07BenyHeisenbergQEC2}
B{\'e}ny C, Kempf A and Kribs D~W 2007 {\em Phys. Rev. Lett.\/} {\bf 98} 100502
  (\textit{Preprint} \eprint{quant-ph/0608071})

\bibitem{07BenyHeisenbergQEC3}
B{\'e}ny C, Kempf A and Kribs D~W 2009 {\em Journal of Mathematical Physics\/}
  {\bf 50} 062108

\bibitem{07HewittHeisenbergEntanglement}
{Hewitt-Horsman} C and {Vedral} V 2007 {\em \pra\/} {\bf 76} 062319
  (\textit{Preprint} \eprint{quant-ph/0611237})

\bibitem{99MessiahQuantumTextbook}
Messiah A 1999 {\em Quantum Mechanics: Two Volumes Bound as One\/} Dover books
  on physics (Dover Publications)
  \urlprefix\url{https://books.google.ca/books?id=KMK4xwEACAAJ}

\bibitem{17PanineSGEP}
{Panine} M and {Kempf} A 2017 {\em International Journal of Geometric Methods
  in Modern Physics\/} {\bf 14} 1750157 (\textit{Preprint} \eprint{1607.00396})

\bibitem{95Katoperturbation}
Kato T 1995 {\em Perturbation Theory for Linear Operators\/} Classics in
  Mathematics (Springer Berlin Heidelberg) ISBN 9783540586616

\bibitem{18KuzmakHeisenbergSWAP}
{Kuzmak} A~R 2018 {\em International Journal of Quantum Information\/} {\bf 16}
  1850044-108 (\textit{Preprint} \eprint{1809.07171})

\end{thebibliography}

\appendices

\section{Eigenvalue Perturbation}\label{sec:EP} 

To establish the notation and adaptation for our purposes, we here review basic perturbation theoretic tools that we will use. To this end, consider a family of self-adjoint operators $H(t)$ on a separable Hilbert space $\mathcal{H}$. Assuming $H(t)$ is analytic with respect to $t$, we consider its Taylor expansion upto the second order with respect to $t$: 
\begin{equation}
H(t)=H^{(0)}+tH^{(1)}+t^2H^{(2)}+O(t^3),
\label{eq:matrix perturbation}
\end{equation}
where the $f^{th}$ order perturbation $H^{(f)}$ is also self-adjoint. The eigenvalues $E_i(t)$ and the corresponding eigenvectors $\ket{i(t)}$ of $H(t)$ can be expanded as:
\begin{align}
    E_i(t)&=E_i^{(0)}+tE_i^{(1)}+t^2E_i^{(2)}+O(t^3)\\
    \ket{i(t)}&=\ket*{i^{(0)}}+t\ket*{i^{(1)}}+t^2\ket*{i^{(2)}}+O(t^3),
\end{align}
where $H^{(0)}\ket*{i^{(0)}}=E_i^{(0)}\ket*{i^{(0)}}$. For our purpose of calculating negativity perturbation as formulated in Sec.~\ref{sec:PoN}, we want to find $E_i^{(1)}$ and $E_i^{(2)}$ in terms of $\{E_i^{(0)}\}$, $\{\ket{i^{(0)}}\}$, $H^{(1)}$, and $H^{(2)}$. This is a slightly more general problem than the time-independent perturbation theory in quantum mechanics, where only the first-order perturbation $H=H^{(0)}+tH^{(1)}$ is considered \cite{99MessiahQuantumTextbook}. Here we summarize results presented in \cite{17PanineSGEP} with some notation changes. See \cite{95Katoperturbation} for a comprehensive treatment on the subject of eigenvalue perturbation to all orders. 

\subsection{Non-degenerate Case}
\label{sec:NDSOMP}

We first assume that the spectrum of $H^{(0)}$ is non-degenerate, that is the eigenspace of each $E_i^{(0)}$ is of rank 1, and each eigenvector $\ket{i^{(0)}}$ is well-defined with no degeneracy. Following \cite{17PanineSGEP}, we have
\begin{align}
    E_i^{(1)}&=\bra*{i^{(0)}}H^{(1)}\ket*{i^{(0)}}
    \label{eq:first-order-eigenvalue}\\
    E_i^{(2)}&=\bra*{i^{(0)}}H^{(2)}\ket*{i^{(0)}}+\sum_{q\neq i}\frac{H^{(1)}_{iq}H^{(1)}_{qi}}{E_i^{(0)}-E_q^{(0)}}\label{eq:second-order-eigenvalue},\\
    &=\bra*{i^{(0)}}\left[H^{(1)}\left(\sum_{q\neq i}\frac{\hat{P}_q^{(0)}}{E_i^{(0)}-E_q^{(0)}}\right)H^{(1)}+H^{(2)}\right]\ket*{i^{(0)}}\label{eq:second-order-eigenvalue-2},
\end{align}
where $H^{(1)}_{qi}=\bra*{q^{(0)}}H^{(1)}\ket*{i^{(0)}}$ and $\hat{P}_{q}^{(0)}=\ket*{q^{(0)}}\bra*{q^{(0)}}$. When $H^{(2)}=0$, Eq.~\ref{eq:second-order-eigenvalue} recovers the standard second-order eigenvalue correction in the time-independent perturbation theory \cite{99MessiahQuantumTextbook}.

\subsection{Degenerate Case}\label{sec:DSOMP}

We now extend our results in Eq.~\ref{eq:first-order-eigenvalue} and \ref{eq:second-order-eigenvalue} to the case where $H^{(0)}$ in Eq.~\ref{eq:matrix perturbation} has a degenerate spectrum. In general, finding the perturbative corrections to the eigenvalues involves a series of diagonalizations of certain operators on nested subspaces, which is known as the reduction process \cite{95Katoperturbation}. Here we summarize results to the second-order. 

Consider a degenerate eigenvalue $E_i^{(0)}$ of $H^{(0)}$. We denote the projection to the eigenspace of $E_i^{(0)}$ be $\hat{P}^{(0)}_i$, and the eigenspace of $E_i^{(0)}$ can then be denoted as $\hat{P}^{(0)}_i\mathcal{H}$. An arbitrary orthonormal basis $\{\ket{i^{(0)},j}\}$ of the eigenspace $\hat{P}^{(0)}_i\mathcal{H}$ is guaranteed to diagonalize $\hat{P}_i^{(0)}H^{(0)}\hat{P}_i^{(0)}$ due to the degeneracy. According to \cite{17PanineSGEP}, the first-order corrections $\{E_{ij}^{(1)}\}$ to $E_i^{(0)}$ are eigenvalues to the operator
\begin{equation}
    \Lambda_{i}^{(1)}=H^{(1)}|_{\hat{P}^{(0)}_i\mathcal{H}}=\hat{P}_i^{(0)} H^{(1)} \hat{P}_i^{(0)},
    \label{eq:1-reduction-matrix}
\end{equation}
where $H^{(1)}|_{\hat{P}^{(0)}_i\mathcal{H}}$ represents $H^{(1)}$ restricted in the eigenspace $\hat{P}^{(0)}_i\mathcal{H}$ of $E_i^{(0)}$. The additional index $j$ tracks the different eigenvalues to the operator $\Lambda_{i}^{(1)}$, and the original degeneracy in the eigenspace $\hat{P}^{(0)}_i\mathcal{H}$ of $H^{(0)}$ will be broken or partially broken by $\Lambda_{i}^{(1)}$ depending on whether the spectrum $\{E_{ij}^{(1)}\}$ of $\Lambda_{i}^{(1)}$ is degenerate or not. 

Let $\hat{P}_{ij}^{(1)}$ be the projection operator onto the eigenspace of $E_{ij}^{(1)}$, which is nested in the eigenspace of $E_{i}^{(0)}$. The second-order corrections $\{E_{ijk}^{(2)}\}$ to $E_i^{(0)}+E_{ij}^{(1)}t$ are then the eigenvalues to the operator
\begin{align}
    \Lambda_{ij}^{(2)}&=\sum_{q\neq i}\frac{\hat{P}_{ij}^{(1)}H^{(1)}\hat{P}_{q}^{(0)}H^{(1)}\hat{P}_{ij}^{(1)}}{E_i^{(0)}-E_q^{(0)}}+\hat{P}_{ij}^{(1)}H^{(2)}\hat{P}_{ij}^{(1)} \label{eq:2-reduction-matrix-1}\\
    &=\hat{P}_{ij}^{(1)}\left[H^{(1)}\left(\sum_{q\neq i}\frac{\hat{P}_{q}^{(0)}}{E_i^{(0)}-E_q^{(0)}}\right)H^{(1)}+H^{(2)}\right]\hat{P}_{ij}^{(1)}
    \label{eq:2-reduction-matrix-2}\\
    &=\left[H^{(1)}\hat{M}_iH^{(1)}+H^{(2)}\right]|_{\hat{P}_{ij}^{(1)}\mathcal{H}},\label{eq:2-reduction-matrix-3}
\end{align}
where we defined $\hat{M}_i:=\sum_{q\neq i}\frac{\hat{P}_{q}^{(0)}}{E_i^{(0)}-E_q^{(0)}}$. Notice that the eigenvalue corrections in Eq.~\ref{eq:first-order-eigenvalue} and \ref{eq:second-order-eigenvalue-2} in the non-degenerate case are indeed eigenvalues to the operators in Eq.~\ref{eq:1-reduction-matrix} and \ref{eq:2-reduction-matrix-2} respectively. In the non-degenerate case, both $\hat{P}_{i}^{(0)}$ and $\hat{P}_{ij}^{(1)}$ are of rank $1$ (the $j$ index vanishes), and Eq.~\ref{eq:1-reduction-matrix} and \ref{eq:2-reduction-matrix-2} directly give the eigenvalue correction results identical to Eq.~\ref{eq:first-order-eigenvalue} and \ref{eq:second-order-eigenvalue-2}. 

In the degenerate case, we can express the eigenvalues to operators of Eq.~\ref{eq:1-reduction-matrix} and \ref{eq:2-reduction-matrix-2} in forms similar to Eq.~\ref{eq:first-order-eigenvalue} and \ref{eq:second-order-eigenvalue-2} by choosing an appropriate basis. Choose the basis $\{\ket{i^{(0)}}\}$ such that it diagonalizes $H^{(0)}$ and all of $\{\Lambda^{(1)}_i\}$ and $\{\Lambda^{(2)}_{ij}\}$ simultaneously. Choosing such basis is indeed possible due to the nesting of relevant eigenspaces: $\hat{P}_{ij}^{(1)}\mathcal{H}\subset\hat{P}_{i}^{(0)}\mathcal{H}\subset\mathcal{H}$. The choice of orthonormal eigenvectors of the $\{\Lambda^{(2)}_{ij}\}$ fixes the choice of eigenbasis of $\{\Lambda^{(1)}_i\}$, which, in turn, fixes the choice of eigenbasis of $H^{(0)}$. With such $\{\ket{i^{(0)}}\}$, Eq.~\ref{eq:first-order-eigenvalue} and \ref{eq:second-order-eigenvalue-2} give the eigenvalues to Eq.~\ref{eq:1-reduction-matrix} and \ref{eq:2-reduction-matrix-2} respectively. Therefore, the results for the degenerate and non-degenerate case can be cast into similar forms, even though the degenerate case is in fact more complicated by requiring diagonalization of some operators to find the appropriate eigenbasis $\ket{i^{(0)}}$.

When $\Lambda_i^{(1)}$ vanishes, which is the case for our calculation of negativity perturbation, all the first-order eigenvalue corrections $E_{i}^{(1)}=0$ with the index $j$ vanishing. Therefore, the projection operator $\hat{P}_{ij}^{(1)}=\hat{P}_{i}^{(0)}$, then Eq.~\ref{eq:2-reduction-matrix-3} becomes
\begin{align}
    \Lambda_{i}^{(2)}&=\left[H^{(1)}\hat{M}_iH^{(1)}+H^{(2)}\right]|_{\hat{P}_{i}^{(0)}\mathcal{H}}\label{eq:2-reduction-matrix-3-first0}.
\end{align}

\section{Perturbative Calculation of Density Matrices}
\label{sec:perturbation calculation density matrix}

We calculate the first and second-order perturbations of the total density matrix in Eq.~\ref{eq:time-evolution-perturbed} for the tripartite system described in Fig.~\ref{fig:system-setup} and the perturbations of $\rho_{AB}(t)$, $\rho_{\tlA B}(t)$, $\rho_{\tlA A}(t)$, the density matrices for the three bipartite subsystems. 

According to Eq.~\ref{eq:time-evolution-perturbed}, we know the zeroth-order term $\rho^{(0)}=\rho_{\rm tri}$, where $\rho_{\rm tri}$ is the initial total density matrix given in Eq.~\ref{eq:tripartite-rho0}. We evaluate the first-order perturbation $\rho^{(1)}=i(\rho_{\rm tri}\hat{H}_{\rm tri}-\hat{H}_{\rm tri}\rho_{\rm tri})$ by first expressing the term $\rho_{\rm tri}\left(\mathbb{I}\ot \hat{A}^p\ot \hat{B}^p \right)$:
\begin{align}
    &\left(\sum_{i,j}\alpha_i\alpha_{j}\ket*{\tla_i}\bra*{\tla_j}\ot\ket*{a_i}\bra*{a_j}\hat{A}^p\right)\ot\left(\sum_{k}\lambda_B^k\ket*{b_k}\bra*{b_k}\hat{B}^p\right)\nonumber\\
    &=\left(\sum_{i,j,l}\alpha_i\alpha_{j}A^p_{jl}\ket*{\tla_i}\bra*{\tla_j}\ot\ket*{a_i}\bra*{a_l}\right)\ot\left(\sum_{k,v}\lambda_B^kB^p_{kv}\ket*{b_k}\bra*{b_v}\right),
    \label{eq:tri-rho0H}
\end{align}
where we defined $A^p_{jl}=\bra{a_j}\hat{A}^p\ket{a_l}$ and $B^p_{kv}=\bra{b_k}\hat{B}^p\ket{b_v}$.
Therefore, 
\begin{align}
\rho_{\rm tri}\hat{H}_{\rm tri}&=\sum_{p,i,j,l,k,v}\alpha_i\alpha_{j}\lambda_B^kA^p_{jl}B^p_{kv}\ket*{\tla_i}\bra*{\tla_j}\ot\ket*{a_i}\bra*{a_l}\ot\ket*{b_k}\bra*{b_v}
\end{align}
Similarly, we can compute $\hat{H}_{\rm tri}\rho_{\rm tri}$, and obtain the final expression for $\rho_{\rm tri}^{(1)}$:
\begin{align}
\rho_{\rm tri}^{(1)}=i\sum_{p,i,j,l,k,u,v}\alpha_i\alpha_jB^p_{uv}\left(A^p_{jl}\delta_{ik}\lambda_B^u-A^p_{ki}\delta_{jl}\lambda_B^v\right)\ket*{\tla_i}\bra*{\tla_j}\ot\ket*{a_k}\bra*{a_l}\ot\ket*{b_u}\bra*{b_v}
\label{eq:tri-total-1}
\end{align}

We can then find the first-order density matrices for the three bipartite subsystems by taking the respective partial trace. We show the calculation of $\rho_{AB}^{(1)}$ with more detail:
\begin{align}
    \rho_{AB}^{(1)}&=\Tr_{\tlA}[\rho_{\rm tri}^{(1)}]\nonumber\\
    &=i\sum_{p,i,j,l,k,u,v,m}\alpha_i\alpha_jB^p_{uv}\left(A^p_{jl}\delta_{ik}\lambda_B^u-A^p_{ki}\delta_{jl}\lambda_B^v\right)\bra*{\tla_m}\ket*{\tla_i}\bra*{\tla_j}\ket*{\tla_m}\ket*{a_k}\bra*{a_l}\ot\ket*{b_u}\bra*{b_v}\nonumber\\
    &=i\sum_{p,l,k,u,v}A^p_{kl}B^p_{uv}\left(\lambda_A^k\lambda_B^u-\lambda_A^l\lambda_B^v\right)\ket*{a_k}\bra*{a_l}\ot\ket*{b_u}\bra*{b_v}
    \label{eq:A-B-density-1}\\
\rho_{\tlA B}^{(1)}&=i\sum_{p}\left[\sum_{i,j}\alpha_i\alpha_jA^p_{ji}\ket*{\tla_i}\bra*{\tla_j}\right]\ot\left[\sum_{u,v}B^p_{uv}\left(\lambda_B^u-\lambda_B^v\right)\ket*{b_u}\bra*{b_v}\right]\label{eq:tlA-B-density-1}
\\
\rho_{\tlA A}^{(1)}&=i\sum_{p}\Tr[\hat{B}^p\rho_B]\left[\sum_{i,j,l,k}\alpha_i\alpha_j\left(A^p_{jl}\delta_{ik}-A^p_{ki}\delta_{jl}\right)\ket*{\tla_i}\bra*{\tla_j}\ot\ket*{a_k}\bra*{a_l}\right]
\label{eq:tlA-A-density-1}
\end{align}

Similarly, we can express the second-order perturbation term $\rho_{\rm tri}^{(2)}$ in  Eq.~\ref{eq:time-evolution-perturbed}: 
\begin{align}
&\rho_{\rm tri}^{(2)}=\hat{H}_{\rm tri}\rho_{\rm tri}\hat{H}_{\rm tri}-\frac{1}{2}\rho_{\rm tri}\hat{H}_{\rm tri}^2-\frac{1}{2}\hat{H}_{\rm tri}^2\rho_{\rm tri}\nonumber\\
&=\sum_{p,q}\Bigg\{\Big(\sum_{i,j,k,l}A^p_{ki}\alpha_i\alpha_{j}A^q_{jl}\ket*{\tla_i}\bra*{\tla_j}\ot \ket*{a_k}\bra*{a_l}\Big)\ot\Big(\sum_{t,u,v}B^p_{tu}\lambda_B^uB^q_{uv}\ket*{b_t}\bra*{b_v}\Big)\nonumber\\
&\qquad-\frac{1}{2}\Big(\sum_{i,j,k,l}\alpha_i\alpha_{j}A^p_{jl}A^q_{lk}\ket*{\tla_i}\bra*{\tla_j}\ot\ket*{a_i}\bra*{a_k}\Big)\ot\Big(\sum_{t,u,v}\lambda_B^uB^p_{uv}B^q_{vt}
\ket*{b_u}\bra*{b_t}\Big)\nonumber\\
&\qquad-\frac{1}{2}\Big(\sum_{i,j,k,l}\alpha_i\alpha_{j}A^p_{kl}A^q_{li}\ket*{\tla_i}\bra*{\tla_j}\ot\ket*{a_k}\bra*{a_j}\Big)\ot\Big(\sum_{t,u,v}\lambda_B^vB^p_{tu}B^q_{uv}\ket*{b_t}\bra*{b_v}\Big)\Bigg\}.
\label{eq:tri-total-2}
\end{align}
Taking the respective partial trace of the above Eq.~\ref{eq:tri-total-2}, we obtain:
\begin{align}
&\rho_{AB}^{(2)}=\Tr_{\tlA}[\rho_{\rm tri}^{(2)}]\nonumber\\
&=\sum_{p,q,k,m,l,u,t,v} A^p_{km}A^q_{ml}B^p_{ut}B^q_{tv}\left( \lambda_A^{m}\lambda_B^{t}-\frac{1}{2}\lambda_A^{k}\lambda_B^{u}-\frac{1}{2}\lambda_A^{l}\lambda_B^{v}\right)\ket{a_k}\bra{a_l}\otimes\ket{b_u}\bra{b_v},
\label{eq:A-B-density-2}
\\
&\rho_{\tlA B}^{(2)}=\Tr_{A}[\rho_{\rm tri}^{(2)}]\nonumber\\
&=\sum_{p,q,i,j,m,t,u,v} \alpha_i\alpha_{j}B^p_{ut}B^q_{tv}\left\{A^p_{mi}A^q_{jm}\lambda_B^t-\frac{1}{2}A^q_{mi}A^p_{jm}(\lambda_B^u+\lambda_B^v)\right\}\ket*{\tla_i}\bra*{\tla_j}\otimes\ket{b_u}\bra{b_v},
\label{eq:tlA-B-density-2}
\\
&\rho_{\tlA A}^{(2)}=\Tr_{B}[\rho_{\rm tri}^{(2)}]\nonumber\\
    &=\sum_{p,q,i,j,k,l}\Tr[\hat{B}^p\rho_B\hat{B}^q]A^p_{ki}\alpha_i\alpha_{j}A^q_{jl}\ket*{\tla_i}\bra*{\tla_j}\ot\ket*{a_k}\bra*{a_l}-\frac{1}{2}\sum_{p,q,i,j,k,l}\Tr[\hat{B}^p\rho_B\hat{B}^q]\nonumber\\
    &\qquad\qquad\alpha_i\alpha_{j}\left(\delta_{ik}\sum_fA^p_{fl}A^q_{jf}+\delta_{lj}\sum_fA^p_{fi}A^q_{kf}\right)\ket*{\tla_i}\bra*{\tla_j}\ot\ket*{a_k}\bra*{a_l}.
    \label{eq:tlA-A-density-2}
\end{align}
These perturbative expressions of density matrices are used to calculate the perturbations of negativity for bipartite subsystems $A;B$, $\tlA;B$, and $\tlA,A$ respectively.

\section{Perturbative Calculation of Negativity}\label{sec:appendix-perturb-negativity}

\subsection{Negativity for \texorpdfstring{$A;B$}{A;B}}
\label{sec:negativity-A-B-calculation}

Following the analysis and notations introduced in Sec.~\ref{sec:negativity-perturbed-A-B}, we continue to calculate the first and second-order perturbation terms for the negativity of $A;B$ with the assumption of ${\rm det}(\rho_B)=0$ and ${\rm det}(\rho_A)\neq0$. According to Eq.~\ref{eq:1-reduction-matrix} in Appendix~\ref{sec:DSOMP}, we first need to find $\hat{H}^{(1)}$ restricted in the degenerate subspace ${\hat{P}^{(0)}_i\mathcal{H}}$. Here the first-order operator correction $\hat{H}^{(1)}=\rho_{A;B}^{T_1(1)}$ is given in Eq.~\ref{eq:pt-A-B-density-1}, and the eigenspace to the zero eigenvalue of $\rho_{A;B}^{T_1(0)}$ is $\hA\ot \mathcal{D}_B$ where $\mathcal{D}_B$ is given in Eq.~\ref{eq:degenerate-B}. Therefore,
\begin{align}
\rho^{T_1(1)}_{A;B}|_{\hA\ot \mathcal{D}_B}=i\sum_{p,i,k}\sum_{j>N,l>N}A^p_{ki}B^p_{jl}\left\{\lambda_A^{k}\lambda_B^{j}-\lambda_A^{i}\lambda_B^{l}\right\}\ket{a^*_i}\bra{a^*_k}\otimes\ket{b_j}\bra{b_l}=0,
\label{eq:1-perturb-density-pt-pure-degenerate}
\end{align}
since $\lambda_B^j=\lambda_B^l=0$ for any $j,l>N$. Therefore, $\Lambda^{(1)}=\rho^{T_1(1)}_{A;B}|_{\hA\ot \mathcal{D}_B}$ vanishes. Since the eigenvalue corrections will be the eigenvalues to the vanishing operator $\Lambda^{(1)}$ as described in Appendix~\ref{sec:DSOMP}, we can conclude that
\begin{equation}
    \lambda_{ij,A;B}^{T_1(1)}=0, \text{ for } j>N.
\end{equation}
As a result, no initial eigenvalues of $\rho_{A;B}^{T_1(1)}$ can contribute to $\mathcal{N}(\rho_{AB}(t))$ at the first-order, so 
\begin{align}
\mathcal{N}_{A;B}^{(1)}=0.
\label{eq:negativity-A-B-1}
\end{align}

We next compute $\mathcal{N}_{A;B}^{(2)}$. According to Eq.~\ref{eq:negativity-perturb-negative} and the analysis in Sec.~\ref{sec:PoN}, 
\begin{align}
\mathcal{N}(\rho_{AB}(t))&=\sum_{\lambda_{n,A;B}^{T_1}(t)<0}-(\lambda_{n,A;B}^{T_1}+\lambda_{n,A;B}^{T_1(1)}t+\lambda_{n,A;B}^{T_1(2)}t^2)+O(t^3),\\
&=\left(\sum_{i,j>N,\lambda_{ij,A;B}^{T_1(2)}<0}-\lambda_{ij,A;B}^{T_1(2)}\right)t^2+O(t^3).
\end{align}
$\lambda_{n,A;B}^{T_1}(t)$ can only contribute to negativity at the onset of interaction if initially $\lambda_{n,A;B}^{T_1}\leq 0$, so only the eigenvalues $\{\lambda_{ij,A;B}^{T_1}(t)|j>N,i\}$ of $\rho^{T_1}_{AB}(t)$ in the vanishing subspace $\hA\ot \mathcal{D}_B$ can contribute to the perturbation of negativity. With $\lambda_{ij,A;B}^{T_1(1)}=0$, the leading order behavior of $\lambda_{ij,A;B}^{T_1}(t)$ will be decided by $\lambda_{ij,A;B}^{T_1(2)}$: if $\lambda_{ij,A;B}^{T_1(2)}<0$, then $\lambda_{ij,A;B}^{T_1}(t)$ drops below 0 and contributes to $\mathcal{N}_{AB}^{(2)}$. Therefore,
\begin{align}
\mathcal{N}_{A;B}^{(2)}&= -\sum_{i,j>N,\lambda_{ij,A;B}^{T_1(2)}<0}\lambda_{ij,A;B}^{T_1(2)}.
\label{eq:negativity-A-B-2-intermediate}
\end{align}

According to Eq.~\ref{eq:2-reduction-matrix-3-first0} of Appendix~\ref{sec:DSOMP} with $\Lambda^{(1)}$ vanishing, $\{\lambda_{ij,A;B}^{T_1(2)}|j>N,i\}$ will be the eigenvalues of the operator $\hat{F}_{A;B}$ where we define 
\begin{equation}
 \hat{F}_{A;B}:=\rho_{A;B}^{T_1(1)}\hat{M}\rho_{A;B}^{T_1(1)}+\rho_{A;B}^{T_1(2)}|_{\hA\ot\mathcal{D}_B}
 \label{eq:F-A-B-definition},
\end{equation}
and 
\begin{align}
\hat{M}&=\sum_{E_u^{(0)}\neq E_n^{(0)}}\frac{\ket*{u^{(0)}}\bra*{u^{(0)}}}{E_n^{(0)}-E_u^{(0)}}=\sum_{u}\sum_{t=1}^{N}\frac{\ket*{a^*_u}\ot\ket*{b_t}\bra*{a^*_u}\ot\bra*{b_t}}{0-\lambda_A^u\lambda_B^t}
\label{eq:M-pt-AB-1}\\
&=-(\sum_{u}\frac{\ket*{a^*_u}\bra*{a^*_u}}{\lambda_A^u})\ot(\sum_{t\leq N}\frac{\ket*{b_t}\bra*{b_t}}{\lambda_B^t}),
\label{eq:M-pt-AB-2}
\end{align}
where $E_n^{(0)}=0$ for the subspace $\hA\ot \mathcal{D}_B$ with $\rho_B$ having zero eigenvalue in $\mathcal{D}_B$. $\ket*{a^{*}_u}\ot\ket*{b_t}$ for $1\leq t\leq N$ are the eigenvectors of $\rho_{A;B}^{T_1(0)}$ within the subspace $\hA\ot N_B$ with eigenvalues $\lambda_A^{u}\lambda_B^{t}\neq 0$.

Since $\mathcal{N}_{A;B}^{(2)}$ is simply the absolute value of the sum of all negative eigenvalues of $\hat{F}_{A;B}$ according to Eq.~\ref{eq:negativity-A-B-2-intermediate}, we can write
\begin{equation}
\mathcal{N}_{A;B}^{(2)}=\frac{1}{2}\left(\Tr\left[\sqrt{\hat{F}_{A;B}^{\dagger} \hat{F}_{A;B}}\right]-\Tr\left[ \hat{F}_{A;B}\right]\right)\label{eq:negativity-A-B-2},
\end{equation}
where $\Tr\left[\sqrt{\hat{F}_{A;B}^{\dagger} \hat{F}_{A;B}}\right]$ represents the sum of singular values of $\hat{F}_{A;B}$. Eq.~\ref{eq:negativity-A-B-2} is similar to the definition of negativity in Eq.~\ref{eq:negativity-norm}.

We now proceed to give an explicit expression for $\hat{F}_{A;B}$ in Eq.~\ref{eq:F-A-B-definition}. We first evaluate the term $\rho_{A;B}^{T_1(1)}\hat{M}\rho_{A;B}^{T_1(1)}$:
\begin{align}
    &\rho_{A;B}^{T_1(1)}M\rho_{A;B}^{T_1(1)}\\
    &=\sum_{p,i,j,k,l}A^p_{ki}B^p_{jl}\left\{\lambda_A^{k}\lambda_B^{j}-\lambda_A^{i}\lambda_B^{l}\right\}\ket{a^*_i}\bra{a^*_k}\otimes\ket{b_j}\bra{b_l} (\sum_{u}\frac{\ket*{a^*_u}\bra*{a^*_u}}{\lambda_A^u})\ot(\sum_{t\leq N}\frac{\ket*{b_t}\bra*{b_t}}{\lambda_B^t}) \nonumber\\
        &\qquad\qquad\sum_{q,w,x,y,z}A^q_{yw}B^q_{xz}\left\{\lambda_A^{y}\lambda_B^{x}-\lambda_A^{w}\lambda_B^{z}\right\}\ket{a^*_w}\bra{a^*_y}\otimes\ket{b_x}\bra{b_z}\\
    &=\sum_{t\leq N,p,i,j,u,q,y,z}\frac{1}{\lambda_A^u\lambda_B^t}A^p_{ui}B^p_{jt}\left\{\lambda_A^{u}\lambda_B^{j}-\lambda_A^{i}\lambda_B^{t}\right\}A^q_{yu}B^q_{tz}\left\{\lambda_A^{y}\lambda_B^{t}-\lambda_A^{u}\lambda_B^{z}\right\}\nonumber\\
        &\qquad\qquad\ket{a^*_i}\bra{a^*_y}\ot\ket{b_j}\bra{b_z}
    \label{eq:H1MH1-A-B}
\end{align}
Restricting the above operator to the subspace $\hA\ot \mathcal{D}_B$, we have 
\begin{align}
    &\rho_{A;B}^{T_1(1)}M\rho_{A;B}^{T_1(1)}|_{\hA\ot \mathcal{D}_B}\nonumber\\
    &=\sum_{t\leq N,j>N,z>N,p,i,u,q,y}\frac{1}{\lambda_A^u\lambda_B^t}A^p_{ui}B^p_{jt}\left\{\lambda_A^{u}\lambda_B^{j}-\lambda_A^{i}\lambda_B^{t}\right\}A^q_{yu}B^q_{tz}\left\{\lambda_A^{y}\lambda_B^{t}-\lambda_A^{u}\lambda_B^{z}\right\}\nonumber\\
        &\qquad\qquad\ket{a^*_i}\bra{a^*_y}\ot\ket{b_j}\bra{b_z}\\
    &=-\sum_{t\leq N,j>N,z>N,p,i,u,q,y}\frac{1}{\lambda_A^u\lambda_B^t}A^p_{ui}B^p_{jt}\lambda_A^{i}\lambda_B^{t}A^q_{yu}B^q_{tz}\lambda_A^{y}\lambda_B^t\ket{a^*_i}\bra{a^*_y}\ot\ket{b_{j}}\bra{b_{z}}\\
    &=-\sum_{p,q}\left(\sum_{i,k,m}\frac{\lambda_A^{i}\lambda_A^{k}}{\lambda_A^m}A^q_{km}A^p_{mi}\ket{a^*_i}\bra{a^*_k}\right)\ot\left(\sum_{j>N,l>N,t\leq N}B^p_{jt}\lambda_ B^tB^q_{tl}\ket{b_{j}}\bra{b_{l}}\right),
    \label{eq:H1MH1-A-B-restricted}
\end{align}
since $\lambda_B^{j}=\lambda_B^{z}=0$ for $j,z>N$. 

We next find $\rho_{A;B}^{T_1(2)}|_{\hA\ot \mathcal{D}_B}$ using Eq.~\ref{eq:pt-A-B-density-2}:
\begin{align}
    &\rho_{A;B}^{T_1(2)}|_{\hA\ot \mathcal{D}_B}\\
    &=\sum_{j>N,l>N,p,q,i,k,m,u} A^p_{km}A^q_{mi}B^p_{ju}B^q_{ul}\left( \lambda_A^{m}\lambda_B^{u}-\frac{1}{2}\lambda_A^{i}\lambda_B^{l}-\frac{1}{2}\lambda_A^{k}\lambda_B^{j}\right)\ket{a^*_i}\bra{a^*_k}\otimes\ket{b_j}\bra{b_l}\\
    &=\sum_{p,q}\left(\sum_{i,k,m}\lambda_A^{m}A^p_{km}A^q_{mi}\ket{a^*_i}\bra{a^*_k}\right)\otimes \left(\sum_{j>N,l>N,u}B^p_{ju}\lambda_B^uB^q_{ul}\ket{b_j}\bra{b_l}\right).
    \label{eq:H2-A-B-restricted}
\end{align}
We define
\begin{align}
    \hat{F}_{B}^{p,q}&:=\sum_{j>N,l>N,t\leq N}B^p_{jt}\lambda_ B^tB^q_{tl}\ket{b_{j}}\bra{b_{l}}=\sum_{j>N,l>N,t}B^p_{jt}\lambda_ B^tB^q_{tl}\ket{b_{j}}\bra{b_{l}}\label{eq:FAB-B-dirac}\\
    &=\left(\sum_{j>N}\ket*{b_j}\bra{b_j}\right)\hat{B}^p\left(\sum_t\lambda_B^t\ket*{b_t}\bra*{b_t}\right)\hat{B}^q\left(\sum_{l>N}\ket*{b_l}\bra{b_l}\right)\\
    &=\hat{P}_B^D\hat{B}^p\rho_B\hat{B}^q\hat{P}_B^D\label{eq:FAB-B-matrix},
\end{align}
where $\hat{P}_B^D$ is the projection operator into the subspace of $\hB$ defined in Eq.~\ref{eq:degenerate-projection} where $\rho_B$ has zero eigenvalue. 

Combining Eq.~\ref{eq:H1MH1-A-B-restricted} and \ref{eq:H2-A-B-restricted}, we obtain the the final expression for $\hat{F}_{A;B}$ using the definition of Eq.~\ref{eq:FAB-B-matrix}:
\begin{align} 
\hat{F}_{A;B}&=\sum_{p,q}\left[\sum_{i,k,m}\left(\lambda_A^{m}A^p_{km}A^q_{mi}-\frac{\lambda_A^{i}\lambda_A^{k}}{\lambda_A^m}A^q_{km}A^p_{mi}\right)\ket{a^*_i}\bra{a^*_k}\right]\otimes \hat{F}_B^{p,q}
\label{eq:H2-H1MH1-A-B-restricted}\\
&=\sum_{p,q}\hat{F}_A^{p,q}\ot\hat{F}_B^{p,q},\label{eq:F-A-B}
\end{align}
where we can further define
\begin{align}
\hat{F}_A^{p,q}&:=\sum_{i,k,m}\left(\lambda_A^{m}A^p_{km}A^q_{mi}-\frac{\lambda_A^{i}\lambda_A^{k}}{\lambda_A^m}A^q_{km}A^p_{mi}\right)\ket{a^*_i}\bra{a^*_k}\label{eq:FAB-A-dirac}\\
&=\sum_{i}\ket*{a^*_i}\bra*{a^*_i}\hat{A}^{q*}\sum_m\lambda_A^{m}\ket*{a^*_m}\bra*{a^*_m}\hat{A}^{p*}\sum_{k}\ket*{a^*_k}\bra{a^*_k}\nonumber\\
&\qquad-\sum_{i}\lambda_A^{i}\ket*{a^*_i}\bra*{a^*_i}\hat{A}^{p*}\sum_m\frac{1}{\lambda_A^m}\ket*{a^*_m}\bra*{a^*_m}\hat{A}^{q*}\sum_k\lambda_A^{k}\ket*{a^*_k}\bra{a^*_k}\\
&=\hat{A}^{q*}\rho_A^*\hat{A}^{p*}-\rho_A^* \hat{A}^{p*}(\rho_A^{-1})^*\hat{A}^{q*}\rho_A^*\\
&=(\hat{A}^{q}\rho_A\hat{A}^{p}-\rho_A \hat{A}^{p}\rho_A^{-1}\hat{A}^{q}\rho_A)^*
\label{eq:FAB-A-matrix}
\end{align}
where we have used the following relation
\begin{align}
A^p_{ij}=\bra*{a_i}\hat{A}^p\ket*{a_j}=\bra*{a_j}\hat{A}^p\ket*{a_i}^*=\bra*{a^*_j}\hat{A}^{p*}\ket*{a^*_i}
\label{eq:conjugate-relation}.
\end{align}

Supplementing the above Eq.~\ref{eq:F-A-B} to Eq.~\ref{eq:negativity-A-B-2}, we have finally found an expression for $\mathcal{N}_{A;B}^{T_1(2)}$, which will be used in Sec.~\ref{sec:negativity-perturbed-A-B}. According to Eq.~\ref{eq:negativity-A-B-2}, we need to find negative eigenvalues of $\hat{F}_{A;B}$ in order to calculate $\mathcal{N}_{A;B}^{(2)}$. In general, there are no simple analytical expressions for the diagonalization of $\hat{F}_{A;B}$, which contains multiple tensor product terms. The eigenbasis of $\hat{F}_{A;B}$ in $\hA\ot \mathcal{D}_B$ will generally be entangled. 

\subsection{Negativity for \texorpdfstring{$\tilde{A};B$}{A;B}}
\label{sec:negativity-tlA-B-calculation} We calculate the first and second-order perturbation terms for the negativity of $\tlA;B$ assuming ${\rm det}(\rho_B)=0$ and ${\rm det}(\rho_A)\neq 0$. The overall procedure is similar to Appendix~\ref{sec:negativity-A-B-calculation} where we applied the formulas of the degenerate eigenvalue perturbation theory reviewed in Appendix~\ref{sec:DSOMP}. We show the main steps of the calculation.

Under the formalism introduced in Sec.~\ref{sec:PoN}, we find the first three perturbative terms for the partial transpose density matrix $\rho^{T_1}_{\tlA;B}(t)$. The zeroth-order term $\rho_{\tlA ;B}^{T_1(0)}$ and its associated eigenvalues are $\lambda_{ik,\tlA;B}^{T_1(0)}$ already given in Eq.~\ref{eq:pt-tlA-B-density-0} and \ref{eq:pt-tlA-B-eigen-0}. The first and second-order terms $\rho_{\tlA ;B}^{T_1(1)}$ and $\rho_{\tlA;B}^{T_1(2)}$ can be obtained by taking the partial transpose of Eq.~\ref{eq:tlA-B-density-1} and \ref{eq:tlA-B-density-2} derived in Appendix~\ref{sec:perturbation calculation density matrix}. We summarize the results below:
\begin{align}
    &\lambda_{ik,A;B}^{T_1(0)}=\lambda_A^{i}\lambda_B^{k}
\label{eq:pt-tlA-B-eigen-0-summary}\\
&\rho_{\tlA ;B}^{T_1(0)}=\left(\sum_{i}\lambda_A^i\ket*{\tla^{*}_i}\bra*{\tla^{*}_i}\right)\ot\left(\sum_{k}\lambda_B^k\ket*{b_k}\bra*{b_k}\right)\label{eq:pt-tlA-B-density-0-summary}\\
&\rho_{\tlA;B}^{T_1(1)}=i\sum_{p}\left[\sum_{i,j}\alpha_i\alpha_jA^p_{ij}\ket*{\tla^*_i}\bra*{\tla^*_j}\right]\ot\left[\sum_{u,v}B^p_{uv}\left(\lambda_B^u-\lambda_B^v\right)\ket*{b_u}\bra*{b_v}\right]\label{eq:pt-tlA-B-density-1}\\
&\rho_{\tlA ;B}^{T_1(2)}=\sum_{p,q,i,j,m,t,u,v} \alpha_i\alpha_{j}B^p_{ut}B^q_{tv}\left\{A^p_{mj}A^q_{im}\lambda_B^t-\frac{1}{2}A^q_{mj}A^p_{im}(\lambda_B^u+\lambda_B^v)\right\}\ket*{\tla^*_i}\bra*{\tla^*_j}\otimes\ket{b_u}\bra{b_v}\label{eq:pt-tlA-B-density-2}
\end{align} 

With the assumption ${\rm det}(\rho_B)=0$, we separate the spectrum of $\rho_B$ into two parts: $\lambda_B^{k}\neq 0$ for $1\leq k\leq N$ and $\lambda_B^{k}=0$ for $k
\geq N+1$, similar to Sec.~\ref{sec:negativity-perturbed-A-B}. We use the projection operators $\hat{P}_B^N$ and  $\hat{P}_B^D$, as well as their corresponding subspaces $\mathcal{N}_B$ and $\mathcal{D}_B$ of $\hB$ defined in Eq.~\ref{eq:non-degenerate-projection}-\ref{eq:degenerate-B}. Similar to Sec.~\ref{sec:negativity-perturbed-A-B}, we know $\lambda_{ik,\tlA;B}^{T_1(0)}> 0$ for any $i$ and $1\leq k\leq N$, which will not contribute to the perturbation of negativity according to Sec.~\ref{sec:PoN}. In comparison, $\lambda_{ik,\tlA;B}^{T_1(0)}=0$ for any $i$ and $k\geq N+1$ can contribute to the negativity if its leading order change is negative, which can be determined by results in Appendix~\ref{sec:DSOMP}.

According to Eq.~\ref{eq:1-reduction-matrix} in Appendix~\ref{sec:DSOMP}, the first-order eigenvalue perturbations $\lambda_{ik,\tlA;B}^{T_1(1)}$ will be the eigenvalues of $\rho_{\tlA ;B}^{T_1(1)}|_{\hlA\ot \mathcal{D}_B}$:
\begin{align}
\rho_{\tlA;B}^{T_1(1)}|_{\hlA\ot \mathcal{D}_B}&=i\sum_{p}\left[\sum_{i,j}\alpha_i\alpha_jA^p_{ij}\ket*{\tla^*_i}\bra*{\tla^*_j}\right]\ot\left[\sum_{u>N,v>N}B^p_{uv}\left(\lambda_B^u-\lambda_B^v\right)\ket*{b_u}\bra*{b_v}\right]\\
&=0,
\end{align}
since $\lambda_B^u=\lambda_B^v=0$ for any $u,v> N$. We see that $\rho_{\tlA ;B}^{T_1(1)}|_{\hlA\ot \mathcal{D}_B}$ vanishes, so do the first-order eigenvalue perturbations:
\begin{equation}
    \lambda_{ij,\tlA;B}^{T_1(1)}=0, \text{ for } j>N.
\end{equation}
Therefore, no eigenvalues of $\rho_{A;B}^{T_1(1)}(t)$ can initially contribute to $\mathcal{N}(\rho_{AB}(t))$ at the first-order:
\begin{align}
\mathcal{N}_{\tlA;B}^{(1)}=0
\label{eq:negativity-tlA-B-1}
\end{align}

For the second-order perturbation of negativity, we can find the following expressions similar to Eq.~\ref {eq:F-A-B-definition}, \ref{eq:negativity-A-B-2}, and \ref{eq:M-pt-AB-2} in Appendix~\ref{sec:negativity-A-B-calculation}:
\begin{align}
\mathcal{N}_{\tlA ;B}^{(2)}&=\frac{1}{2}\left(\Tr\left[\sqrt{\hat{F}_{\tlA;B}^{\dagger} \hat{F}_{\tlA;B}}\right]-Tr\left[ \hat{F}_{\tlA ;B}\right]\right)\label{eq:negativity-tlA-B-2}\\
\hat{F}_{\tlA;B}&=(\rho_{\tlA ;B}^{T_1(1)}\hat{M}\rho_{\tlA ;B}^{T_1(1)}+\rho_{\tlA ;B}^{T_1(2)})|_{\hlA\ot \mathcal{D}_B}
\label{eq:F-tlA-B-definition}\\
\hat{M}&=-(\sum_{u}\frac{\ket*{\tla^*_u}\bra*{\tla^*_u}}{\lambda_A^u})\ot(\sum_{t\leq N}\frac{\ket*{b_t}\bra*{b_t}}{\lambda_B^t}),
\label{eq:M-pt-tlAB}
\end{align}
We next explicitly express the two terms of $\hat{F}_{\tlA ;B}$ in Eq.~\ref{eq:F-tlA-B-definition}: 
\begin{align}
&\rho_{\tlA;B}^{T_1(1)}\hat{M}\rho_{\tlA ;B}^{T_1(1)}=-\sum_{p}\left[\sum_{i,j}\alpha_i\alpha_jA^p_{ij}\ket*{\tla^*_i}\bra*{\tla^*_j}\right]\ot\left[\sum_{u,v}B^p_{uv}\left(\lambda_B^u-\lambda_B^v\right)\ket*{b_u}\bra*{b_v}\right]\nonumber\\
&\qquad\qquad\left(\sum_{e}\frac{\ket*{\tla^*_e}\bra*{\tla^*_e}}{\lambda_A^e}\right)\ot\left(\sum_{t\leq N}\frac{\ket*{b_t}\bra*{b_t}}{\lambda_B^t}\right)\sum_{q}\left[\sum_{x,y}\alpha_x\alpha_yA^q_{xy}\ket*{\tla^*_x}\bra*{\tla^*_y}\right]\ot\nonumber\\
&\qquad\qquad\left[\sum_{z,w}B^q_{zw}\left(\lambda_B^z-\lambda_B^w\right)\ket*{b_z}\bra*{b_w}\right]\\
&=-\sum_{t\leq N,p,q,e,i,y,u,w}\frac{\alpha_i\alpha_y}{\lambda_B^t}A^p_{ie}A^q_{ey}B^p_{ut}B^q_{tw}\left(\lambda_B^t-\lambda_B^u\right)\left(\lambda_B^t-\lambda_B^w\right)\ket*{\tla^*_i}\bra*{\tla^*_y}\ot\ket*{b_u}\bra*{b_w}
\label{eq:H1MH1-tlA-B}
\end{align}
Restricting the above operator to the subspace $\hlA\ot \mathcal{D}_B$, we have 
\begin{align}
    &\rho_{\tlA;B}^{T_1(1)}\hat{M}\rho_{\tlA ;B}^{T_1(1)}|_{\hlA\ot \mathcal{D}_B}\nonumber\\
    &=-\sum_{t\leq N,u>N,w>N,p,q,e,i,y}\frac{\alpha_i\alpha_y}{\lambda_B^t}A^p_{ie}A^q_{ey}B^p_{ut}B^q_{tw}\left(\lambda_B^t-\lambda_B^u\right)\left(\lambda_B^t-\lambda_B^w\right)\ket*{\tla^*_i}\bra*{\tla^*_y}\ot\ket*{b_u}\bra*{b_w}\\
    &=-\sum_{p,q}\left(\sum_{e,i,y}\alpha_i\alpha_yA^p_{ie}A^q_{ey}\ket*{\tla^*_i}\bra*{\tla^*_y}\right)\ot\left(\sum_{t\leq N,u>N,w>N}B^p_{ut}\lambda_B^tB^q_{tw}\ket*{b_u}\bra*{b_w}\right)
   \label{eq:H1MH1-tlA-B-restricted}
\end{align}

Using Eq.~\ref{eq:pt-tlA-B-density-2}, we express the second-order term in Eq.~\ref{eq:F-tlA-B-definition}:
\begin{align}
   &\rho_{\tlA;B}^{T_1(2)}|_{\hlA\ot \mathcal{D}_B}=\sum_{u>N,v>N,t\leq N,p,q,i,j,m} \alpha_i\alpha_{j}B^p_{ut}B^q_{tv}\nonumber\\
   &\qquad\qquad\left\{A^p_{mj}A^q_{im}\lambda_B^t-\frac{1}{2}A^q_{mj}A^p_{im}(\lambda_B^u+\lambda_B^v)\right\}\ket*{\tla^*_i}\bra*{\tla^*_j}\otimes\ket{b_u}\bra{b_v}\\
    &=\sum_{p,q}\left(\sum_{i,j,m} \alpha_i\alpha_{j}A^q_{im}A^p_{mj}\ket*{\tla^*_i}\bra*{\tla^*_j}\right)\otimes\left(\sum_{t\leq N,u>N,v>N}B^p_{ut}\lambda_B^tB^q_{tv}\ket{b_u}\bra{b_v}\right)
\label{eq:H2-tlA-B-restricted}
\end{align}
Therefore, we can obtain the final expression for $\hat{F}_{\tlA;B}$ in Eq.~\ref{eq:F-tlA-B-definition}:
\begin{align}
\hat{F}_{\tlA ;B}&:=\sum_{p,q}\left\{\sum_{i,j,m} \alpha_i\alpha_{j}\left(A^q_{im}A^p_{mj}-A^p_{im}A^q_{mj}\right)\ket*{\tla^*_i}\bra*{\tla^*_j}\right\}\nonumber\\
&\qquad\qquad\otimes\left(\sum_{t\leq N,u>N,v>N}B^p_{ut}\lambda_B^tB^q_{tv}\ket{b_u}\bra{b_v}\right)
\label{eq:F-tlA-B-Dirac}\\
&=\sum_{p,q}\hat{F}_{\tlA}^{p,q}\ot\hat{F}_B^{p,q}
\label{eq:F-tlA-B},
\end{align}
where $\hat{F}_B^{p,q}$ has already been defined in Eq.~\ref{eq:FAB-B-dirac} and \ref{eq:FAB-B-matrix}, and we make the following new definition:
\begin{align}
    \hat{F}^{p,q}_{\tlA}&:=\sum_{i,j,m} \alpha_i\alpha_{j}\left(A^q_{im}A^p_{mj}-A^p_{im}A^q_{mj}\right)\ket*{\tla^*_i}\bra*{\tla^*_j}\label{eq:FtlAB-A-dirac}\\
    &=\sum_{i,j,m}\alpha_i\alpha_j\ket*{\tla_i^*}\bra*{a_j^*}\left(\hat{A}^{p*}\ket*{a_m^*}\bra*{a_m^*}\hat{A}^{q*}-\hat{A}^{q*}\ket*{a_m^*}\bra*{a_m^*}\hat{A}^{p*}\right)\ket*{a_i^*}\bra*{\tla_j^*}\\
    &=\left\{\sum_{i,j}\left(\bra*{a_j}\hat{A}^{p}\hat{A}^{q}-\hat{A}^{q}\hat{A}^{p}\ket*{a_i}\right)\alpha_i\alpha_j\ket*{\tla_i}\bra*{\tla_j}\right\}^*\label{eq:FtlAB-A-matrix}.
\end{align}
Substituting Eq.~\ref{eq:F-tlA-B} to Eq.~\ref{eq:negativity-tlA-B-2}, we find the expression for $\mathcal{N}_{A;B}^{(2)}$, which will be discussed and analyzed in Sec.~\ref{sec:negativity-perturbed-tlA-B}. Eq.~\ref{eq:negativity-A-B-2} requires us to find all negative eigenvalues of $\hat{F}_{\tlA;B}$, for which an analytical expression is generally hard to find for an interaction Hamiltonian with multiple terms.

\subsection{Negativity for \texorpdfstring{$\tilde{A};A$}{A;A}}
\label{sec:negativity-tlA-A-calculation}

We here calculate the first and second-order perturbation terms for the negativity of $\tlA;A$ assuming ${\rm det}(\rho_A)\neq 0$ using formulas Eq.~\ref{eq:first-order-eigenvalue} and \ref{eq:second-order-eigenvalue} of the non-degenerate eigenvalue perturbation theory reviewed in Appendix~\ref{sec:NDSOMP}. Under the formalism introduced in Sec.~\ref{sec:PoN}, we find the first three perturbation terms of the partial transpose density matrix $\rho^{T_1}_{\tlA ;A}(t)$ at $t=t_0$. The zeroth-order term of the partial transpose density matrix has already been given in Eq.~\ref{eq:pt-tlA-A-density-0}, and its associated eigenvalues and eigenvectors are given in Eq.~\ref{eq:pt-tlA-A-eigenvalue-0-uv}-\ref{eq:pt-tlA-A-eigenvector-0-u}. The first and second order terms can be obtained by taking the partial transpose of Eq.~\ref{eq:tlA-A-density-1} and \ref{eq:tlA-A-density-2} derived in Appendix~\ref{sec:perturbation calculation density matrix}. We summarize the results below:
\begin{align}
    \rho_{\tlA; A}^{T_1(0)}&=\sum_{i,j}\alpha_i\alpha_{j}\ket*{\tla^*_j}\bra*{\tla^*_i}\ot\ket*{a_i}\bra*{a_j}\\
    \rho_{\tlA; A}^{T_1(1)}&=i\sum_{p}\Tr[\hat{B}^p\rho_B]\left[\sum_{i,j,l,k}\alpha_i\alpha_j\left(A^p_{il}\delta_{jk}-A^p_{kj}\delta_{il}\right)\ket*{\tla_i^*}\bra*{\tla_j^*}\ot\ket*{a_k}\bra*{a_l}\right]\label{eq:pt-tlA-A-density-1}\\
    \rho_{\tlA; A}^{T_1(2)}&=\sum_{p,q,i,j,k,l}\Tr[\hat{B}^p\rho_B\hat{B}^q]\bigg[A^p_{kj}\alpha_i\alpha_{j}A^q_{il}-\frac{1}{2}\alpha_i\alpha_{j}\nonumber\\
    &\qquad\qquad\left(\delta_{jk}\sum_fA^p_{fl}A^q_{if}+\delta_{il}\sum_f A^p_{fj}A^q_{kf}\right)\bigg]\ket*{\tla^*_i}\bra*{\tla^*_j}\ot\ket*{a_k}\bra*{a_l}.
     \label{eq:pt-tlA-A-density-2}
\end{align}

With the built-in assumption of ${\rm det}(\rho_A)\neq 0$ in our setup, we know that all eigenvalues of $\rho_{\tlA;A}^{T_1(0)}$ are non-zero as seen from Eq.~\ref{eq:pt-tlA-A-eigenvalue-0-uv} and \ref{eq:pt-tlA-A-eigenvalue-0-u}, and only perturbations of the negative zeroth-order eigenvalues $\lambda_{uv-,\tlA; A}^{T_1}=-\alpha_u\alpha_v$ will contribute to the perturbation of negativity. Therefore, any $f^{th}$ order perturbation to the the negativity of $\tlA;A$ is
\begin{equation}
    \mathcal{N}^{(f)}_{\tlA;A} = -\sum_{u<v}\lambda_{uv-,\tlA; A}^{T_1(f)}\label{eq:tlA-A-negativity-all-order}.
\end{equation}

We can use results of Appendix~\ref{sec:NDSOMP} to find the first and second-order perturbation of these negative eigenvalues of $\rho_{\tlA;A}^{T_1(0)}$. The perturbative calculation for the negativity of the $\tlA;A$ system is different from the cases for the $A;B$ and $\tlA;B$ systems in the previous Appendix~\ref{sec:negativity-A-B-calculation} and \ref{sec:negativity-tlA-B-calculation} where we need to use the degenerate eigenvalue perturbation theory in Appendix~\ref{sec:DSOMP}. When $\rho^{T_1(0)}$ has multiple zero eigenvalues as the cases of $\rho^{T_1(0)}_{A;B}$ and $\rho^{T_1(0)}_{\tlA;B}$, we need to distinguish whether each of the vanishing, degenerate eigenvalue has positive or negative perturbation, which have different impacts on the perturbation of negativity as discussed in Sec.~\ref{sec:PoN}. In the present case for the $\tlA;A$ system, we simply sum all perturbations to $\lambda_{uv-,\tlA; A}^{T_1}$ when ${\rm det}(\rho_A)\neq 0$. Even in the presence of some degenerate eigenvalue $\lambda_A^u=\lambda_A^v\neq 0$ for $u\neq v$, since we do not need to distinguish the perturbation of each degenerate eigenvalue (we sum over all their perturbative contributions), we can use the non-degenerate theory for calculation.

Using Eq.~\ref{eq:pt-tlA-A-density-1}, we first find
\begin{align}
\bra*{\tla^*_x}\ot\bra{a_y}\rho_{\tlA;A}^{T_1(1)}\ket*{\tla^*_u}\ot\ket{a_v}=i\sum_{p}\Tr[\hat{B}^p\rho_B]\alpha_x\alpha_u\left(A^p_{xv}\delta_{uy}-A^p_{yu}\delta_{xv}\right)
\label{eq:pt-tlA-A-eigen-1-help}
\end{align}

According to Eq.~\ref{eq:first-order-eigenvalue}, we can then find the first-order corrections:
\begin{align}
    &\lambda_{uv-,\tlA ;A}^{T_1(1)}=\bra*{\tla_u^*\alpha_v,-}\rho_{\tlA;A}^{T_1(1)}\ket*{\tla_u^*\alpha_v,-}\\
    &=\frac{1}{2}\left(\bra*{\tla_u^*}\ot\bra{a_v}-\bra*{\tla_v^*}\ot\bra{a_u}\right)\rho_{\tlA;A}^{T_1(1)}\left(\ket*{\tla_u^*}\ot\ket{a_v}-\ket*{\tla_v^*}\ot\ket{a_u}\right)\\
    &=\frac{1}{2}\sum_{p}\Tr[\hat{B}^p\rho_B]\left(0-\left(A_{vv}^p-A_{uu}^p\right)-\left(A_{uu}^p-A_{vv}^p\right)+0\right)=0
    \label{eq:eigenvalue-tlA-A-1}
\end{align}
Since the first-order corrections to all the negative eigenvalues of $\rho_{\tlA;A}^{T_1}$ vanish, then
\begin{align}
 \mathcal{N}_{\tlA;A}^{T_1(1)}=0
\label{eq:negativity-tlA-A-1}
\end{align}

For the second-order correction to the  eigenvalues $\lambda_{uv-,\tlA; A}^{T_1}$, we first consider the term involving the first-order correction of the Hamiltonian in Eq.~\ref{eq:second-order-eigenvalue}: $-\sum_{k\neq n}H_1^{kn}H_1^{nk}/(E_k^{(0)}-E_n^{(0)})$. Here $n$ represents $\ket*{\tla^*_u\alpha_v,-}$ defined in Eq.~\ref{eq:pt-tlA-A-eigenvector-0-u-v}, and $k$ represents all the eigenvectors of $\rho_{\tlA;A}^{T_1(0)}$ which are not $\ket*{\tla^*_u\alpha_v,-}$. We need to consider both types of eigenvectors of $\rho_{\tlA ;A}^{T_1(0)}$ shown in Eq.~\ref{eq:pt-tlA-A-eigenvector-0-u-v} and \ref{eq:pt-tlA-A-eigenvector-0-u}. For the unentangled eigenvector (tracked by single index) in Eq.~\ref{eq:pt-tlA-A-eigenvector-0-u},  we use Eq.~\ref{eq:pt-tlA-A-eigen-1-help} for the calculation:
\begin{align}
   &H_1^{x;uv-}=\bra*{\tla^*_x\alpha_{x}}\rho_{\tlA;A}^{T_1(1)}\ket*{\tla_u\alpha_{v},-}\\
   &=\frac{\sqrt{2}}{2}\bra*{\tla_x}\ot\bra{a_x}\rho_{\tlA;A}^{T_1(1)}\left(\ket*{\tla_u}\ot\ket{a_v}-\ket*{\tla_v}\ot\ket{a_u}\right)\\
   &=i\frac{\sqrt{2}}{2}\sum_{p}\Tr[\hat{B}^p\rho_B]\Big[\alpha_x\alpha_u\left(A^p_{xv}\delta_{ux}-A^p_{xu}\delta_{xv}\right)-\alpha_x\alpha_v\left(A^p_{xu}\delta_{vx}-A^p_{xv}\delta_{xu}\right)\Big]
\end{align}
We see that $H_1^{x;uv-}$ term is only non-zero when $x=u$ or $v$. Therefore,
\begin{align}
H_1^{u;uv-}&=i\frac{\sqrt{2}}{2}\sum_{p}\Tr[\hat{B}^p\rho_B]\alpha_u(\alpha_u+\alpha_v)A^p_{uv}\\
 \frac{H_1^{u;uv-}H_1^{uv-;u}}{\lambda_{u,\tlA; A}^{T_1(0)}-\lambda_{uv-,\tlA; A}^{T_1(0)}}&=\frac{1}{2}\frac{\sum_{p,q}\Tr[\hat{B}^p\rho_B]\Tr[\hat{B}^q\rho_B]\left(\alpha_u(\alpha_u+\alpha_v)\right)^2A^p_{uv}A^q_{vu}}{\alpha_u^2-(-\alpha_u\alpha_v)}\\
 &=\frac{1}{2}\sum_{p,q}\Tr[\hat{B}^p\rho_B]\Tr[\hat{B}^q\rho_B]A^p_{uv}A^q_{vu}\alpha_u(\alpha_u+\alpha_v)
 \label{eq:pt-tlA-A-eigen-2-1-u}
\end{align}
Similarly,
\begin{align}
   \frac{H_1^{v;uv-}H_1^{uv-;v}}{\lambda_{v,\tlA; A}^{T_1(0)}-\lambda_{uv\pm,\tlA; A}^{T_1(0)}}&=\frac{1}{2}\sum_{p,q}\Tr[\hat{B}^p\rho_B]\Tr[\hat{B}^q\rho_B]A^p_{vu}A^q_{uv}\alpha_v(\alpha_u+\alpha_v)
    \label{eq:pt-tlA-A-eigen-2-1-v}
\end{align}
For the entangled eigenvectors (tracked by double indices) of $\rho_{\tlA ;A}^{T_1(0)}$ in Eq.~\ref{eq:pt-tlA-A-eigenvector-0-u-v}, we also make use of  Eq.~\ref{eq:pt-tlA-A-eigen-1-help}:
\begin{align}
    &H_1^{xy\pm;uv-}=\bra*{\tla^*_x\alpha_{y},\pm}\rho_{\tlA;A}^{T_1(1)}\ket*{\tla^*_{u}\alpha_{v},-}\\
    &=\frac{1}{2}\left(\bra*{\tla_x}\ot\bra{a_y}\pm\bra*{\tla_y}\ot\bra{a_x}\right)\rho_{\tlA;A}^{T_1(1)}\left(\ket*{\tla_u}\ot\ket{a_v}-\ket*{\tla_v}\ot\ket{a_u}\right)\\
    &=\frac{i}{2}\sum_{p}\Tr[\hat{B}^p\rho_B]\Big[\alpha_x\alpha_u\left(A^p_{xv}\delta_{uy}-A^p_{yu}\delta_{xv}\right)-\alpha_x\alpha_v\left(A^p_{xu}\delta_{vy}-A^p_{yv}\delta_{xu}\right)\nonumber\\
    &\qquad\pm\alpha_y\alpha_u\left(A^p_{yv}\delta_{ux}-A^p_{xu}\delta_{yv}\right)\mp\alpha_y\alpha_v\left(A^p_{yu}\delta_{vx}-A^p_{xv}\delta_{yu}\right)\Big]\\
    &=\frac{i}{2}\sum_{p}\Tr[\hat{B}^p\rho_B]\Big[\left(\alpha_x\alpha_u\pm\alpha_y\alpha_v\right)\left(A^p_{xv}\delta_{uy}-A^p_{yu}\delta_{xv}\right)\nonumber\\
    &\qquad+(\alpha_x\alpha_v\pm\alpha_y\alpha_u)\left(A^p_{yv}\delta_{ux}-A^p_{xu}\delta_{yv}\right)\Big] \label{eq:pt-tlA-A-eigen-2-1-x-y},
\end{align}
where $x<y$ and $(x,y)\neq(u,v)$ when we take the minus sign. As seen in Eq.~\ref{eq:pt-tlA-A-eigen-2-1-x-y}, the Kronecker delta functions require at least one of $x$ and $y$ needs to take the value of $u$ or $v$ in order for $H_1^{xy\pm;uv-}$ to be non-zero. We next simplify the expressions of non-zero $H_1^{xy\pm;uv-}$ for different values of $x$ and $y$. When $(x,y)=(u,v)$, we can only take the plus sign, and
\begin{align}
    H_1^{uv+;uv-}&=i\sum_{p}\Tr[\hat{B}^p\rho_B]\alpha_u\alpha_v\left(A^p_{vv}-A^p_{uu}\right),\\
    \frac{H_1^{uv+;uv-}H_1^{uv-;uv+}}{\lambda^{T_1(0)}_{uv+,\tlA; A}-\lambda^{T_1(0)}_{uv-,\tlA; A}}&=\frac{1}{2}\sum_{p,q}\Tr[\hat{B}^p\rho_B]\Tr[\hat{B}^q\rho_B]\left(A^p_{vv}-A^p_{uu}\right)\left(A^q_{vv}-A^q_{uu}\right)\alpha_u\alpha_v\label{eq:pt-tlA-A-eigen-2-1-uv+}.
\end{align}
When $x=u$ while $y\neq v$ and $y>u$,
\begin{align}
    H_1^{uy\pm;uv-}&=\frac{i}{2}\sum_{p}\Tr[\hat{B}^p\rho_B]\alpha_u(\alpha_v\pm\alpha_y)A^p_{yv}\label{eq:pt-tlA-A-eigen-2-1-uy-H}\\
    \frac{H_1^{uy\pm;uv-}H_1^{uv-;uy\pm}}{\lambda^{T_1(0)}_{uy\pm,\tlA; A}-\lambda^{T_1(0)}_{uv-,\tlA; A}}&=\frac{1}{4}\sum_{p,q}\Tr[\hat{B}^p\rho_B]\Tr[\hat{B}^q\rho_B]\alpha_u(\alpha_v\pm\alpha_y)A^p_{yv}A^q_{vy}
    \label{eq:pt-tlA-A-eigen-2-1-uy}
\end{align}
The evaluations and results of the other three cases (when $x=v$ while $y>v$, when $y=u$ while $x<u$, and when $y=v$ while $x<v$ and $x\neq u$) are very similar to the above Eq.~\ref{eq:pt-tlA-A-eigen-2-1-uy-H} and \ref{eq:pt-tlA-A-eigen-2-1-uy}. Combining Eq.~\ref{eq:pt-tlA-A-eigen-2-1-u}, \ref{eq:pt-tlA-A-eigen-2-1-v}, \ref{eq:pt-tlA-A-eigen-2-1-uv+}, \ref{eq:pt-tlA-A-eigen-2-1-uy}, and the three other analogues to Eq.~\ref{eq:pt-tlA-A-eigen-2-1-uy}, we obtain the expression for the second term in Eq.~\ref{eq:second-order-eigenvalue}:
\begin{align}
&-\sum_{k\neq n}\frac{H_1^{kn}H_1^{nk}}{E_k^{(0)}-E_n^{(0)}}=-\frac{1}{4}\sum_{p,q}\Tr[\hat{B}^p\rho_B]\Tr[\hat{B}^q\rho_B]\Big\{2\alpha_u\alpha_v\left(A^p_{vv}-A^p_{uu}\right)\left(A^q_{vv}-A^q_{uu}\right)\nonumber\\
&\qquad+2A^p_{uv}A^q_{vu}\alpha_u(\alpha_u+\alpha_v)+2A^p_{vu}A^q_{uv}\alpha_v(\alpha_u+\alpha_v)\nonumber\\
&\qquad+\sum_{y>u,y\neq v}\alpha_u(\alpha_v\pm\alpha_y)A^p_{yv}A^q_{vy}+\sum_{y>v}\alpha_v\left(\alpha_u\pm\alpha_y\right)A^p_{yu}A^q_{uy}\nonumber\\
&\qquad+\sum_{x<u}\alpha_u(\alpha_v\pm\alpha_x)A^p_{xv}A^q_{vx}+\sum_{x<v,x\neq u}\alpha_v(\alpha_u\pm\alpha_x)A^p_{xu}A^{q}_{ux}\Big\}\\
&=-\frac{1}{2}\sum_{p,q}\Tr[\hat{B}^p\rho_B]\Tr[\hat{B}^q\rho_B]\Big\{\alpha_u\alpha_v\left(A^p_{vv}-A^p_{uu}\right)\left(A^q_{vv}-A^q_{uu}\right)+A^p_{uv}A^q_{vu}\alpha_u(\alpha_u+\alpha_v)\nonumber\\
&\qquad +A^p_{vu}A^q_{uv}\alpha_v(\alpha_u+\alpha_v)+\sum_{y\neq u,y\neq v}\alpha_u\alpha_v(A^p_{yv}A^q_{vy}+A^p_{yu}A^q_{uy})\Big\}\\
&=-\frac{1}{2}\sum_{p,q}\Tr[\hat{B}^p\rho_B]\Tr[\hat{B}^q\rho_B]\Big\{\left(\lambda_A^uA^p_{uv}A^q_{vu}+\lambda_A^vA^p_{vu}A^q_{uv}-\alpha_u\alpha_vA^p_{vv}A^q_{uu}-\alpha_u\alpha_vA^p_{uu}A^q_{vv}\right)\nonumber\\
&\qquad+\sum_{y}\alpha_u\alpha_v(A^p_{yv}A^q_{vy}+A^p_{yu}A^q_{uy})\Big\}
\label{eq:pt-tlA-A-eigen-2-1}
\end{align}

We next determine the first term $\bra{n^{(0)}}H_2\ket{n^{(0)}}$ in Eq.~\ref{eq:second-order-eigenvalue} where $H_2=\rho_{\tlA;A}^{T_1(2)}$ is given in Eq.~\ref{eq:pt-tlA-A-density-2} and $\ket*{n^{(0)}}=\ket*{\tla^*_ua_v,-}$ ($u<v$) are the eigenvectors corresponding to negative eigenvalues of $\rho_{\tlA;A}^{T_1(0)}$. Using Eq.~\ref{eq:pt-tlA-A-density-2}, we can find the following expressions:
\begin{align}
\bra*{\tla^*_u}\ot\bra*{\alpha_v}\rho_{\tlA;A}^{T_1(2)}\ket*{\tla^*_u}\ot\ket*{\alpha_v}&=\sum_{p,q}\Tr[\hat{B}^p\rho_B\hat{B}^q]A^p_{vu}\lambda_A^uA^q_{uv}\label{eq:help-tlA-A-eigen-2-2-uv-uv}\\
\bra*{\tla^*_v}\ot\bra*{\alpha_u}\rho_{\tlA;A}^{T_1(2)}\ket*{\tla^*_u}\ot\ket*{\alpha_v}
    &=\sum_{p,q}\Tr[\hat{B}^p\rho_B\hat{B}^q]\bigg[A^p_{uu}\alpha_u\alpha_{v}A^q_{vv}-\frac{1}{2}\alpha_u\alpha_{v}\nonumber\\
    &\qquad\left(\sum_fA^p_{fv}A^q_{vf}+\sum_fA^p_{fu}A^q_{uf}\right)\bigg]\label{eq:help-tlA-A-eigen-2-2-vu-uv}
\end{align}
Using the above Eq.~\ref{eq:help-tlA-A-eigen-2-2-uv-uv} and \ref{eq:help-tlA-A-eigen-2-2-vu-uv}, we have
\begin{align}
    &\bra*{n^{(0)}}H_2\ket*{n^{(0)}}=\bra*{\tla^*_u\alpha_{v},-}\rho_{\tlA;A}^{T_1(2)}\ket*{\tla^*_u\alpha_{v},-}\\
    &=\frac{1}{2}\left(\bra*{\tla^*_u}\ot\bra{a_v}-\bra*{\tla^*_v}\ot\bra{a_u}\right)\rho_{\tlA;A}^{T_1(2)}\left(\ket*{\tla^*_u}\ot\ket{a_v}-\ket*{\tla^*_v}\ot\ket{a_u}\right)\\
    &=\frac{1}{2}\sum_{p,q}\Bigg\{\Tr[\hat{B}^p\rho_B\hat{B}^q](A^p_{vu}\lambda_A^uA^q_{uv}+A^p_{uv}\lambda_A^vA^q_{vu}-A^p_{vv}\alpha_u\alpha_{v}A^q_{uu}-A^p_{uu}\alpha_u\alpha_{v}A^q_{vv})\qquad\nonumber\\
    &\qquad+\Tr[\hat{B}^q\rho_B\hat{B}^p]\alpha_u\alpha_{v}\left(\sum_fA^p_{vf}A^q_{fv}+\sum_fA^p_{uf}A^q_{fu}\right)\Bigg\}\\
    &=\frac{1}{2}\Tr[\hat{B}^p\rho_B\hat{B}^q]\sum_{p,q}\Bigg\{(A^p_{vu}\lambda_A^uA^q_{uv}+A^p_{uv}\lambda_A^vA^q_{vu}-A^p_{vv}\alpha_u\alpha_{v}A^q_{uu}-A^p_{uu}\alpha_u\alpha_{v}A^q_{vv})\qquad\nonumber\\
    &\qquad+\alpha_u\alpha_{v}\left(\sum_fA^p_{fv}A^q_{vf}+\sum_fA^p_{fu}A^q_{uf}\right)\Bigg\}
    \label{eq:pt-tlA-A-eigen-2-2}
\end{align}
Combining Eq.~\ref{eq:pt-tlA-A-eigen-2-1} and \ref{eq:pt-tlA-A-eigen-2-2}, we obtain the second-order correction to any negative eigenvalue $-\alpha_u\alpha_v$ of $\rho_{\tlA;A}^{T_1(0)}$:
\begin{align}
\lambda_{uv-,\tlA; A}^{T_1(2)}&=E_n^{(2)}=\bra*{n^{(0)}}H_2\ket*{n^{(0)}}-\sum_{k\neq n}\frac{H_1^{kn}H_1^{nk}}{E_k^{(0)}-E_n^{(0)}}\\
&=\frac{1}{2}\sum_{p,q}\Big(\Tr[\hat{B}^p\rho_B\hat{B}^q]-\Tr[\hat{B}^p\rho_B]\Tr[\hat{B}^q\rho_B]\Big)\Big\{A^p_{vu}\lambda_A^uA^q_{uv}+A^p_{uv}\lambda_A^vA^q_{vu}\nonumber\\
&\qquad-A^p_{vv}\alpha_u\alpha_{v}A^q_{uu}-A^p_{uu}\alpha_u\alpha_{v}A^q_{vv}+\alpha_u\alpha_v\sum_{y}(A^p_{yv}A^q_{vy}+A^p_{yu}A^q_{uy})\Big\}
\label{eq:pt-tlA-A-eigen-2}
\end{align}
The unsymmetrized covariance between $\hat{B}^p$ and $\hat{B^q}$ operators under the density matrix $\rho_B$ is
\begin{align}
    {\rm ucov}(\hat{B}^p,\hat{B}^q)=\Tr[\hat{B}^p\rho_B\hat{B}^q]-\Tr[\hat{B}^p\rho_B]\Tr[\hat{B}^q\rho_B]
    \label{eq:ucov-Bp-Bq}
\end{align}
Using Eq.~\ref{eq:tlA-A-negativity-all-order} and \ref{eq:pt-tlA-A-eigen-2}, we finally find the second-order perturbation to the negativity of the system $\tlA;A$:
\begin{align}
&\mathcal{N}_{\tlA;A}^{(2)}=-\sum_{u<v}\lambda_{uv-,\tlA; A}^{T_1(2)}\\
&=-\frac{1}{2}\sum_{p,q}{\rm ucov}(\hat{B}^p,\hat{B}^q)\sum_{u<v}\Big[A^p_{vu}\lambda_A^uA^q_{uv}+A^p_{uv}\lambda_A^vA^q_{vu}\nonumber\\
    &\qquad-A^p_{vv}\alpha_u\alpha_{v}A^q_{uu}-A^p_{uu}\alpha_u\alpha_{v}A^q_{vv}+\alpha_u\alpha_v\sum_{y}(A^p_{yv}A^q_{vy}+A^p_{yu}A^q_{uy})\Big]\\
&=-\frac{1}{4}\sum_{p,q}{\rm ucov}(\hat{B}^p,\hat{B}^q)\Bigg\{\sum_{u,v}\Big[A^p_{vu}\lambda_A^uA^q_{uv}+A^p_{uv}\lambda_A^vA^q_{vu}-A^p_{vv}\alpha_u\alpha_{v}A^q_{uu}\nonumber\\
    &\qquad-A^p_{uu}\alpha_u\alpha_{v}A^q_{vv}+\alpha_u\alpha_v\sum_{y}(A^p_{yv}A^q_{vy}+A^p_{yu}A^q_{uy})\Big]-\sum_{u,y}2\lambda_A^uA^p_{yu}A^q_{uy}\Bigg\}\\
&=-\frac{1}{4}\sum_{p,q}{\rm ucov}(\hat{B}^p,\hat{B}^q)\sum_{u,v}\alpha_u\alpha_v\left[\sum_{y}(A^p_{yv}A^q_{vy}+A^p_{yu}A^q_{uy})-A^p_{vv}A^q_{uu}-A^p_{uu}A^q_{vv}\right]\label{eq:negativity-tlA-A-2-dirac-symmetric}\\
&=-\frac{1}{2}\sum_{p,q}{\rm ucov}(\hat{B}^p,\hat{B}^q)\sum_{u,v}\alpha_u\alpha_v\left[\sum_{y}A^p_{yv}A^q_{vy}-A^p_{vv}A^q_{uu}\right]\label{eq:negativity-tlA-A-2-dirac-simple}\\
&=-\frac{1}{2}\sum_{p,q}{\rm ucov}(\hat{B}^p,\hat{B}^q)\left[\left(\sum_{u}\alpha_u\right)\left(\sum_{v,y}\alpha_vA^p_{yv}A^q_{vy}\right)-\left(\sum_{v}\alpha_vA^p_{vv}\right)\left(\sum_{u}\alpha_uA^q_{uu}\right)\right]\\
&=-\frac{1}{2}\sum_{p,q}{\rm ucov}(\hat{B}^p,\hat{B}^q)\left(\Tr[\sqrt{\rho_A}]\Tr[\hat{A}^p\sqrt{\rho_A}\hat{A}^q]-\Tr[\sqrt{\rho_A}\hat{A}^p]\Tr[\sqrt{\rho_A}\hat{A}^q]\right),
\label{eq:negativity-tlA-A-2}
\end{align}
where $\sqrt{\rho_A}=\sum_{u}\alpha_u\ket*{a_u}\bra*{a_u}=\sum_{u}\sqrt{\lambda_A^u}\ket*{a_u}\bra*{a_u}$. Since the first-order perturbation of negativity between $\tlA$ and $A$ vanishes in Eq.~\ref{eq:negativity-tlA-A-1}, the second-order term given in Eq.~\ref{eq:negativity-tlA-A-2} is the leading-order perturbation that characterizes the initial change of the entanglement between $\tlA$ and $A$. We will analyze Eq.~\ref{eq:negativity-tlA-A-2} in Sec.~\ref{sec:negativity-perturbed-tlA-A}. 

\section{Numerical Examples}
We list all the specific numerical examples of density matrices and Hamiltonians used to illustrate entanglement dynamics and verify our perturbative calculation in this work. All matrices are given in the basis of density matrices $\rho_A$ and $\rho_B$. We use either a system of qutrits or a system of qubits for our numerical calculations.

\subsection{Qutrits}\label{sec:qutrit}
\begin{center}
\begin{minipage}[b]{.4\textwidth}
\vspace{-0.2cm}
\begin{equation}
   \rho_A = \begin{pmatrix} 0.6 & 0 & 0\\ 0 & 0.3 & 0\\ 0 & 0 & 0.1 \end{pmatrix}
   \label{eq:A-qutrit}
\end{equation}
\end{minipage}
\hfill
\begin{minipage}[b]{.4\textwidth}
\vspace{-\baselineskip}
\begin{equation}
\rho_B = \begin{pmatrix} 0.25 & 0 & 0\\ 0 & 0.4 & 0 \\ 0 & 0 & 0.35
\end{pmatrix}
\label{eq:qutrit-B-mixed}
\end{equation}
\end{minipage}
\hspace*{1cm}
\end{center}

\begin{center}
\begin{minipage}[b]{.45\textwidth}
\vspace{-\baselineskip}
\begin{equation}
\hat{A}_1 = \begin{pmatrix} 2 & 1+1i & 0.5\\1-1i & 3 & 4+2i \\0.5 & 4-2i &1
\end{pmatrix}
\label{eq:A_1-qutrit}
\end{equation}
\end{minipage}
\hfill
\begin{minipage}[b]{.45\textwidth}
\begin{equation}
\hat{B}_1 = \begin{pmatrix} 3 & 2 & 0 \\ 2 & 1 & 1 \\ 0 & 1 & 4 \end{pmatrix}
\label{eq:B_1-qutrit}
\end{equation}
\end{minipage}
\hspace*{1cm}
\end{center}

\begin{center}
\begin{minipage}[b]{.45\textwidth}
\vspace{-\baselineskip}
\begin{equation}
\hat{A}_2 = \begin{pmatrix} 1 & 3 & -0.25i \\ 3 & 2 &0 \\ 0.25i & 0 & 3 \end{pmatrix}
\label{eq:A_2-qutrit}
\end{equation}
\end{minipage}
\hfill
\begin{minipage}[b]{.45\textwidth}
\begin{equation}
\hat{B}_2 = \begin{pmatrix} 0.8 & 2-1i & 1 \\ 2+1i & 1 & 2i \\ 1 & -2i & 2 \end{pmatrix}
\label{eq:B_2-qutrit}
\end{equation}
\end{minipage}
\hspace*{1cm}
\end{center}

\begin{center}
\begin{minipage}[b]{.45\textwidth}
\vspace{-\baselineskip}
\begin{equation}
\hat{C} = \begin{pmatrix} 1 & 1 & 3 \\ 1 & 0 & 2i \\ 3 & -2i & 0.5 \end{pmatrix}
\label{eq:C-qutrit}
\end{equation}
\end{minipage}
\hfill
\begin{minipage}[b]{.45\textwidth}
\begin{equation}
\hat{D} = \begin{pmatrix} 0.5 & 2+1i & 8+3i \\ 2-1i & 1.5 & -4 \\ 8-3i & -4 & 2.2 \end{pmatrix}
\label{eq:D-qutrit}
\end{equation}
\end{minipage}
\hspace*{1cm}
\end{center}
The above matrices are some generic examples of density states and Hamiltonians for a qutrit. 

\subsection{Qubits}\label{sec:qubit}

\begin{center}
\begin{minipage}[b]{.4\textwidth}
\vspace{-0.2cm}
\begin{equation}
   \rho_A = \begin{pmatrix} 0.8 & 0\\ 0 & 0.2 \end{pmatrix}
   \label{eq:A-qubit}
\end{equation}
\end{minipage}
\hfill
\begin{minipage}[b]{.4\textwidth}
\vspace{-\baselineskip}
\begin{equation}
\rho_B = \begin{pmatrix} 0.6 & 0 \\ 0 & 0.4
\end{pmatrix}
\label{eq:qubit-B-mixed}
\end{equation}
\end{minipage}
\hspace*{1cm}
\end{center}

\begin{equation}
\hat{F} = \begin{pmatrix} 0 & 0.5+0.5i \\ 0.5-0.5i & 1
\end{pmatrix}
\label{eq:free-B-qubit}
\end{equation}

\begin{equation}
\hat{H}_{\rm SWAP} = \hat{\sigma_x}\ot\hat{\sigma_x}+\hat{\sigma_y}\ot\hat{\sigma_y}+\hat{\sigma_z}\ot\hat{\sigma_z}
\label{eq:H-SWAP-qubit}
\end{equation}
$\hat{\sigma}_{x,y,z}$ denote the three Pauli matrices. $\hat{H}_{\rm SWAP}$ in Eq.~\ref{eq:H-SWAP-qubit} is the Heisenberg Hamiltonian that can generate the SWAP gate and interchange the states of two qubits at $t=\frac{\pi}{4}$ \cite{18KuzmakHeisenbergSWAP}. 

\subsection{Verification of Perturbative Calculations}\label{sec:verification}
We verify our perturbative calculations of the negativity for systems $A;B$, $\tlA;B$, and $\tlA;A$ (upto the second order) performed in Appendix.~\ref{sec:appendix-perturb-negativity}. For all three plots, we choose both $A$ and $B$ as qutrits: $\rho_A$ is given in Eq.~\ref{eq:A-qutrit}, and $\rho_B$ is assumed to be pure. As listed in the legend, we choose different forms of interaction Hamiltonians, of which the component matrices are given in Eq.~\ref{eq:A_1-qutrit}-\ref{eq:D-qutrit} in Appendix~\ref{sec:qutrit}. The captions in the legend are ordered as the lines in the plot (from the top to the bottom). Based on the plots below, we confirm that our perturbative expressions summarized and discussed in Sec.~\ref{sec:pure, perturbative} and \ref{sec:negativity-perturbed-tlA-A} are indeed consistent with the numerical calculations at the onset.

\begin{figure}[htbp]
\centerline{\includegraphics[width=0.6\hsize]{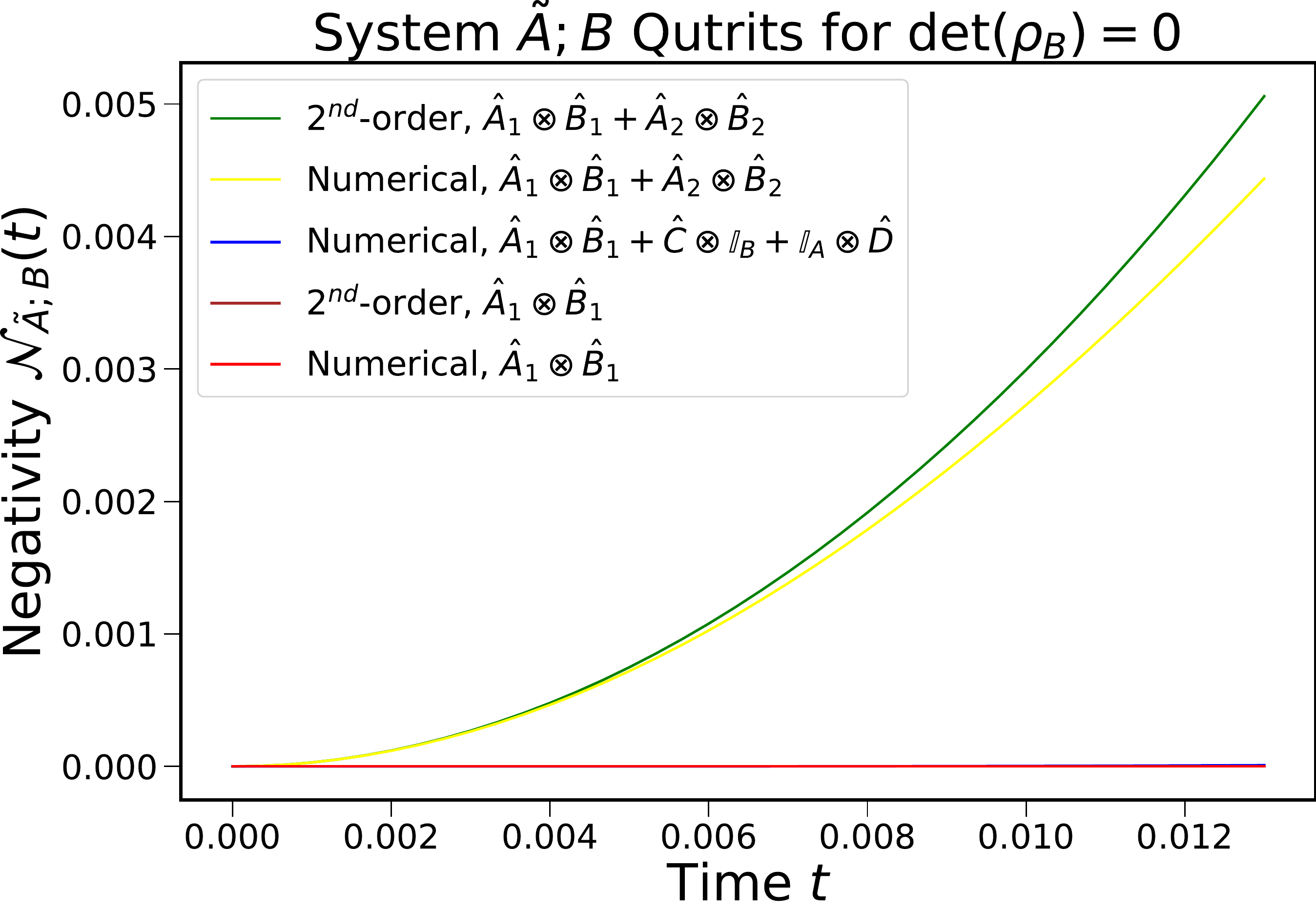}}
\caption{The dynamics of negativity for $\tlA; B$ assuming $\det(\rho_B)=0$ under our setup. We plot both the numerical calculations and the perturbative results to the $2^{nd}$-order using Eq.~\ref{eq:negativity-tlA-B-1-summary}-\ref{eq:FtlAB-tlA-matrix-summary}. The red and blue (third and fifth) curves mostly overlap, which suggests free Hamiltonians $\hat{C}$ and $\hat{D}$ do no contribute to $\mathcal{N}_{\tlA;B}^{(2)}$ and confirms our results in Sec.~\ref{sec:free-negativity-tlA-B}. The red and brown (last two) curves vanish, which confirms our result in Sec.~\ref{sec:pure-pf-tlAB-unentangled-perturbed} that $\hat{H}_{{\rm tot}}=\hat{A}\ot\hat{B}$ can not entangle $\tlA$ and $B$ at the onset.}
\label{fig:pure_qutrit_tlAB_perturb_initial}
\end{figure}

\begin{figure}[htbp!]
\centerline{\includegraphics[width=0.6\hsize]{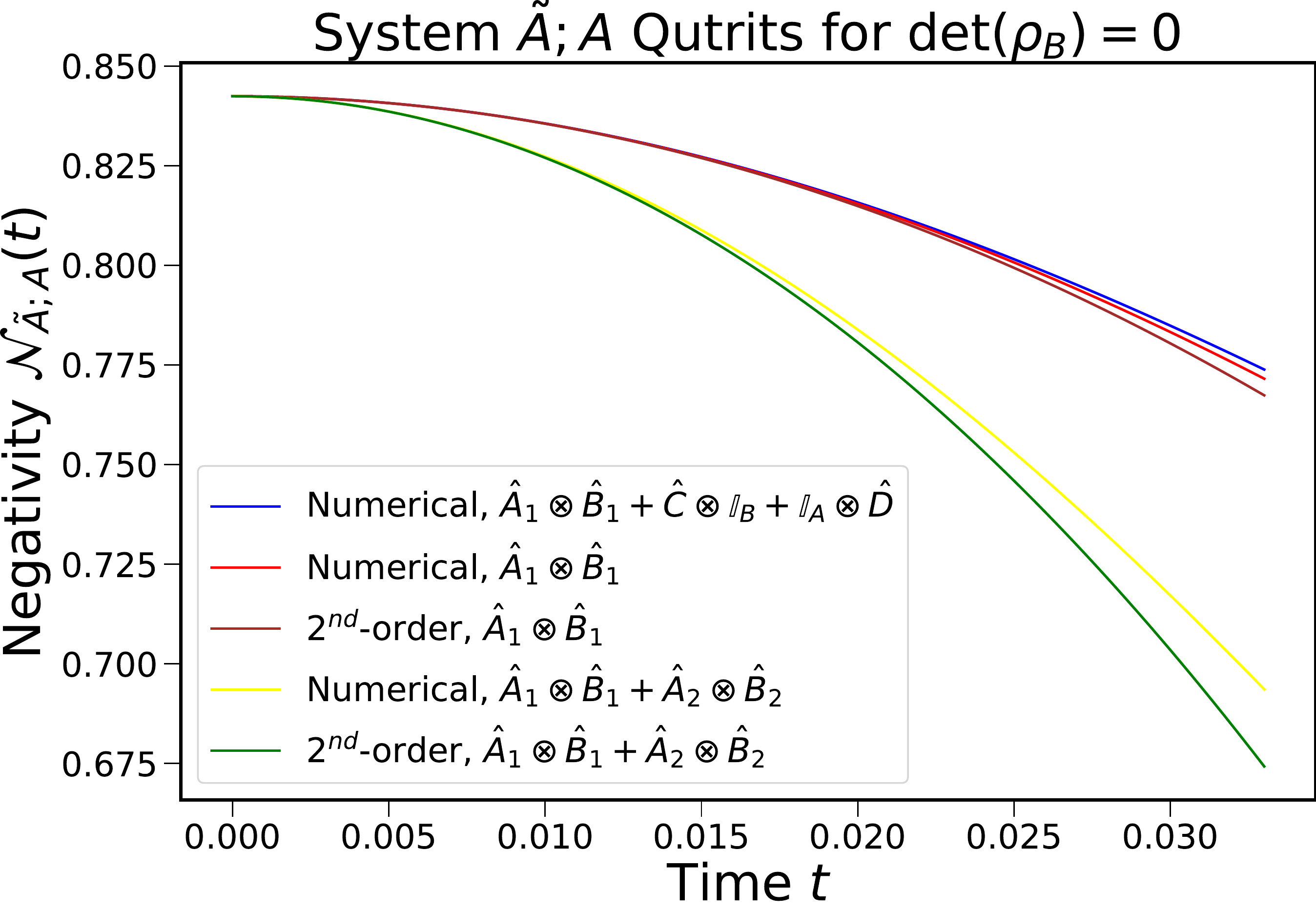}}
\caption{The dynamics of negativity for the $\tlA;A$ system assuming $\det(\rho_A)\neq 0$ under our setup. We plot both the numerical calculations and the perturbative results to the $2^{nd}$-order using Eq.~\ref{eq:negativity-tlA-A-2-summary}. The red and blue (first two) curves mostly overlap, which suggests free Hamiltonians $\hat{C}$ and $\hat{D}$ do no contribute to the negativity vulnerability $\mathcal{N}_{\tlA;A}^{(2)}$ at the onset, confirming our results in Sec.~\ref{sec:free-negativity-tlA-A}.}
\label{fig:pure_qutrit_tlAA_perturb_initial}
\end{figure}

\end{document}